\documentclass[11pt]{article}
\usepackage[a4paper,margin=2.2cm]{geometry}
\usepackage{graphicx}
\usepackage{epstopdf}
\usepackage{amsmath,amssymb,amsthm,amsfonts}
\usepackage{color,xcolor}
\usepackage{cite}
\usepackage{verbatim}
\usepackage{appendix}
\usepackage[pdftex]{hyperref}
\usepackage{mathrsfs}
\usepackage{dsfont}

\newtheorem{thm}{Theorem}[section]

\newtheorem{cor}[thm]{Corollary}
\newtheorem{lemma}[thm]{Lemma}
\newtheorem{lem}[thm]{Lemma}
\newtheorem{proposition}[thm]{Proposition}

\theoremstyle{definition}

\theoremstyle{remark}
\newtheorem{rem}{Remark}[section]

\numberwithin{equation}{section}

\newcommand{\R}{{\mathbb R}}

\newcommand{\om}{\Omega}

\newcommand{\mic}[1]{\textcolor{teal}{#1}}

\renewcommand{\geq}{\geqslant}
\renewcommand{\leq}{\leqslant}
\renewcommand{\ge}{\geqslant}
\renewcommand{\le}{\leqslant}

\newcommand{\xv}{\mathbf{x}}

\newcommand{\lf}{\left}
\newcommand{\ri}{\right}

\newcommand{\braketr}[2]{\lf\langle #1\lf|#2\ri. \ri\rangle}

\newcommand{\meanlrlr}[3]{\lf\langle #1\lf|#2\ri|#3\ri\rangle}
\newcommand{\beq}{\begin{equation}}
\newcommand{\eeq}{\end{equation}}
\newcommand{\bdm}{\begin{displaymath}}
\newcommand{\edm}{\end{displaymath}}
\newcommand{\bdn}{\begin{eqnarray}}
\newcommand{\edn}{\end{eqnarray}}
\newcommand{\bay}{\begin{array}{c}}
\newcommand{\eay}{\end{array}}
\newcommand{\ben}{\begin{enumerate}}
\newcommand{\een}{\end{enumerate}}
\newcommand{\beqn}{\begin{eqnarray}}
\newcommand{\eeqn}{\end{eqnarray}}
\newcommand{\bml}[1]{\begin{multline} #1 \end{multline}}
\newcommand{\bmln}[1]{\begin{multline*} #1 \end{multline*}}
\newcommand{\OO}{\mathcal{O}}
\newcommand{\N}{\mathbb{N}}

\newcommand{\energy}{\mathcal{F}_{a,\om}}
\newcommand{\enom}{\mathcal{F}_{a_{\om}, \om}}
\newcommand{\gse}{E_{a,\Omega}}
\newcommand{\gseom}{E_{a_{\om},\Omega}}
\newcommand{\gs}{u_{a, \Omega}}
\newcommand{\gsom}{u_{\Omega}}
\newcommand{\domain}{\mathscr{H}}
\newcommand{\form}{\mathfrak{q}_{\om,\nu}}

\begin{document}
\title{From Giant Vortex States to Bright Solitons in Rapidly Rotating Bose--Einstein Condensates with Attractive Interactions}
\author{
Michele Correggi\thanks{Dipartimento di Matematica, Politecnico di Milano, P.zza Leonardo da Vinci, 32, 20133, Milano, Italy.  Email:  michele.correggi@gmail.com},\ \
Yujin Guo\thanks{School of Mathematics and Statistics,  Key Laboratory of Nonlinear Analysis $\&$ Applications (Ministry of Education), Central China Normal University, Wuhan 430079, P. R. China. Email: yguo@ccnu.edu.cn. },\ \
Yong Luo\thanks{School of Mathematics and Statistics, and Hubei Key Laboratory of Mathematical Sciences,  Central China Normal University, Wuhan 430079, P. R. China. Email: yluo@ccnu.edu.cn.}}
\date{}
\maketitle

\begin{abstract}
We study the ground state of a Bose--Einstein condensate (BEC)
with attractive interaction of strength \(a_\Omega\), confined by a
two-dimensional anharmonic trap rotating at angular velocity \(\om\).
We prove the occurrence of a phase transition, as
\(\om\to\infty\), from an annular giant vortex state to an angularly localized bright soliton.  More precisely, we distinguish three regimes according to the relative sizes of the radial and angular concentration scales.
In the weak coupling regime \(a_\om=\OO({\om}^{-1/2})\),
the limiting angular profile is governed by a variational problem on the
unit circle showing rotational symmetry breaking:  below a given sharp threshold, the
condensate remains uniformly distributed in the annulus and has a
giant vortex profile with a vortex-free bulk and macroscopic circulation;
above it, rotational symmetry is broken.  In the intermediate coupling regime
\(\om^{-1/2}\ll a_\om\ll1\), the condensate localizes at a point on the unit circle with radial and angular widths of order
\(\om^{-1/2}\) and
\((a_\om\sqrt{\om})^{-1}\), respectively.
After rescaling, the limiting profile factorizes into the ground state of the
one-dimensional harmonic oscillator in the radial variable and a
one-dimensional bright soliton in the angular variable.  In the regime
\(a_\om= a\in(0,a^*)\), corresponding to stable attraction below the
collapse threshold \(a^*\), both the radial and angular widths
are of order \(\om^{-1/2}\); consequently, the limiting profile is no longer a product but rather a genuinely two-dimensional localized state.
\end{abstract}

\section{Introduction}

The physical phenomenon of superfluidity is responsible for some of the most striking manifestations of quantum mechanics at the mesoscopic or macroscopic level \cite{Leggett}. Such a quantum phenomenon often occurs together with Bose-Einstein condensation in cold interacting Bose gases \cite{Bloch2008,PethickSmith,PitaevskiiStringari}, although it is known that the two phase transitions are not directly linked to one another. The experimental realization of Bose--Einstein condensates (BECs) in dilute atomic gases \cite{Anderson1995,DavisEtAl1995} has provided a highly controllable setting for testing superfluidity and its macroscopic features, typically by setting BECs under rotation: any superfluid can indeed store angular momentum only through the nucleation of quantized vortices, i.e., isolated singularities of the phase of the condensate wave function, where the condensate density is depleted \cite{Fetter2009}. The occurrence of such topological defects has also been observed in the experiments \cite{Madison2000}, as well as their spectacular arrangement in periodic regular structures (vortex lattices) \cite{AboShaeer2001}, and the emergence of a quantum Hall regime for angular velocities close to the harmonic trap frequency \cite{Schweikhard2004}.

For all the above experimental observations, the sign of the interaction potential plays a crucial role, since stability of the condensate can be achieved for all values of the parameters only for repulsive systems. On the opposite, attractive BECs collapse above a finite critical particle number \cite{Bradley1997,Sackett1999}. When the attraction combines with the imposed rotation or a global quantization of the circulation, new phenomena may occur \cite{Adi01,Adi02,CaCl06,KJB04}, depending on the interaction strength and confinement, such as vortex splitting or vortex state instability \cite{SaUe02,SaUe04} and center-of-mass rotation \cite{Lundh2004}. Even more intriguing is the transition from vortex states to bright (Townes) solitons predicted for condensates confined to annular traps \cite{KSU03} and observed experimentally when the interaction undergoes a fast change from repulsion to attraction \cite{BZTT+25}, which is the phase transition we are analytically deriving in this work. In this last paper, indeed, in an experiment with Cesium atoms, it is observed that a sudden change of the interatomic interaction from repulsive to attractive is responsible for the formation of an annular vortex state. Such a state, however, shows angular instabilities leading for stronger attraction to the fragmentation of the wave function into different components with the typical Townes soliton behavior. These features are also confirmed by numerical experiments in the framework of the Gross-Pitaevskii (GP) theory.

In the theoretical physics and mathematical physics literature, many works have been devoted to the analysis of rotating BECs within the Gross-Pitaevskii theory. Most of the works however deal with the {\it repulsive} case, where a clear picture of different regimes and phase transitions as the angular velocity increases has been derived (see, e.g., \cite{Ka26,PeKa26} for a heuristic description of such phases). To summarize, for slow rotation, no effect is observed at the level of the (GP) minimizer, since a BEC can store angular momentum only in multiples of a fundamental quantum. As a first threshold is crossed, corresponding to the first critical velocity, quantized vortices are nucleated in the sample and their number and distribution are determined by the minimization of suitable renormalized interaction energies \cite{Ignat2006,Ignat2006b}. For angular velocities of the same order of the first critical velocity, the vortices cover a significant portion of the sample and their distribution is to leading order uniform but there are corrections due to the trap inhomogeneity \cite{CoRu13}. In anharmonic traps, as the rotational speed gets higher, the vortices occupy the whole sample and arrange in the famous Abrikosov lattice until a two subsequent thresholds are reached corresponding to the formation of a macroscopic hole and of a giant vortex state, where the bulk becomes annular and vortex free \cite{Correggi2011,Correggi2012, Rougerie2011}. Harmonically trapped BECs are in this respect special, because the angular velocity can not reach arbitrarily large values without destroying the trap, and when it approaches the harmonic trap frequency, a lowest Landau level regime sets in \cite{Aftalion2006b,NgRu22}.

The rigorous picture for attractive BECs is less complete and has
mainly been developed near the collapse regime. It is indeed very well known that beyond a certain threshold, i.e., for $ a > a^{*} $ for some given $ a^* $, the system collapses and the GP energy is unbounded from below \cite{GuSe14}. However, when $ a $ approaches the collapse threshold $ a^*$ from below and the system is set under rotation, some interesting phenomena may appear \cite{Dinh2024,GuoLuoPeng2023}. In fact, for slow rotation in radially symmetric harmonic traps, the collapse occurs at the trap center and any ground state remains vortex-free \cite{Lewin2018,Guo2020}. Vortex-free behavior also persists
for certain angular velocities
tending to infinity in anharmonic traps \cite{Guo2024}.
In the opposite regime of weak interaction, i.e., for $ |a| \ll 1 $, when the BEC is set under very rapid rotation, ground states may be given by giant vortex states with annular support \cite{Luo2026}.
 
These results describe two qualitatively different effective behaviors for fast rotating BECs: weak
focusing may preserve a delocalized annular giant vortex, whereas attraction
near the collapse threshold favours strongly localized, vortex-free states.
What is however not fully understood  is the transition between these two opposite behaviors:
as the effective attraction increases within the stable range, does
the system break rotational symmetry through the nucleation of additional
vortices, as in the repulsive case, or does a  different mechanism
emerge?

The present work focuses precisely on such a question and investigates the asymptotic behavior as $ \Omega \to +\infty $ of a rotating attractive BEC trapped on an annulus. We prove the occurrence of a different mechanism leading to rotational symmetry breaking: the attraction may drive indeed an angular
concentration of the density along the annulus where the bulk of the condensate is confined. The energy competition behind is between the radial confinement and the angular localization induced by the attraction.  As $ \Omega \to + \infty $, depending on the asymptotics of the coupling \(a_\om = \OO(1) \), we identify three different regimes:
\begin{itemize}
	\item for small coupling $ a_{\Omega} = o(1) $, there is an approximate factorization of the ground state into a radial and angular component, where the latter minimizes a suitable one-dimensional model on the circle and is constant for small enough $ a_{\Omega} $; such states carry a large winding phase proportional to $ \Omega $ itself ({\it giant vortex});
	\item for larger but still infinitesimal values of $ a_{\Omega}  $, the optimal angular profile breaks the rotational symmetry, assuming a Townes soliton profile concentrated at a point of the circle ({\it bright soliton});
	\item finally, when $ a \in (0, a^*)  $ stays fixed below the collapse threshold, the limiting profile is determined by a genuinely two-dimensional variational problem. 
\end{itemize}
Therefore, we observe that attraction drives a symmetry breaking
transition from an annular a giant vortex state to an angularly localized soliton.

\bigskip

\noindent 
\begin{footnotesize}
\textbf{Acknowledgments.} M. Correggi acknowledges the support of {\it INdAM-GNFM}, of {\it MUR} grant ``Dipartimento di Eccellenza 2023-2027'' of Dipartimento di Matematica, Politecnico di Milano, and of {\it PNRR Italia Domani} and {\it Next Generation Eu} through the {\it ICSC National Research Centre for High Performance Computing, Big Data and Quantum Computing}. Y.  Guo is partially supported by  {\it NSF of China}
	(Grants 12225106 and 12371113) and {\it National Key R $\&$ D Program of China} (Grant 2023YFA1010001). Y. Luo is partially supported by {\it NSF of China} (Grant  12571119) and the {\it Program of China Scholarship Council} (Grant: 202506770070). 
\end{footnotesize}

\medskip

\noindent
\begin{footnotesize}
 \textbf{Declaration of AI-assisted tools.}
During the preparation of this manuscript, the authors used OpenAI models as auxiliary tools. All AI-assisted material was critically reviewed and substantially revised by the authors, who take full responsibility for the manuscript.
\end{footnotesize}

\section{Main Results}

We aim to study a two-dimensional rapidly rotating BEC with attractive interaction confined in an annular trap in the framework of the GP theory. The explicit form of the GP energy in the non-inertial frame rotating at the same angular velocity as the trap reads then \cite{Rougerie2011,Luo2026} (see also \cite{GZZ,HJ23} for further results on non-rotating BECs in ring-shaped traps):

\begin{equation}\label{eq:model-energy}
	\energy(u)
	=\int_{\mathbb R^2}|(\nabla-i\om x^\perp)u|^2\,dx
	+\frac{\om^2}{8}\int_{\mathbb R^2}(|x|^2-1)^2|u|^2\,dx
	-\frac{a}{2}\int_{\mathbb R^2}|u|^4\,dx.
\end{equation}
Here \(x^\perp:=(-x_2,x_1)\) for \(x=(x_1,x_2)\),
\(\om>0\) is the angular velocity,
\(a\geq0\) is the possibly \(\om\)-dependent attractive
coupling strength satisfying
\beq
	\label{eq: stability}
	0 \leq a < a^{*},
\eeq  
where  \(a^*\) is the optimal constant in the two-dimensional
Gagliardo--Nirenberg inequality
\begin{equation}\label{eq:critical-mass}
\frac{a^*}{2}\int_{\mathbb R^2}|v|^4\,dx
\leq \|v\|_{L^2(\mathbb R^2)}^2\|\nabla v\|_{L^2(\mathbb R^2)}^2.
\end{equation}

We denote the ground state energy of $ \energy $ as
\begin{equation}\label{eq:ground-state-energy}
\gse:=
\inf_{\substack{u\in\domain, \lf\|u\ri\|_{2}=1}}
\energy(u),
\end{equation}
where $ \mathscr{H} $ is the functional domain:
\begin{equation}\label{eq:energy-space}
\domain:= \lf\{u\in H^1(\mathbb R^2;\mathbb C):
\int_{\mathbb R^2}|x|^4|u|^2\,dx<\infty \ri\}.
\end{equation}
For fixed $ \Omega $, condition \eqref{eq: stability} ensures the existence of at least one minimizer $ \gs $ of $ \energy $, i.e., such that
\beq
	\gse = \energy[\gs],
\eeq
via direct variational arguments \cite{Guo2020}, since \(a_\Omega<a^*\) and the anharmonic trap is confining. By a simple rescaling of the distance from  the circle of radius $ |\xv| = 1 $, it is easy to show that any state with energy of order at most $ \Omega $ must be essentially supported in an annulus of width $ \Omega^{-1/2} $. Any normalized profile supported in this annulus has size \(\om^{1/4}\), so that its quartic energy is of order \(\sqrt\om\). This suggests that, as $ \Omega \to + \infty $, the relevant parameter is
\begin{equation}\label{eq:sigma-main}
\sigma_\om:=a_\om\sqrt\om,
\qquad \sigma_\om=\OO(\sqrt\om),
\end{equation}
and we are going to distinguish two asymptotic regimes
\beq
	\sigma_{\om} =
	\begin{cases}	
		o(\sqrt{\om}) 	& \Longleftrightarrow \quad a_{\om} = o(1),	\\
		a \sqrt{\om}	&	\mbox{with } a \in (0, a^*) \mbox{ fixed}.
	\end{cases}
\eeq
In these regimes we use the compact notation
\beq
	\gsom : = u_{a_{\om}, \Omega} \in \domain,
\eeq
for any minimizer of $ \enom $.

We first discuss the regime
\beq
	\label{eq: weak int attraction}
	\sigma_\om = o(\sqrt{\om}).
\eeq
In this case, in order to determine the asymptotics $ \om \to + \infty $ of the GP energy functional minimization and to find the optimal giant vortex winding number, it turns out to be convenient to use the angular Fourier decomposition of the linear part of the energy, that leads to the radial
quadratic form
\begin{equation}\label{eq:fiber-quadratic-form}
\form(f)
:=\int_0^\infty
\Big\{|f'(r)|^2+
\Big(\frac \nu r-\om r\Big)^2|f(r)|^2
+\frac{\om^2}{8}(r^2-1)^2|f(r)|^2\Big\}r\,dr,
\qquad \nu\in\mathbb R,
\end{equation}
with form domain
\begin{equation}\label{eq:fiber-form-domain}
	\mathscr D(\form)
	:=\Big\{
	f\in L^2(\mathbb R_+,r\,dr)\cap H^1_{\mathrm{loc}}(\mathbb R_+):
	\form(f)<\infty
	\Big\}.
\end{equation}
We denote by \(\mathcal L_{\om,\nu}\) the associated self-adjoint operator, whose differential expression is
\begin{equation}\label{eq:continuous-radial-operator}
\mathcal L_{\om,\nu}=-\frac1r\partial_r(r\partial_r)
+\Big(\frac{\nu}{r}-\om r\Big)^2
+\frac{\om^2}{8}(r^2-1)^2.
\end{equation}
Since, thanks to the trapping potential, \(\mathcal L_{\om,\nu}\) has compact resolvent, its spectrum is given by eigenvalues $ \lf\{ \lambda_{\om,\nu}^{(j)} \ri\}_{j \in \N_+} $ that we order in a non-decreasing way (including multiplicity):
\[
\lambda_{\om,\nu}^{(1)}
\leq\lambda_{\om,\nu}^{(2)}
\leq\cdots,
\]
and we choose the corresponding orthonormal eigenfunctions $ \lf\{  \phi_{\om,\nu}^{(j)} \ri\}_{j \in \N_+} $ in such a way that 
\(\phi_{\om,\nu}^{(1)}\) is positive. We now optimize the ground state energy $ \lambda_{\om,\nu}^{(1)} $ with respect to $  \nu $ in $ \R $ and $ \mathbb Z $, obtaining the quantities
\beq
	\label{eq:E-int-def}
	E_\om^{\rm cont} :=\inf_{\nu\in\mathbb R}\lambda_{\om,\nu}^{(1)}, \qquad	
E_\om^{\rm int} :=\min_{n\in\mathbb Z}\lambda_{\om,n}^{(1)},
\eeq
and we denote by  $ \tau_\om \in \R $ and $ n_\om \in \mathbb Z $ the associated minimizers (in case of non-uniqueness, we fix a specific choice), such that
\beq
E_\om^{\rm cont} = \lambda_{\om,\tau_\om}^{(1)}, 	\qquad		E_\om^{\rm int} =\lambda_{\om,n_\om}^{(1)},
\eeq
\begin{equation}\label{eq:alpha-def}
	\alpha_\om:=\tau_\om-n_\om.
\end{equation}
The integer \(n_\om\) is the giant vortex winding number, whereas
\(\alpha_\om\) is the residual flux in the effective angular problem. Indeed, as we are going to see $ n_\om \sim \om $ and (see Proposition \ref{prop:first-band-expansion})
\begin{equation}\label{eq:alpha-asymptotic-bound}
\limsup_{\om\to\infty}|\alpha_\om|\leq\tfrac12.
\end{equation}

The effective model on the circle is however given by the following energy
\begin{equation}\label{eq:circle-functional}
	\mathcal E_{\alpha,\sigma}(h)
	=\frac19\int_{S^1}|(-i\partial_\theta-\alpha)h|^2\,d\theta
	-\frac{\alpha^2}{9}
	-\frac{\Gamma\sigma}{2}\int_{S^1}|h|^4\,d\theta,
\end{equation}
where we identify \(S^1 = \mathbb R/(2\pi\mathbb Z)\), and we set
\begin{equation}\label{eq:e-circle}
	e_{\alpha,\sigma}:=
	\inf_{\|h\|_{2}=1}\mathcal E_{\alpha,\sigma}(h).
\end{equation}

Before stating our first result, we have also to introduce the effective radial problem which turns out to be the simple harmonic oscillator on the line
\begin{equation}\label{eq:limiting-HO}
H_0:=-\partial_t^2+\beta^2t^2 \mbox{ on }L^2(\mathbb R),
\qquad \beta:=\tfrac{3}{\sqrt2},
\end{equation}
with normalized ground state $ \chi_0 $ satisfying
\beq
	\label{eq:chi0-properties}
	H_0 \chi_0 = \beta \chi_0, \qquad \chi_0(t):=\lf(\tfrac{\beta}{\pi}\ri)^{1/4}
e^{-\beta t^2/2},\qquad
	\Gamma:=\int_{\mathbb R}\chi_0^4(t)\,dt
	=\sqrt{\tfrac{\beta}{2\pi}}.
\end{equation}

The first result reduces the two-dimensional energy uniformly to the circle
problem in the whole regime \eqref{eq: weak int attraction}.

\begin{thm}[Weak and intermediate coupling $ \sigma_{\om} = o(\sqrt{\om}) $]\label{thm:A}
\mbox{}	\\
Assume that \(\sigma_\Omega\) satisfies \eqref{eq: weak int attraction}. Then, as $ \om \to \infty $,
\begin{equation}\label{eq:A-energy-expansion}
\gseom=E_\om^{\rm int}
+e_{\alpha_\om,\sigma_\om}
+o(1 + \sigma_\om^2),
\end{equation}
where \(E_\om^{\rm int}\) and
\(e_{\alpha_\om,\sigma_\om}\) are defined in
\eqref{eq:E-int-def} and \eqref{eq:e-circle}, respectively.
Moreover, for every minimizer \(\gsom\) of \( \enom\), there exists
\( f_\Omega \in H^1(S^1)\) satisfying \(\| f_\Omega\|_{L^2(S^1)}=1\) and
\begin{equation}\label{eq:A-angular-almost-minimizer}
\mathcal E_{\alpha_\Omega,\sigma_\Omega}(f_\Omega)
=e_{\alpha_\Omega,\sigma_\Omega}
+o(1+ \sigma_\om^2),
\end{equation}
such that
\begin{equation}\label{eq:A-L2-reduction}
\lf\|\gsom
-e^{in_\Omega\theta}\phi_{\Omega,\tau_\om}^{(1)}(r) f_\Omega(\theta)
\ri\|_{L^2(\mathbb R^2)}
\xrightarrow[\Omega \to \infty]{}0.
\end{equation} 
\end{thm}

Theorem~\ref{thm:A} separates the universal radial confinement from
the nonlinear angular response.  Since
\beq
r \lf| \phi_{\om,\tau_\om}^{(1)}(r) \ri|^2  \xrightarrow[\om \to \infty]{C^*_0(\R^+)}  \delta_1(r), 
\eeq
the mass of every ground state concentrates radially on the unit
circle.  Its distribution along the circle, however, need not be
uniform: by \eqref{eq:A-L2-reduction}, it is determined to leading
order by \(|f_\Omega(\theta)|^2\), where
\( f_\Omega \) is an
almost minimizer of the effective circle functional.  The factor
\(e^{in_\om\theta}\) carries the macroscopic circulation, whereas
\(  f_\Omega \) describes the angular density profile.

The same separation appears at the energy level.  The radial spectral
estimates in Lemma~\ref{lem:standard-spectral-input} below show that
\(E_\om^{\rm int}\) is of order \(\om\).  After this leading
linear contribution is subtracted, Theorem~\ref{thm:A} gives
\[
\gse-E_\om^{\rm int}
=e_{\alpha_\om,\sigma_\om}
+o(1+ \sigma_\om^2).
\]
Since \(1+\sigma_\om^2=o(\om)\), the circle energy is a
lower-order correction to the total energy, but it is this first order correction
that distinguishes different angular profiles.  It therefore governs
the transition from an angularly uniform giant vortex to an angularly
localized state.

Now, we extract more details about the asymptotics for 
\beq
	\label{eq: weak attraction}
	\sigma_{\om} = \OO(1).
\eeq
As we are going to see, below a sharp coupling threshold, the constant
profile
\beq
h_0(\theta)=\tfrac1{\sqrt{2\pi}}
\eeq
is the unique circle minimizer of \eqref{eq:e-circle} up to a constant phase.  Consequently,
the mass remains uniformly distributed along the annulus.  The next theorem provides the full asymptotics of any GP minimizer and shows that it is given by a giant vortex profile with angular momentum of order \(\om\).

\begin{thm}[Weak coupling $ \sigma_{\om} = \OO(1) $]
\label{thm:B}
\mbox{}	\\
Let the assumptions of Theorem~\ref{thm:A} be satisfied, and assume in addition that
\begin{equation}\label{eq:bounded-regime}
\alpha_\om \xrightarrow[\Omega \to \infty]{} \alpha\in(-1/2,1/2),
\qquad \sigma_\om \xrightarrow[\Omega \to \infty]{} \sigma\in
\Big[0,\tfrac{\pi}{9\Gamma}(1-4\alpha^2)\Big),
\end{equation}
then, up to the extraction of a subsequence, there exists 
\(\gamma_\om\in\mathbb R\) such that, for every fixed \(R>0\),
\begin{equation}\label{eq:giant-local-profile}
\sup_{\substack{|t|\leq R,\,\theta\in S^1}}
\lf|
\tfrac{1}{\om^{1/4}}e^{-i(\gamma_\om+n_\om\theta)}
u_\om\lf(1+ \tfrac{t}{\sqrt\om},\theta \ri)
-\tfrac{1}{\sqrt{2\pi}}\chi_0(t)
\ri| \xrightarrow[\om \to \infty]{} 0.
\end{equation}
Moreover, with \(L_z:=-i\partial_\theta\) denoting the angular momentum operator,
\begin{equation}\label{eq:giant-momentum-limit}
	\meanlrlr{\gs}{L_z}{\gs}_{L^2(\mathbb R^2)}
=n_\Omega+o(1)
=\Omega+o(\sqrt{\Omega}).
\end{equation} 
\end{thm}

\begin{rem}[Rotational symmetry breaking]
\mbox{}	\\
The quantity
\(
\sigma_{\rm c}(\alpha)
:=\frac{\pi}{9\Gamma}(1-4\alpha^2)
\)
identifies the sharp threshold for the breaking of the rotational symmetry for the fast rotation effective model: by Lemma~\ref{lem:sharp-circle-interpolation}
and Proposition~\ref{prop:homogeneous-circle}, indeed, any minimizer of the circle energy for \(\sigma<\sigma_{\rm c}(\alpha)\) is given by the uniform profile 
\(
|h(\theta)|=\frac{1}{\sqrt{2\pi}}
\), whereas it has
nonconstant modulus when \(\sigma>\sigma_{\rm c}(\alpha)\). Moreover, Theorem~\ref{thm:A} shows that the modulus of any GP minimizer on the circle is close to an approximate minimizer of the energy functional  \eqref{eq:e-circle} on the circle. 
Consequently, the GP minimizers are asymptotically rotationally
symmetric below this threshold and break rotational symmetry above it.
The threshold is maximal at \(\alpha=0\) and decreases to zero as
\(|\alpha|\to1/2\).  Hence, the mismatch between the optimal continuous
angular momentum and its integer quantization promotes symmetry
breaking: near half-integer residual flux, even a weak effective
attraction can destabilize the uniform angular profile.
\end{rem}

\begin{rem}[Giant vortex state]
\mbox{}	\\
The locally uniform convergence in \eqref{eq:giant-local-profile}
implies that, for every fixed \(R>0\), the ground state has no zeros in
the bulk annulus
\[
\bigl\{x:\bigl||x|-1\bigr|\leq R\om^{-1/2}\bigr\}
\]
for all sufficiently large \(\om\).  At the same time,
\eqref{eq:giant-momentum-limit} shows that its angular
momentum is of order \(\om\).  Thus the state contains no vortices in the
bulk but retains macroscopic circulation, and these two properties
characterize the giant vortex structure.
\end{rem}
When \(\sigma=0\), Theorem~\ref{thm:B} recovers the weak interaction
giant vortex regime of \cite{Luo2026}.  The present result
extends that regime to every finite limiting effective coupling
\(0<\sigma<\sigma_{\rm c}(\alpha)\) and identifies the sharp boundary
at which the uniform angular density loses its minimizing property.

\bigskip
Theorem~\ref{thm:B} treats the weak coupling regime, in which
the limiting mass remains uniformly distributed along the annulus.
We now turn to the intermediate coupling regime
\beq
	\label{eq: ass thm C}
	\sigma_\om\to\infty,	\qquad	\sigma_\om = o(\sqrt{\om}).
\eeq  
In this regime, the attractive term
dominates the effective circle problem and forces its minimizers to
concentrate at a point
of \(S^1\) on a scale \( \sim \sigma_\Omega^{-1}\). As a consequence, we have that
\[
e_{\alpha_\om,\sigma_\om}
=-\tfrac{3\Gamma^2}{16}\sigma_\om^2
+o(\sigma_\om^2)
\]
as $ \om \to \infty $. After recentering and rescaling, the corresponding universal angular
profile is the normalized one-dimensional bright soliton
\begin{equation}\label{eq:line-profile-main}
	H(s):=\sqrt{\tfrac{9\Gamma}{8}}
	\operatorname{sech}\lf(\tfrac{9\Gamma}{4}s\ri),
\end{equation}
so that $ \|H\|_{L^2(\mathbb R)}=1 $.
To lift this angular localization to the original two-dimensional
problem, given \( \gsom \in\domain\),
\(\theta_\om\in S^1\), and \(\gamma_\om\in\mathbb R\), we set
\begin{equation}\label{eq:strong-rescaled-definition}
V_\om(t,s):= 
	\begin{cases}
		\tfrac{1}{\om^{1/4}\sigma_\om^{1/2}}
\lf(1+\tfrac{t}{\sqrt\om} \ri)^{1/2}
e^{-i\gamma_\om-i n_\om (\theta_\om+ \frac{s}{\sigma_\om} )} u_\om \lf(1+\tfrac{t}{\sqrt\om},
\theta_\om+\tfrac{s}{\sigma_\om}\ri),	&	\mbox{in } \mathcal{D}^{(1)}_\om,	\\
		0,							&	\mbox{otherwise},
	\end{cases}
\end{equation}
where 
\beq
	\label{eq: dom om} 
	\mathcal{D}^{(1)}_{\om} = \lf\{ (s,t) \in \R^2 \: : \:  t \geq -\sqrt\om, -\pi \sigma_\om <s \leq \pi \sigma_\om \ri\}.
\eeq

\begin{thm}[Intermediate coupling $ \sigma_{\om} \to \infty $, $\sigma_\om = o(\sqrt{\om}) $]\label{thm:C}
\mbox{}	\\
Let the assumptions of Theorem~\ref{thm:A} be satisfied, and assume in addition \eqref{eq: ass thm C}. Then, as $ \Omega \to \infty $,
\begin{equation}\label{eq:strong-main-energy}
\gseom=E_\om^{\rm int}
-\tfrac{3\Gamma^2}{16}\sigma_\om^2
+o(\sigma_\om^2).
\end{equation}
Furthermore, there exist $ \theta_*, \theta_{\om} \in S^1 $ and \(\gamma_\om\in\mathbb R\) such that, up to the extraction of a subsequence,
\beq
	\label{eq:strong-rescaled-profile}
	 \theta_\om \xrightarrow[\om \to \infty]{} \theta_*,	\qquad	V_\om \xrightarrow[\om  \to \infty]{L^2(\R^2)}\chi_0(t)H(s).
\end{equation}
\end{thm}

\begin{rem}[Residual flux]
\mbox{}	\\
Unlike Theorem~\ref{thm:B}, Theorem~\ref{thm:C} above requires no limiting
assumption on the residual flux \(\alpha_\om\), since in this intermediate coupling regime, \(\alpha_\om\)
contributes only at lower order relative to the
balance on the \(\sigma_\om^2\) scale that determines angular
localization.  Consequently, the leading energy correction and the
soliton profile \(H\) are independent of \(\alpha_\om\).  Every sequence of minimizers concentrates, up
to a subsequence, at some point of the unit circle, thereby
spontaneously breaking rotational symmetry.  
\end{rem}

The energy expansion in Theorem~\ref{thm:C} shows that the first
nonlinear correction to \(E_\om^{\rm int}\) is of order
\(
\sigma_\om^2=\om a_\om^2.
\)
Throughout the regime \eqref{eq: weak int attraction}, this correction is
\(o(\om)\) and therefore subleading with respect to the leading term in $ \gseom $ which is of order $ \om $.  We next consider the strong interaction regime
\beq
	\label{eq: strong attraction}
a_\om=a \in(0,a^*),
\qquad \sigma_\om=a\sqrt\om.
\eeq
At this scale, the nonlinear interaction and the local linear energy
contribute at the same order.  The radial and angular variables can no
longer be separated since both live on a scale of order
\(\om^{-1/2}\), and the circle reduction is replaced by a genuinely
two-dimensional local variational problem. In this regime, we define
\begin{equation}\label{eq:endpoint-local-form}
	\mathfrak q(v):=\int_{\mathbb R^2}\Big(|\partial_t v|^2+
	|(-i\partial_s-2t)v|^2+\tfrac12t^2|v|^2-
	\tfrac{3\sqrt2}{2}|v|^2\Big)\,dt\,ds
\end{equation}
on
\begin{equation}\label{eq:endpoint-form-domain}
	\mathscr{D}(\mathfrak q):=\lf\{v\in L^2(\mathbb R^2,\mathbb C):
	\int_{\mathbb R^2}\Big(|\partial_t v|^2+|(-i\partial_s-2t)v|^2+
	t^2|v|^2\Big)\,dt\,ds<\infty \ri\},
\end{equation}
and, for \(0\leq b<a^*\), set
\begin{equation}\label{eq:endpoint-lge}
	\mathfrak e_b:=\inf_{v\in\mathscr D(\mathfrak q), 
			\|v\|_{2}=1}
	\lf\{\mathfrak q(v)-\frac b2\int_{\mathbb R^2}|v|^4\,dt\,ds\ri\}.
\end{equation}
We denote by $ V_b \in \mathscr{D}(\mathfrak q) $ any associated normalized minimizer, i.e., such that 
\beq
	\mathfrak{e}_b = \mathfrak q(V_b)-\frac b2\int_{\mathbb R^2}|V_b|^4\,dt\,ds.
\eeq
Furthermore, we suitably rescale the GP minimizer by defining for some $ \theta_\om \in S^1 $
\begin{equation}\label{eq:endpoint-rescaling}
W_\om(t,s):=
	\begin{cases}
		\tfrac{1}{\om^{1/2}} \lf( 1+\tfrac{t}{\sqrt\om} \ri)^{1/2}
e^{-i n_\om(\theta_\om+\frac{s}{\sqrt\om})} \gsom \lf(1+\tfrac{t}{\sqrt\om},
\theta_\om+\tfrac{s}{\sqrt\om} \ri),	&	\mbox{in } \mathcal{D}^{(2)}_\om,	\\
		0,							&	\mbox{otherwise}.
	\end{cases}
\end{equation}
where analogously to \eqref{eq: dom om} 
\beq
	\mathcal{D}^{(2)}_{\om} = \lf\{ (s,t) \in \R^2 \: : \:  t \geq -\sqrt\om, -\pi \sqrt{\om} <s \leq \pi \sqrt{\om} \ri\}.
\eeq.

\begin{thm}[Strong coupling $ a \in (0,a^*) $]\label{thm:ep-concentration}
\mbox{}	\\
Assume that \(a\in(0,a^*)\) is fixed.  Then, as $ \om \to \infty $,
\begin{equation}\label{eq:endpoint-main-energy}
 \gse = E_\om^{\rm int}+\om\mathfrak e_a +o(\om).
\end{equation}
Furthermore, there exist $ \theta_*, \theta_{\om} \in S^1 $ such that, up to the extraction of a subsequence, 
\begin{equation}\label{eq:endpoint-rescaled-profile}
 \theta_\om \xrightarrow[\om \to \infty]{} \theta_*,	\qquad W_\om
\xrightarrow[\om \to \infty]{L^2(\R^2)} V_a.
\end{equation}
\end{thm}

\begin{rem}[Two-dimensional effective model]
\mbox{}	\\
Compared with Theorem~\ref{thm:C}, the radial and angular variables in
Theorem~\ref{thm:ep-concentration} have the same concentration scale
\(\om^{-1/2}\).  Consequently, the separated profile
\(\chi_0(t)H(s)\) is replaced by a minimizer \(V_a\) of the genuinely
two-dimensional effective model \eqref{eq:endpoint-lge}, where the kinetic term involving
\((-i\partial_s-2t)v\) is the main responsible of the coupling between the two coordinates.
\end{rem}

\begin{rem}[Concentration in terms of measures]
\mbox{}	\\
The convergences in Theorems~\ref{thm:C} and
\ref{thm:ep-concentration} both imply that, up to the extraction of a subsequence,
\beq
	|u_\om(x)|^2 \xrightarrow[\om \to \infty]{C^*_0(\R^2)} \delta_{x_*}(x) \quad\hbox{for some}\,\ x_* \in \R^2, |x_*| = 1.
\eeq
Rotational invariance leaves the concentration point \(\theta_*\) on the circle, or equivalently the associated angle $ \theta_* $, undetermined.
\end{rem}

%

\subsection{Heuristics and Outline of the Proofs}

We first explain the reduction used to prove Theorem~\ref{thm:A}.  The basic idea is that for small enough $ \sigma_{\om} $, the nonlinear term is just a higher order correction to the GP energy and perturbation theory around the linear model provides the correct asymptotics of the energy. The linear
part of \eqref{eq:model-energy} is actually the quadratic form
\begin{equation}\label{eq:linear-form}
\mathcal Q_\om(u)=\int_{\mathbb R^2}|(\nabla-i\om x^\perp)u|^2\,dx
+\frac{\om^2}{8}\int_{\mathbb R^2}(|x|^2-1)^2|u|^2\,dx,
\qquad  \mathscr{D}(\mathcal Q_\om)=\domain,
\end{equation}
which in polar coordinates becomes
\begin{equation}\label{eq:linear-fpr}
\mathcal Q_\om(u)=\int_0^\infty\int_0^{2\pi}
\Big\{|\partial_r u|^2+
\Big|\Big(\frac{-i\partial_\theta}{r}-\om r\Big)u\Big|^2
+\frac{\om^2}{8}(r^2-1)^2|u|^2\Big\}r\,d\theta\,dr .
\end{equation}
Using the angular Fourier decomposition
\begin{equation}\label{eq:angular-Fourier-expansion}
u(r,\theta)=\frac1{\sqrt{2\pi}}
\sum_{n\in\mathbb Z}u_n(r)e^{in\theta}
\quad\hbox{in }L^2(\mathbb R^2),
\end{equation}
we split the quadratic form above into radial fibers as
\begin{equation}\label{eq:fiber-form-decomposition}
\mathcal Q_\om(u)
=\sum_{n\in\mathbb Z}\mathfrak q_{\om,n}(u_n),
\end{equation}
where the form $ \form $ is defined in \eqref{eq:fiber-quadratic-form}.

For any normalized ground state $ \gsom $ of the GP functional we may define the coefficients 
\beq
a_{n,m}:=
\int_0^\infty
\overline{\phi_{\om,n}^{(m)}(r)} u_{\om,n}(r)\,r\,dr,
\eeq
of the expansion of the radial components \( \lf\{ u_{\om,n}(r) \ri\}_{n \in \N} \) into the orthonormal basis of eigenfunctions of the operator \eqref{eq:continuous-radial-operator}. Hence,
\begin{equation}\label{eq:full-spectral-decomp}
u_\om(r,\theta)=\frac1{\sqrt{2\pi}}
\sum_{n\in\mathbb Z}\sum_{m\geq1}
a_{n,m}\phi_{\om,n}^{(m)}(r)e^{in\theta},
\end{equation}
with
\(
\sum_{n\in\mathbb Z}\sum_{m \in \N_+}|a_{n,m}|^2=1.
\)
Plugging such an expansion in the GP energy, we obtain the splitting
\begin{equation}\label{eq:energy-excess-spectral}
\gseom -E_\om^{\rm int}
=
\sum_{n\in\mathbb Z}\sum_{m \in \N_+}
\bigl(\lambda_{\om,n}^{(m)}-E_\om^{\rm int}\bigr)
 |a_{n,m}|^2-\frac{a_\om}{2}
\int_{\mathbb R^2}|u_\om(x)|^4\,dx.
\end{equation}
Since the spectral gap between the ground state of the radial operator \eqref{eq:continuous-radial-operator} and its excited states is bounded from below by \(c\om\), for some \(c>0\), whereas
the energy correction in Theorem~\ref{thm:A} is of order
\(1+\sigma_\om^2=o(\om)\), we deduce that the occupation of excited state is negligible, and the leading part of any minimizer lies in the ground state eigenspace.

In order to capture the first order corrections one needs to explore more closely the Fourier decomposition \eqref{eq:fiber-form-decomposition} to determine which frequencies contribute to the energy. A refined analysis performed in Proposition~\ref{prop:first-band-expansion} shows that, uniformly in a growing
neighborhood of \(n_\om\),
\beq
	\label{eq: second term}
\lambda_{\om,n_\om+j}^{(1)}-E_\om^{\rm int}
=\tfrac19\bigl((j-\alpha_\om)^2-\alpha_\om^2\bigr)
+o\bigl(1+(j-\alpha_\om)^2\bigr).
\eeq
The quadratic growth in \(j-\alpha_\om\) confines the relevant lowest energy
modes to an angular window \(\mathcal W_\om\) defined in
\eqref{eq:unified-effective-window}. Normalizing the retained Fourier sum
gives the profile in \eqref{eq:unified-normalized-profile}:
\[
  g_\Omega(\theta)
=\bigg(2\pi\sum_{j\in\mathcal W_\Omega}|a_{n_\Omega+j,1}|^2 \bigg)^{-1/2} \sum_{j\in\mathcal W_\Omega}a_{n_\Omega+j,1}e^{i j \theta}
.
\]
The radial component of the GP minimizer can then be replaced by \(\phi_{\om,\tau_\om}^{(1)}\) exploiting the uniform
eigenfunction estimates in Lemma~\ref{lem:radial-profile-comparison}.  Thus,
\[
\gs(r,\theta)
\approx
e^{in_\Omega\theta}
\phi_{\Omega,\tau_\om}^{(1)}(r)  g_\Omega(\theta).
\]
Moreover, since
\[
\int_0^\infty
\bigl|\phi_{\om,\tau_\om}^{(1)}(r)\bigr|^4r\,dr
\sim\Gamma\sqrt\om,
\]
the quartic interaction term becomes 
\[
\Gamma\sigma_\Omega
\int_{S^1}| g_\Omega (\theta)|^4\,d\theta.
\]
Combining this with \eqref{eq:energy-excess-spectral} and the estimate \eqref{eq: second term}, we heuristically recover the effective energy \(\mathcal E_{\alpha_\om,\sigma_\om}\) introduced in \eqref{eq:e-circle}.  The proof of Theorem~\ref{thm:A} follows exactly these lines for the lower bound part, while the upper bound is obtained by a suitable trial state in the lowest energy space.

For Theorem~\ref{thm:B}, assume that
\(\sigma_\om\to\sigma<\sigma_{\rm c}(\alpha)\).  The sharp magnetic
interpolation inequality in
Lemma~\ref{lem:sharp-circle-interpolation} shows that the constant angular profile
\((2\pi)^{-1/2}\) is the unique minimizer of the limiting circle problem, up
to a constant phase.  The circle reduction therefore implies that, up to a constant phase, the
angular profile converges to the constant.
Combining this with the radial convergence in  Corollary~\ref{lem:uniform-two-mode-HO} yields the giant
vortex profile.  Angular rigidity and the spectral bounds for the discarded
modes yield the angular momentum estimate.

For Theorem~\ref{thm:C}, the effective coupling satisfies
\(\sigma_\om\to\infty\).  Balancing the angular kinetic and nonlinear
terms gives the concentration scale \(\sigma_\om^{-1}\).  We recenter a
circle almost minimizer at \(\theta_\om\) and set
\(s=\sigma_\om(\theta-\theta_\om)\).  After division by
\(\sigma_\om^2\), the circle functional converges to
\[
\frac19\int_{\mathbb R}|h'|^2\,ds
-\frac{\Gamma}{2}\int_{\mathbb R}|h|^4\,ds.
\]
Concentration compactness identifies its normalized minimizer with
the bright soliton \(H\).  Together with the radial convergence in
Corollary~\ref{lem:uniform-two-mode-HO}, this yields
the product profile \(\chi_0(t)H(s)\).

For Theorem~\ref{thm:ep-concentration}, one has
\(\sigma_\om=a\sqrt\om\), so the angular scale
\(\sigma_\om^{-1}\) has the same order as the radial scale
\(\om^{-1/2}\).  The preceding separation of variables therefore no
longer applies.  Once the mass is localized near a point on the unit circle,
rescaling the radial and angular variables simultaneously leads to the
two-dimensional limiting problem \eqref{eq:endpoint-lge} and its ground state energy $ \mathfrak{e}_a $.  The main difficulty is
to prove that asymptotically the associated minimizer $ V_a $ carries all the mass.  The key input
is the strict subadditivity inequality
\[
\mathfrak e_a
<
\mu\mathfrak e_{a\mu}
+(1-\mu)\mathfrak e_{a(1-\mu)},
\qquad 0<\mu<1,
\]
which follows from the concavity of \(b\mapsto\mathfrak e_b\) and the strict
negativity \(\mathfrak e_a<0\).  This negativity is a consequence of the
attractive interaction and is essential here; the corresponding inequality
fails for repulsive interaction.  The strict inequality excludes more than one
nontrivial component, so one component carries asymptotically all the mass.
An interior cutoff and compactness
argument then yield the limiting profile \(V_a\) in
Theorem~\ref{thm:ep-concentration}.

\medskip
Throughout the paper, the letters \(c,C>0\) denote constants independent of \(\Omega\),
whose values may change
from line to line. Unless otherwise stated, all limits and asymptotic notation refer
to \(\Omega\to\infty\). For \(g_\Omega>0\), we write
\bdm
	\begin{aligned}
	f_\Omega &=\OO(g_\Omega) \Longleftrightarrow \limsup_{\om \to \infty} \frac{f_\Omega}{g_\Omega} \leq C,	\\
	f_\Omega &=o(g_\Omega) \Longleftrightarrow \limsup_{\om \to \infty} \frac{f_\Omega}{g_\Omega} = 0, 	\\
	f_\Omega & \sim g_\Omega \Longleftrightarrow \lim_{\om \to \infty} \frac{f_\Omega}{g_\Omega} = c.
	\end{aligned}
\edm

The rest of the paper is organized as follows: Sections~\ref{sec:radial-spectrum} and~\ref{sec:uniform-circle-reduction}
establish the radial spectral estimates and the effective circle reduction,
respectively.  Section~\ref{sec:circle-consequences} analyzes the circle
problem, and Section~\ref{subsec:endpoint} treats the fixed attraction
regime.  The auxiliary radial and circle estimates are proved in the
Appendixes.

\section{Spectral Estimates of the Radial Operator \(\mathcal L_{\om,\nu}\)}
\label{sec:radial-spectrum}

Recall the linear radial operator $\mathcal L_{\om,\nu}$ defined in \eqref{eq:continuous-radial-operator}, and its eigenvalue $\lambda^{(m)}_{\om,\nu}$, $ m \in \N_+ $. Recall also the definitions \eqref{eq:E-int-def}, \eqref{eq:alpha-def} of $\tau_\om$, $n_\om$ and $\alpha_\om$. The purpose of this section is to  study the asymptotic behavior of   $\lambda^{(1)}_{\om,\nu}$  for 
\beq
	|\nu-\om|=o(\sqrt\om),
\eeq
as \(\om\to\infty\). 

\begin{proposition}\label{prop:first-band-expansion}
Let  \(L_\om\to\infty\) with \(L_\om=o(\sqrt\om)\). Then, 
\begin{equation}\label{eq:uniform-curvature-new}
\lim_{\om\to\infty}\sup_{|\nu-\tau_\om|\le L_\om}
\Big|
\partial_{\nu\nu}\lambda_{\om,\nu}^{(1)}-\tfrac29
\Big|
=0,
\end{equation}
and
\begin{equation}\label{eq:int-window-expansion}
\lim_{\om\to\infty}\sup_{|j-\alpha_\om|\le L_\om}
\frac{
\Big|
\lambda_{\om,n_\om+j}^{(1)}-E_\om^{\rm int}
-\frac19\bigl((j-\alpha_\om)^2-\alpha_\om^2\bigr)
\Big|
}{1+(j-\alpha_\om)^2}
=0.
\end{equation}
Moreover, we have
\begin{equation}\label{eq:alpha-error}
\limsup_{\om\to\infty}|\alpha_\om|\leq\tfrac12,
\end{equation}
and there are \(c,C>0\) such that, for all sufficiently large
\(\om\) and for $ j \in \mathbb Z $,
\begin{equation}\label{eq:angular-gap-new}
\lambda_{\om,n_\om+j}^{(1)}-E_\om^{\rm int}
\geq c\min\{(j-\alpha_\om)^2,\om\}-C.
\end{equation}
\end{proposition}

Since the proof of Proposition~\ref{prop:first-band-expansion} is
lengthy, we begin by outlining its main steps. By the definition of \(\mathcal L_{\om,\nu}\) in  \eqref{eq:continuous-radial-operator}, the first eigenfunctions \(\phi_{\om,\nu}^{(1)}\)
of \(\mathcal L_{\om,\nu}\) are localized near \(r=1\) in a layer of
width \(\OO(\om^{-1/2})\).  Using the unitary rescaling
\eqref{eq:radial-unitary-map}, we consider
\[
H_{\om,\nu}
=\om^{-1}U_\om\mathcal L_{\om,\nu}U_\om^{-1}.
\]
Along sequences for which
\(
(\nu-\om)/\sqrt\om\to q
\).
\(H_{\om,\nu}\) converges on compact intervals to
the following operator
\[
-\partial_t^2+(q-2t)^2+\tfrac12t^2
=
-\partial_t^2
+\tfrac92\left(t-\tfrac{4}{9}q\right)^2+\tfrac{1}{9}q^2.
\]
For every fixed integer \(m\geq1\),
Lemma~\ref{lem:bounded-q-radial-control} proves that the first \(m\)
eigenvalues of \(H_{\Omega,\nu}\) converge to the corresponding
eigenvalues of the above limiting operator.  When \(q=0\),
it also gives convergence of the first two normalized eigenfunctions of \(H_{\Omega,\nu}\) to
the first two normalized   eigenfunctions of
\(-\partial_t^2+\frac92t^2\).  On this basis,
Lemma~\ref{lem:standard-spectral-input} gives
\[
\tau_\om-\om=o(\sqrt\om),
\qquad
n_\om-\om=o(\sqrt\om),
\]
and combines the rescaled spectral information with a global potential
estimate to obtain \mic{for $ \nu \in \R $}
\begin{equation}\label{eq:oct34}
\lambda_{\om,\nu}^{(1)}
\geq
\om e_0+c\min\{(\nu-\om)^2,\om\}-o(\om).
\end{equation}

The key step is to establish the uniform curvature estimate
\eqref{eq:uniform-curvature-new}, which does not follow from
the preceding eigenvalue asymptotics.
A separate analysis of the \(r^{-2}\) singularity at \(r=0\), together
with the   spectral sum result in
Lemma~\ref{lem:FH-spectral-sum-strict}  yields  
\(
\partial_{\nu\nu}\lambda_{\om,\nu}^{(1)}
=\frac29+o(1)\)  uniformly for 
\(|\nu-\tau_\om|=o(\sqrt\om)\).
Since $\lambda_{\om,\tau_\om}^{(1)}=E_\om^{\rm cont}$ and $
\left.\partial_\nu\lambda_{\Omega,\nu}^{(1)}
\right|_{\nu=\tau_\om}=0$, Taylor's formula then yields
\[
\lambda_{\om,\tau_\om+s}^{(1)}
=E_\om^{\rm cont}+\tfrac19s^2+o(1+s^2)\quad\hbox{uniformly for}\,\ 
|s|=o(\sqrt\om).
\]
Taking \(s=-\alpha_\om\) and \(s=j-\alpha_\om\), and using
\(n_\om=\tau_\om-\alpha_\om\), gives
\begin{equation}\label{eq:oct35}
\lambda_{\om,n_\om+j}^{(1)}-E_\om^{\rm int}
=\tfrac19\bigl((j-\alpha_\om)^2-\alpha_\om^2\bigr)
+o\bigl(1+(j-\alpha_\om)^2\bigr).
\end{equation}
Since \(n_\Omega\) minimizes $\lambda^{(1)}_{\om,n}$ over the integers $n\in\mathbb Z$, we have
$
\lambda_{\Omega,n_\Omega}^{(1)}
\leq \lambda_{\Omega,n_\Omega\pm1}^{(1)},
$
which together with \eqref{eq:oct35} gives \eqref{eq:alpha-error}.
Combining \eqref{eq:oct34} with \eqref{eq:oct35} proves
\eqref{eq:angular-gap-new}.

\subsection{The radial operator \(H_{\om,\nu}\) and its properties}
Let
\[
I_\om:=(-\sqrt\om,\infty),
\qquad
\rho_\om(t):=1+\frac{t}{\sqrt\om},\quad t\in I_\om.
\]
For \(f=f(r)\in L^2(\mathbb R_+,r\,dr)\), define
\begin{equation}\label{eq:radial-unitary-map}
	(U_\om f)(t):=\om^{-1/4}f(\rho_\om(t)),
	\qquad t\in I_\om .
\end{equation}
Then \(U_\om\) is unitary from \(L^2(\mathbb R_+,r\,dr)\) onto
\(L^2(I_\om,\rho_\om(t)\,dt)\).
We work in the rescaled Hilbert space
\begin{equation}\label{eq:rescaled-hilbert-space}
\domain_\om:=L^2(I_\om,\rho_\om(t)\,dt),
\end{equation}
and define the rescaled self-adjoint operator on \(\domain_\om\) by
\begin{equation}\label{eq:rescaled-op}
	H_{\om,\nu}:=\om^{-1}U_\om\mathcal L_{\om,\nu}U_\om^{-1},
	\qquad
	\mathscr D(H_{\om,\nu})
	:=U_\om\mathscr D(\mathcal L_{\om,\nu}).
\end{equation}
A direct computation from \eqref{eq:continuous-radial-operator}, with
\(r=\rho_\Omega(t)\), gives
\begin{equation}\label{eq:rescaled-radial-operator}
\begin{aligned}
H_{\Omega,\nu}
=-\frac1{\rho_\Omega}\partial_t(\rho_\Omega\partial_t)
+V_{\Omega,\nu},\quad\hbox{where}\,\ 
V_{\Omega,\nu}(t)
:=\frac{(\nu-\Omega\rho_\Omega(t)^2)^2}{\Omega\rho_\Omega(t)^2}
+\frac{\Omega}{8}(\rho_\Omega(t)^2-1)^2.
\end{aligned}
\end{equation}
Define the rescaled eigenpairs by
\begin{equation}\label{eq:rescaled-eigenpairs}
\psi_{\om,\nu}^{(m)}:=U_\om\phi_{\om,\nu}^{(m)},\qquad
e_{\om,\nu}^{(m)}:=\om^{-1}\lambda_{\om,\nu}^{(m)}.
\end{equation}
Since \(U_\om\) is unitary from \(L^2(\mathbb R_+,r\,dr)\) onto
\(\domain_\om\), we have
\begin{equation}\label{sep25-1}
\mathcal L_{\om,\nu}\phi_{\om,\nu}^{(m)}
=\lambda_{\om,\nu}^{(m)}\phi_{\om,\nu}^{(m)}
\quad \mbox{in } L^2(\R_+, r dr) \quad\Longleftrightarrow\quad
H_{\om,\nu}\psi_{\om,\nu}^{(m)}
=e_{\om,\nu}^{(m)}\psi_{\om,\nu}^{(m)}
\quad\hbox{in }\domain_\om .
\end{equation}
Accordingly, we henceforth work with the rescaled eigenvalue problem for
\(H_{\om,\nu}\).

Denote
\begin{equation}\label{eq:def-qom}
q_{\om,\nu}:=\frac{\nu-\om}{\sqrt\om}. 
\end{equation}
In terms of
\(q_{\Omega,\nu}\), the potential in
\eqref{eq:rescaled-radial-operator} takes the form
\begin{equation}\label{eq:rescaled-potential-q}
V_{\Omega,\nu}(t)
=\lf(\frac{q_{\Omega,\nu}-2t-t^2/\sqrt\Omega}{\rho_\Omega(t)}\ri)^2
+\frac18\lf(2t+\frac{t^2}{\sqrt\Omega}\ri)^2.
\end{equation}
If \(q_{\om,\nu}\to0\), then
\eqref{eq:rescaled-radial-operator} and \eqref{eq:rescaled-potential-q}
formally yield
\[
H_{\om,\nu}\longrightarrow
-\partial_t^2+\tfrac92t^2=H_0,
\]
where \(H_0\) is defined in \eqref{eq:limiting-HO}.  Its first two normalized
eigenfunctions and eigenvalues are
\begin{equation}\label{eq:first-two-HO-modes}
e_0:=\beta=\tfrac{3}{\sqrt2},\qquad e_1:=3\beta,\qquad
\chi_1(t):=\sqrt{2\beta}\,t\chi_0(t),\qquad
H_0\chi_j=e_j\chi_j\quad (j=0,1).
\end{equation}
Whenever \(\chi_0\) or \(\chi_1\) is compared with a function in
\(\domain_\om\), it is restricted from \(\mathbb R\) to \(I_\om\).

We begin with the following lemma.
\begin{lemma}\label{lem:bounded-q-radial-control}
Fix \(A>0\) and \(\ell \in \N_+ \).  Let \(q_{\om,\nu}\) be defined by
\eqref{eq:def-qom}, and assume that \(|q_{\om,\nu}|\leq A\) uniformly in
\(\om\).  Then there exists a constant \(C_{A,\ell}>0\) such that, as $ \om \to \infty $,
\begin{equation}\label{eq:bounded-q-low-energy}
e_{\om,\nu}^{(\ell)}\leq C_{A,\ell},
\qquad
\int_{I_\om}\lf( \lf|(\psi_{\om,\nu}^{(\ell)})' \ri|^2
+t^2 \lf|\psi_{\om,\nu}^{(\ell)} \ri|^2 \ri)
\rho_\om\,dt\leq C_{A,\ell}.
\end{equation}
Moreover,
\begin{equation}\label{eq:bounded-q-eigenlimit}
\lim_{\om\to\infty}\sup_{|q_{\om,\nu}|\leq A}
\lf|
e_{\om,\nu}^{(\ell)}-\lf((2\ell-1)\beta
+\tfrac{1}{9} q_{\om,\nu}^2 \ri)
\ri|=0.
\end{equation}
If \(q_{\om,\nu} \xrightarrow[\om\to\infty]{} 0 \), then 
\begin{equation}\label{eq:q-zero-low-mode-conv}
\lim_{\om\to\infty}e_{\om,\nu}^{(1)}= e_0 
\quad\hbox{and}\quad \lim_{\om\to\infty}e_{\om,\nu}^{(2)}= e_1.
\end{equation}
Choose the signs of the real eigenfunctions so that
\[
\int_{I_\om}
\psi_{\om,\nu}^{(m)}(t)\chi_{m-1}(t)\rho_\om(t)\,dt\geq0
\qquad(m=1,2),
\]
then
\begin{equation}\label{eq:q-zero-low1}
\lim_{\om\to\infty} \lf[ \lf\|\psi_{\om,\nu}^{(1)}-\chi_0 \ri\|_{\domain_\om}
+ \lf\|\psi_{\om,\nu}^{(2)}-\chi_1 \ri\|_{\domain_\om} \ri]
=0.
\end{equation}
Moreover, for every fixed \(R>0\),
\begin{equation}\label{eq:oct30}
\psi_{\om,\nu}^{(1)} \xrightarrow[\om \to \infty]{H^1(-R,R)} \chi_0,\qquad
\psi_{\om,\nu}^{(2)} \xrightarrow[\om \to \infty]{H^1(-R,R)} \chi_1.
\end{equation}
\end{lemma}
Since the proof is independent of the subsequent analysis and rather
lengthy, we defer its proof to
Appendix~\ref{app:bounded-q-radial}.

\begin{lemma}
\label{lem:standard-spectral-input}
We have that
\begin{equation}\label{eq:E-size-new}
\lim_{\om\to\infty}\frac{E_\om^{\rm cont}}{\om}
=\lim_{\om\to\infty}\frac{E_\om^{\rm int}}{\om}=e_0,
\end{equation}
\begin{equation}\label{eq:E-size-new1}
\lim_{\om\to\infty}\frac{\tau_\om-\om}{\sqrt\om}
=\lim_{\om\to\infty}
\frac{n_\om-\om}{\sqrt\om}=0,
\end{equation}
and there are \(c>0\) and a sequence \(\varepsilon_\om\to0\) as \(\om\to\infty\) such
that, for all sufficiently large \(\om\),
\begin{equation}\label{eq:proof-coarse-fiber-bound}
\lambda_{\om,\nu}^{(1)}
\geq
\om e_0+c\min\{(\nu-\om)^2,\om\}
-\varepsilon_\om\om
\qquad(\nu\in\mathbb R).
\end{equation}
Moreover, there exists a constant \(c>0\) such that, for all sufficiently large
\(\om\),
\begin{equation}\label{eq:radial-gap-new}
\lambda_{\om,n}^{(m)}-E_\om^{\rm int}\geq c\om
\quad(n\in\mathbb Z,\ m\geq2).
\end{equation}
\end{lemma}
\begin{proof}[\normalfont\bfseries Proof]
Let \(\widehat n_\om\) be an integer nearest to \(\om\).  Since
\(q_{\om,\widehat n_\om}=O(\om^{-1/2})\),
\eqref{eq:bounded-q-eigenlimit} with \(\ell=1\) gives
\begin{equation}\label{eq:proof-minimum-upper-bounds}
\begin{aligned}
E_\om^{\rm cont}
\leq
E_\om^{\rm int}
&\leq\lambda_{\om,\widehat n_\om}^{(1)}
=\om e_0+o(\om).
\end{aligned}
\end{equation}
We first prove
\begin{equation}\label{eq:rough-local}
	|\tau_\om-\om|+|n_\om-\om|\leq C\sqrt\om.
\end{equation}
Indeed, for
\(|r-1|\leq1/2\), one has \(1/4\leq r^2\leq9/4\), and the radial
potential satisfies
\begin{equation}\label{eq:radial-potential}
\begin{aligned}
\Big(\frac{\nu}{r}-\om r\Big)^2
+\frac{\om^2}{8}(r^2-1)^2
&=\frac{\bigl((\nu-\om)-\om(r^2-1)\bigr)^2}{r^2}
+\frac{\om^2}{8}(r^2-1)^2\\
&\geq c\Big(
\bigl((\nu-\om)-\om(r^2-1)\bigr)^2
+\om^2(r^2-1)^2
\Big)\geq c(\nu-\om)^2.
\end{aligned}
\end{equation}
For \(|r-1|>1/2\), the trapping term is at least \(c\om^2\).
Consequently, pointwise for every \(r>0\),
\begin{equation}\label{eq:global-potential-bound}
\Big(\frac{\nu}{r}-\om r\Big)^2
+\frac{\om^2}{8}(r^2-1)^2
\geq c_*\min\{(\nu-\om)^2,\om^2\}.
\end{equation}
Dropping the kinetic term and applying this bound at
\(\nu=\tau_\om\) and \(\nu=n_\om\), we obtain from
\eqref{eq:proof-minimum-upper-bounds}
\[
c_*\Big(
\min\{(\tau_\om-\om)^2,\om^2\}
+\min\{(n_\om-\om)^2,\om^2\}
\Big)
\leq E_\om^{\rm cont}+E_\om^{\rm int}
\leq C\om.
\]
For large \(\om\), neither minimum can equal \(\om^2\), hence \eqref{eq:rough-local} holds true.

Based on \eqref{eq:rough-local}, we next prove \eqref{eq:E-size-new1}.  After passing to a subsequence \(\Omega_k\to\infty\) as $k\to\infty$ if necessary, there exist some \(q_{\rm cont},q_{\rm int}\in\mathbb R\) such that
\[
\frac{\tau_{\Omega_k}-\Omega_k}{\sqrt{\Omega_k}}
\xrightarrow[k \to \infty]{} q_{\rm cont} 
\quad\hbox{and}\quad
\frac{n_{\Omega_k}-\Omega_k}{\sqrt{\Omega_k}}
\xrightarrow[k \to \infty]{} q_{\rm int}.
\]
Thus, we obtain from \eqref{eq:bounded-q-eigenlimit}  that
\[
\frac{E_{\Omega_k}^{\rm cont}}{\Omega_k}
\xrightarrow[k \to \infty]{} e_0+\frac{q_{\rm cont}^2}{9} 
\quad\hbox{and}\quad
\frac{E_{\Omega_k}^{\rm int}}{\Omega_k}
\xrightarrow[k \to \infty]{} e_0+\frac{q_{\rm int}^2}{9}.
\]
On the other hand, \eqref{eq:proof-minimum-upper-bounds} gives
\[
\limsup_{k\to\infty}\frac{E_{\Omega_k}^{\rm cont}}{\Omega_k}
\leq e_0,
\qquad
\limsup_{k\to\infty}\frac{E_{\Omega_k}^{\rm int}}{\Omega_k}
\leq e_0.
\]
Consequently,
\(
e_0+\frac{1}{9}q_{\rm cont}^2\leq e_0\),
\(e_0+\frac{1}{9}q_{\rm int}^2\leq e_0,
\)
which implies that \(q_{\rm cont}=q_{\rm int}=0\).  Since the original sequence
was arbitrary, this proves \eqref{eq:E-size-new1}.  Applying
\eqref{eq:bounded-q-eigenlimit} at
\(\nu=\tau_\om\) and \(\nu=n_\om\) then gives
\[
\frac{E_\om^{\rm cont}}{\om} \xrightarrow[\om \to \infty]{} e_0
\quad\hbox{and}\quad
\frac{E_\om^{\rm int}}{\om} \xrightarrow[\om \to \infty]{} e_0,
\]
which completes the proof of
\eqref{eq:E-size-new} and \eqref{eq:E-size-new1}.

We next prove \eqref{eq:proof-coarse-fiber-bound}.
Choose \(A\geq2\) so large that \(c_*A^2>e_0+1\), where \(c_*\) is the
constant in \eqref{eq:global-potential-bound}, and set
\begin{equation}\label{eq:def-varie}
\varepsilon_\om:=
\sup_{|\nu-\om|\leq A\sqrt\om}
\lf|
e_{\om,\nu}^{(1)}
-e_0-\frac{(\nu-\om)^2}{9\om}
\ri|
+\lf|\frac{E_\om^{\rm int}}{\om}-e_0\ri|.
\end{equation}
Equations~\eqref{eq:bounded-q-eigenlimit} and
\eqref{eq:E-size-new} imply that \(\varepsilon_\om\to0\).  In the
region \(|\nu-\om|\leq A\sqrt\om\), its definition gives
\[
\lambda_{\om,\nu}^{(1)}
\geq\om e_0+\tfrac19(\nu-\om)^2-\varepsilon_\om\om.
\]
In the complementary region \(|\nu-\om|> A\sqrt\om\), \eqref{eq:global-potential-bound}
and the choice of \(A\) give
\[
\lambda_{\om,\nu}^{(1)}\geq\om(e_0+c_0)
\]
for some \(c_0>0\).  Decreasing \(c>0\) if necessary proves
\eqref{eq:proof-coarse-fiber-bound}.

We finally prove \eqref{eq:radial-gap-new}.  Following
\eqref{eq:bounded-q-eigenlimit}, we have
\[
e_{\om,\nu}^{(2)}-e_{\om,\nu}^{(1)}
\xrightarrow[\om \to \infty]{} 2\beta\quad\hbox{uniformly for}\,\ 
|\nu-\om|\leq A\sqrt\om.\]  
Hence, for all
sufficiently large \(\om\),
\begin{equation}\label{eq:bounded-window-radial-gap}
\lambda_{\om,n}^{(m)}-E_\om^{\rm int}\geq\lambda_{\om,n}^{(m)}-\lambda_{\om,n}^{(1)}\geq\lambda_{\om,n}^{(2)}-\lambda_{\om,n}^{(1)}\geq\beta\om\quad\hbox{uniformly for}\,\ 
|n-\om|\leq A\sqrt\om.
\end{equation}
%
If \(|n-\om|>A\sqrt\om\), then
\eqref{eq:proof-coarse-fiber-bound},
and the fact that
\(\varepsilon_\om\to0\) give
\begin{equation}\label{eq:oct36}
\lambda_{\om,n}^{(m)}-E_\om^{\rm int}\geq\lambda_{\om,n}^{(1)}-E_\om^{\rm int}\geq c\om\quad\hbox{uniformly for}\,\ 
|n-\om|\geq A\sqrt\om.
\end{equation}
We conclude from \eqref{eq:bounded-window-radial-gap} and \eqref{eq:oct36} that
\eqref{eq:radial-gap-new} holds true.
\end{proof}
\begin{cor}\label{lem:uniform-two-mode-HO}
Let \(L_\om\to\infty\) and \(L_\om=o(\sqrt\om)\) as \(\om\to\infty\).  Choose the
eigenfunction signs by the convention in
Lemma~\ref{lem:bounded-q-radial-control}.  Then, uniformly for
\(|\nu-\tau_\om|\leq L_\om\),
\begin{equation}\label{eq:low-mode-eigenvalue-conv}
e_{\om,\nu}^{(1)}\to e_0,
\qquad e_{\om,\nu}^{(2)}\to e_1\quad\hbox{as}\,\ \om\to\infty,
\end{equation}
\begin{equation}\label{eq:low-mode-eigenfunction-conv}
\lf\|\psi_{\om,\nu}^{(1)}-\chi_0\ri\|_{\domain_\om}
+\lf\|\psi_{\om,\nu}^{(2)}-\chi_1\ri\|_{\domain_\om}
\xrightarrow[\om \to \infty]{}0.
\end{equation}
Moreover, for every fixed \(R>0\) and \(m=1,2\),
\begin{equation}\label{eq:Aug-5}
\sup_{|\nu-\tau_\om|\leq L_\om}
\lf\|\psi_{\om,\nu}^{(m)}-\chi_{m-1}\ri\|_{H^1(-R,R)}
\xrightarrow[\om \to \infty]{} 0.
\end{equation}
In particular,
\begin{equation}\label{eq:low-mode-gap}
e_{\om,\nu}^{(m)}-e_{\om,\nu}^{(1)}\geq c
\quad(m\geq2)\quad\hbox{uniformly  for}\,\ 
|\nu-\tau_\om|\leq L_\om.
\end{equation}
\end{cor}

\begin{proof}[\normalfont\bfseries Proof]
By \eqref{eq:E-size-new1},
\begin{equation*}
\sup_{|\nu-\tau_\om|\leq L_\om}
|q_{\om,\nu}|
\leq
\frac{L_\om+|\tau_\om-\om|}{\sqrt\om}
\longrightarrow0\quad\hbox{as}\,\ \om\to\infty,
\end{equation*}
so every family \(\nu=\nu_\om\) in the stated window
satisfies \(q_{\om,\nu_\om}\to0\).
Equation~\eqref{eq:bounded-q-eigenlimit} therefore gives
\eqref{eq:low-mode-eigenvalue-conv}.  To prove the uniform eigenfunction
convergence, suppose that \eqref{eq:low-mode-eigenfunction-conv} fails.
Then there are \(\varepsilon>0\), a sequence \(\Omega_k\to\infty\), and
\(\nu_k\) with \(|\nu_k-\tau_{\Omega_k}|\leq L_{\Omega_k}\) such that
\[
\lf\|\psi_{\Omega_k,\nu_k}^{(1)}-\chi_0 \ri\|_{\domain_{\Omega_k}}
+ \lf\|\psi_{\Omega_k,\nu_k}^{(2)}-\chi_1 \ri\|_{\domain_{\Omega_k}}
\geq\varepsilon.
\]
The estimate for
\(\sup_{|\nu-\tau_\om|\leq L_\om}|q_{\om,\nu}|\) gives
\(q_{\Omega_k,\nu_k}\to0\), whereas
Lemma~\ref{lem:bounded-q-radial-control} gives convergence of the
left-hand side to zero with the prescribed signs.  This contradiction
proves \eqref{eq:low-mode-eigenfunction-conv}.  

We finally prove \eqref{eq:Aug-5} and \eqref{eq:low-mode-gap}. If  \eqref{eq:Aug-5} failed, there must exist fixed \(R>0\),
\(m\in\{1,2\}\), and \(\varepsilon>0\), together with
\(\Omega_k\to\infty\) and
\(|\nu_k-\tau_{\Omega_k}|\leq L_{\Omega_k}\), such that
\begin{equation*}
	\lf\|\psi_{\om_k,\nu_k}^{(m)}-\chi_{m-1}\ri\|_{H^1(-R,R)}
	\geq \varepsilon>0\quad\hbox{for all}\,\ k.
\end{equation*}
Since \(L_{\Omega_k}=o(\sqrt{\Omega_k})\) as \(\om_k\to\infty\), we have
\(q_{\Omega_k,\nu_k}\to0\).  Thus the local \(H^1\) convergence in
Lemma~\ref{lem:bounded-q-radial-control} holds true, a contradiction. This gives \eqref{eq:Aug-5}.
Since
\(e_{\om,\nu}^{(m)}\geq e_{\om,\nu}^{(2)}\) for \(m\geq2\), we have
\begin{equation*}
e_{\om,\nu}^{(m)}-e_{\om,\nu}^{(1)}\geq e_{\om,\nu}^{(2)}-e_{\om,\nu}^{(1)}
\longrightarrow e_1-e_0=2\beta
\end{equation*}
uniformly  for
\(|\nu-\tau_\om|\leq L_\om\),  and  \eqref{eq:low-mode-gap} follows.
\end{proof}

\subsection{Proof of Proposition~\ref{prop:first-band-expansion}}

\begin{lemma}\label{lem:FH-spectral-sum-strict}
Let \(L_\om\to\infty\),  \(L_\om=o(\sqrt\om)\) as \(\om\to\infty\), and
\begin{equation*} 
S_{\om,\nu}:=\sum_{m\ge2}\frac{1}{\lambda_{\om,\nu}^{(m)}-\lambda_{\om,\nu}^{(1)}} \lf|
	\int_0^\infty\Big(\frac{2\nu}{r^2}-2\om\Big)
	\phi_{\om,\nu}^{(m)}(r)
	\phi_{\om,\nu}^{(1)}(r)r\,dr
	\ri|^2,
\end{equation*}
where $(\phi_{\om,\nu}^{(m)},\lambda_{\om,\nu}^{(m)})$ is the $m$-th eigenpair of $\mathcal L_{\om,\nu}$.
Then we have 
\begin{equation}\label{eq:series-convergence}
\lim_{\om\to\infty}S_{\om,\nu}=\tfrac89\quad\hbox{uniformly for}\,\ |\nu-\tau_\om|\leq L_\om.
\end{equation}
\end{lemma}

\begin{proof}[\normalfont\bfseries Proof]
Set
\beq
D_{\om,\nu}(t):=\frac1{\sqrt\om}
\Big(\frac{2\nu}{\rho_\om(t)^2}-2\om\Big),
\qquad
b_{\om,\nu}:=D_{\om,\nu}\psi_{\om,\nu}^{(1)},
\qquad
b(t):=-4t\chi_0(t),
\eeq
where $\rho_\om(t)=1+t/\sqrt{\om}$ and $\chi_0(t)$ is given by \eqref{eq:limiting-HO}. Let \(b_{\om,\nu}^\perp\) be the orthogonal
projection of \(b_{\om,\nu}\) onto
\(\{\psi_{\om,\nu}^{(1)}\}^\perp\) in \(\domain_\om\). By the definition of $\psi_{\om,\nu}^{(m)}$ in \eqref{eq:rescaled-eigenpairs},
we obtain that
\[
S_{\om,\nu}=\sum_{m\geq2}
\frac{1}{e_{\om,\nu}^{(m)}-e_{\om,\nu}^{(1)}} \lf| \braketr{b_{\om,\nu}^\perp}{
	\psi_{\om,\nu}^{(m)}}_{\domain_\om} \ri|^2,
\]

We claim that
\begin{equation}\label{eq:proof-b-convergence}
\sup_{|\nu-\tau_\om|\leq L_\om}
\Big(
\int_{I_\om}|b_{\om,\nu}(t)-b(t)|^2
\rho_\om(t)\,dt
\Big)^{1/2}
\xrightarrow[\om \to \infty]{} 0.
\end{equation}
Assuming \eqref{eq:proof-b-convergence}, we prove
\eqref{eq:series-convergence} as follows.  
Since \(b=-4t\chi_0\) and \(\chi_0\) is even, we have
\[
\int_{\mathbb R}b(t)\chi_0(t)\,dt
=
-4\int_{\mathbb R}t\chi_0(t)^2\,dt
=0,
\]
which implies that \(b\perp\chi_0\) in \(L^2(\mathbb R)\).
Thus \eqref{eq:proof-b-convergence} and the fact that
\(\psi_{\om,\nu}^{(1)}\to\chi_0\) in \eqref{eq:low-mode-eigenfunction-conv} imply
\begin{equation}\label{eq:proof-bperp-convergence}
	\lf\|b_{\om,\nu}^\perp-b\ri\|_{\domain_\om}\xrightarrow[\om \to \infty]{}0.
\end{equation}
Rewrite
\begin{equation*}
	S_{\om,\nu}
	=\frac{1}{e_{\om,\nu}^{(2)}-e_{\om,\nu}^{(1)}} \lf| \braketr{b_{\om,\nu}^\perp}{
		\psi_{\om,\nu}^{(2)}}_{\domain_\om} \ri|^2
	+T_{\om,\nu},
\end{equation*}
where \(T_{\om,\nu}\) is defined by
\[
T_{\om,\nu}:=\sum_{m\geq3}
\frac{1}{e_{\om,\nu}^{(m)}-e_{\om,\nu}^{(1)}} \lf| \braketr{b_{\om,\nu}^\perp}{
	\psi_{\om,\nu}^{(m)}}_{\domain_\om} \ri|^2.
\]
By \eqref{eq:first-two-HO-modes},
\(t\chi_0=(2\beta)^{-1/2}\chi_1\) and \(e_1-e_0=2\beta\).  Hence we obtain from \eqref{eq:proof-bperp-convergence} and \eqref{eq:low-mode-eigenfunction-conv} that the first term in $S_{\om,\nu}$
satisfies
\begin{equation*}
	\lim_{\om\to\infty}\frac{1}
	{e_{\om,\nu}^{(2)}-e_{\om,\nu}^{(1)}} \lf| \braketr{b_{\om,\nu}^\perp}{
		\psi_{\om,\nu}^{(2)}}_{\domain_\om} \ri|^2 =\frac{\lf| \braketr{-4t\chi_0}{\chi_1} \ri|^2}{e_1-e_0}
	=\frac4{\beta^2}=\frac89.
\end{equation*}
By \eqref{eq:low-mode-gap},
\begin{equation*}
	0\leq T_{\om,\nu}
	\leq C\sum_{m\geq3}
	\lf| \braketr{b_{\om,\nu}^\perp}{
	\psi_{\om,\nu}^{(m)}}_{\domain_\om} \ri|^2.
\end{equation*}
Since \(b_{\om,\nu}^\perp\perp
\psi_{\om,\nu}^{(1)}\), Parseval's identity gives
\begin{equation*}
	0\leq T_{\om,\nu}\leq C \lf(
	\lf\|b_{\om,\nu}^\perp \ri\|_{\domain_\om}^2
	-\lf| \braketr{b_{\om,\nu}^\perp}{
		\psi_{\om,\nu}^{(2)}}_{\domain_\om} \ri|^2 \ri).
\end{equation*}
The expression in parentheses converges, by
\eqref{eq:proof-bperp-convergence} 
and \eqref{eq:low-mode-eigenfunction-conv}, to
\begin{equation*}
	\lf\|b\ri\|_{L^2(\R)}^2-\lf| \braketr{b}{\chi_1} \ri|^2=0.
\end{equation*}
 Hence
\(T_{\om,\nu}=o(1)\) uniformly in the stated window.  Thus
\eqref{eq:proof-b-convergence} implies
\eqref{eq:series-convergence}.

It remains to prove \eqref{eq:proof-b-convergence}.  The exact identity
\begin{equation}\label{eq:proof-D-exact}
D_{\om,\nu}(t)
=
\frac{2(q_{\om,\nu}-2t-t^2/\sqrt\om)}
{\rho_\om(t)^2}\quad\hbox{where}\,\ q_{\om,\nu}=\frac{\nu-\om}{\sqrt{\om}}\,\ \hbox{is given by}\,\ \eqref{eq:def-qom},
\end{equation}
shows that \(D_{\om,\nu}\to-4t\) uniformly on every compact interval
and uniformly for \(|\nu-\tau_\om|\leq L_\om\).
Corollary~\ref{lem:uniform-two-mode-HO} \eqref{eq:low-mode-eigenfunction-conv} therefore gives, for every fixed
\(R>0\),
\begin{equation}\label{eq:oct37}
\sup_{|\nu-\tau_\om|\leq L_\om}
\lf\|b_{\om,\nu}-b\ri\|_{L^2(-R,R)}
\xrightarrow[\om \to \infty]{} 0.
\end{equation}
To upgrade this local convergence to
\eqref{eq:proof-b-convergence}, we next prove the uniform tail bound
\begin{equation}\label{eq:global-conv}
\lim_{R\to\infty}\limsup_{\om\to\infty}
\sup_{|\nu-\tau_\om|\leq L_\om}
\int_{I_\om\cap\{|t|>R\}}
\lf|D_{\om,\nu}\psi_{\om,\nu}^{(1)} \ri|^2
\rho_\om\,dt=0.
\end{equation}

Since \(D_{\om,\nu}\) is singular where \(\rho_\om=0\), we split
the integration region in \eqref{eq:global-conv} as
\[
I_\om\cap\{|t|>R\}
=
\{t\in I_\om:|t|>R,\ \rho_\om(t)\geq1/2\}
\cup
\{t\in I_\om:|t|>R,\ 0<\rho_\om(t)<1/2\}.
\]
We first consider the region
\(\{t\in I_\Omega:|t|>R,\ \rho_\Omega(t)\geq1/2\}\).
Equation~\eqref{eq:proof-D-exact} gives
\begin{equation}\label{eq:proof-D-bulk-bound}
|D_{\om,\nu}(t)|^2\rho_\om(t)\leq C(1+t^2)\quad\hbox{uniformly in}\,\  |\nu-\tau_\om|\leq L_\om.
\end{equation}
Since
\(\sup_{|\nu-\tau_\om|\leq L_\Omega}|q_{\Omega,\nu}|\to0\),
\eqref{eq:rescaled-potential-q} gives
\(V_{\Omega,\nu}\to9t^2/2\)
and \(\rho_\Omega\to1\)
uniformly on \((-R,R)\) as \(\om\to\infty\).  The eigenvalue and local \(H^1\)
convergences \eqref{eq:low-mode-eigenvalue-conv} and \eqref{eq:Aug-5}
in Corollary~\ref{lem:uniform-two-mode-HO} therefore give, for each fixed
\(R>0\),
\bmln{
\int_{I_\Omega\cap\{|t|>R\}}
\lf( \lf|(\psi_{\Omega,\nu}^{(1)})' \ri|^2
+V_{\Omega,\nu} \lf|\psi_{\Omega,\nu}^{(1)} \ri|^2 \ri)\rho_\Omega\,dt =e_{\Omega,\nu}^{(1)}
-\int_{-R}^{R} \lf( \lf| (\psi_{\Omega,\nu}^{(1)})' \ri|^2
+V_{\Omega,\nu} \lf|\psi_{\Omega,\nu}^{(1)} \ri|^2 \ri)\rho_\Omega\,dt\\
\xrightarrow[\om \to \infty]{}
e_0-\int_{-R}^{R} \lf(|\chi_0'|^2+\frac92t^2|\chi_0|^2 \ri)\,dt
=\int_{|t|>R} \lf(|\chi_0'|^2+\frac92t^2|\chi_0|^2 \ri)\,dt,
}
uniformly for \(|\nu-\tau_\om|\leq L_\Omega\).  The last integral tends to zero as
\(R\to\infty\).  Since \(2+t/\sqrt{\Omega}\geq1\) on \(I_\Omega\),
\eqref{eq:rescaled-potential-q} gives
\(V_{\Omega,\nu}\geq t^2/8\).  Thus, for \(R\geq1\),
\bmln{
\int_{\{t\in I_\om:|t|>R,\ \rho_\om(t)\geq1/2\}}
(1+t^2) \lf|\psi_{\Omega,\nu}^{(1)} \ri|^2\rho_\Omega\,dt\\
\leq 16\int_{\{t\in I_\om:|t|>R,\ \rho_\om(t)\geq1/2\}}
\lf( \lf|(\psi_{\Omega,\nu}^{(1)})' \ri|^2
+V_{\Omega,\nu} \lf|\psi_{\Omega,\nu}^{(1)} \ri|^2 \ri)\rho_\Omega\,dt.
}
Combining these estimates with \eqref{eq:proof-D-bulk-bound},  we obtain
\begin{equation}\label{eq:proof-D-bulk-tail}
\lim_{R\to\infty}\limsup_{\om\to\infty}
\sup_{|\nu-\tau_\om|\leq L_\om}
\int_{\{t\in I_\om:|t|>R,\ \rho_\om(t)\geq1/2\}}
\lf|D_{\om,\nu}\psi_{\om,\nu}^{(1)} \ri|^2
\rho_\om\,dt=0.
\end{equation}

It remains to consider the region \(\{t\in I_\om:|t|>R,\ 0<\rho_\om(t)<1/2\}\).  Uniformly for
\(|\nu-\tau_\om|\leq L_\Omega\),
Lemmas \ref{lem:bounded-q-radial-control} and \ref{lem:standard-spectral-input} give
\[
\nu=\Omega+o(\sqrt{\Omega}),
\qquad e_{\Omega,\nu}^{(1)}=O(1).
\]
Thus 
\(\nu-\Omega\rho_\Omega^2\geq\Omega/4\) when
\(0<\rho_\Omega\leq3/4\) and \(\Omega\) is sufficiently large.  Consequently, by the definition of \(V_{\Omega,\nu}\) in  \eqref{eq:rescaled-radial-operator}, there exists \(c>0\) such that
\begin{equation}\label{eq:proof-origin-coercivity}
V_{\Omega,\nu}-e_{\Omega,\nu}^{(1)}=\frac{(\nu-\Omega\rho_\Omega^2)^2}{\Omega\rho_\Omega^2}
+\frac{\Omega}{8}(\rho_\Omega^2-1)^2-e_{\Omega,\nu}^{(1)}
\geq\frac{c\Omega}{\rho_\Omega^2}
\qquad(0<\rho_\Omega\leq3/4).
\end{equation}

Choose \(\chi\in C^\infty([0,\infty);[0,1])\) such that \(\chi=1\) on 
\([0,1/2]\) and \(\chi=0\) on \([3/4,\infty)\). Set
\[
f_{\Omega,\varepsilon}(t):=
\frac{\chi(\rho_\Omega(t))}{\rho_\Omega(t)+\varepsilon},
\qquad \varepsilon>0.
\]
For fixed \(\Omega\) and \(\varepsilon\), both
\(f_{\Omega,\varepsilon}\) and \(f'_{\Omega,\varepsilon}\)  are bounded.  Testing the eigenvalue equation \eqref{sep25-1}
with   \(f_{\Omega,\varepsilon}^2\psi_{\Omega,\nu}^{(1)}\)
and expanding derivatives gives
\bmln{
\int_{I_\Omega} \lf|f_{\Omega,\varepsilon}' \ri|^2
\lf|\psi_{\Omega,\nu}^{(1)} \ri|^2\rho_\Omega\,dt
=\int_{I_\Omega} \lf(
\lf|(f_{\Omega,\varepsilon}\psi_{\Omega,\nu}^{(1)})' \ri|^2
+(V_{\Omega,\nu}-e_{\Omega,\nu}^{(1)})
f_{\Omega,\varepsilon}^2 \lf|\psi_{\Omega,\nu}^{(1)} \ri|^2
\ri)\rho_\Omega\,dt\\
\geq c\Omega\int_{I_\Omega}
\frac{1}{\rho_\Omega} f_{\Omega,\varepsilon}^2\lf|\psi_{\Omega,\nu}^{(1)}\ri|^2
\,dt,
}
where we used \eqref{eq:proof-origin-coercivity} in the last inequality.
On the other hand, since \(\rho_\Omega'=\Omega^{-1/2}\), there exists  \(C>0\), independent of \(\Omega,\nu,\varepsilon\), such that 
\[
|f_{\Omega,\varepsilon}'|^2
\leq\frac{C}{\Omega}\mathds1_{\{1/2<\rho_\Omega<3/4\}}
+\frac{2}{\Omega}\frac{f_{\Omega,\varepsilon}^2}{\rho_\Omega^2}.
\]
Combining the above two inequalities and using
\( \lf\|\psi_{\Omega,\nu}^{(1)} \ri\|_{\domain_\Omega}=1\), we obtain
\[
\left(c\Omega-\frac2\Omega\right)
\int_{I_\Omega}
\frac{1}{\rho_\Omega} f_{\Omega,\varepsilon}^2 \lf|\psi_{\Omega,\nu}^{(1)} \ri|^2
\,dt\leq\frac C\Omega.
\]
As \(\varepsilon\searrow0\), we have
\[
f_{\Omega,\varepsilon}(t)
\nearrow \frac{\chi(\rho_\Omega(t))}{\rho_\Omega(t)}
\qquad\hbox{for every }t\in I_\Omega.
\]
Since \(\chi=1\) on \([0,1/2]\), the monotone convergence theorem gives
\begin{equation}\label{eq:proof-origin-weighted-estimate}
	\int_{\rho_\Omega<1/2}
	\frac{1}{\rho_\Omega^3} \lf|\psi_{\Omega,\nu}^{(1)}\ri|^2\,dt
	\leq\frac{C}{\Omega^2}
	\quad\hbox{uniformly for }|\nu-\tau_\om|\leq L_\Omega.
\end{equation}
By \eqref{eq:proof-D-exact} and \(\nu\leq C\Omega\),
\[
|D_{\Omega,\nu}|^2\rho_\Omega
\leq C\Omega\rho_\Omega^{-3}
\qquad(0<\rho_\Omega<1/2).
\]
Thus \eqref{eq:proof-origin-weighted-estimate} implies, uniformly for \(|\nu-\tau_\om|\leq L_\Omega\),
\begin{equation}\label{eq:proof-D-origin-tail}
\int_{\rho_\Omega<1/2}
\lf|D_{\Omega,\nu}\psi_{\Omega,\nu}^{(1)}\ri|^2\rho_\Omega\,dt\leq C\om\int_{\rho_\Omega<1/2}
\frac{1}{\rho_\Omega^3} \lf|\psi_{\Omega,\nu}^{(1)}\ri|^2\,dt
\leq\frac C\Omega\longrightarrow0\quad\hbox{as}\,\ \om\to\infty.
\end{equation}
  Combining
\eqref{eq:proof-D-origin-tail} with \eqref{eq:proof-D-bulk-tail}, we obtain
\begin{equation}\label{eq:proof-D-global-tail}
\lim_{R\to\infty}\limsup_{\om\to\infty}
\sup_{|\nu-\tau_\om|\leq L_\om}
\int_{I_\om\cap\{|t|>R\}}
\lf|D_{\om,\nu}\psi_{\om,\nu}^{(1)}\ri|^2
\rho_\om\,dt=0.
\end{equation}
Since \(b=-4t\chi_0\) decays exponentially at infinity, the local \(L^2\) convergence \eqref{eq:oct37}
and \eqref{eq:proof-D-global-tail} prove
\eqref{eq:proof-b-convergence}. This completes the proof of Lemma \ref{lem:FH-spectral-sum-strict}.
\end{proof}
\smallskip
\begin{proof}[\normalfont\bfseries Proof of Proposition~\ref{prop:first-band-expansion}]
We first show that the first eigenpair depends smoothly on \(\nu\) and
derive a formula for
\(\partial_{\nu\nu}\lambda_{\om,\nu}^{(1)}\).  Since
\(\tau_\om=\om+o(\sqrt\om)\), every window
\(|\nu-\tau_\om|=o(\sqrt\om)\) considered below satisfies
\(\nu>0\) for all sufficiently large \(\om\). For each fixed sufficiently large \(\Omega\), the form domain
\(\mathscr D(\mathfrak q_{\Omega,\nu})\) defined in
\eqref{eq:fiber-form-domain} is independent of \(\nu\) in this
window. 
Since the first eigenvalue is simple, analytic
perturbation theory
\cite[Chapter~VII, Sects.~1 and~4]{Kato1995} allows
\(\phi_{\om,\nu}^{(1)}\) to be chosen real, normalized, and smoothly
depending on \(\nu\).  Moreover,
\eqref{eq:proof-origin-weighted-estimate} and
\eqref{eq:radial-unitary-map} give
\[
\int_0^{1/2}
\frac{1}{r^3} \lf|\phi_{\Omega,\nu}^{(1)}(r) \ri|^2\,dr
=\int_{\rho_\Omega<1/2}
\frac{1}{\rho_\Omega^3} \lf|\psi_{\Omega,\nu}^{(1)} \ri|^2\,dt
\leq\frac C{\Omega^2}.
\]
On \([1/2,\infty)\), \(r^{-4}\leq16\), so
\(r^{-2}\phi_{\Omega,\nu}^{(1)}\in L^2(\mathbb R_+,r\,dr)\).
Thus the differentiations and the \(L^2(\mathbb R_+,r\,dr)\) eigenfunction
expansion below are justified.  Since
\(\partial_\nu\mathcal L_{\om,\nu}=2\nu/r^2-2\om\), differentiating
the eigenvalue equation with respect to \(\nu\) gives
\begin{equation}\label{eq:partial-lam1}
\partial_\nu\lambda_{\om,\nu}^{(1)}
=
\braketr{
\phi_{\om,\nu}^{(1)}}{
\Big(\frac{2\nu}{r^2}-2\om\Big)
\phi_{\om,\nu}^{(1)}}.
\end{equation}

Differentiating the eigenvalue equation $\bigl(\mathcal L_{\om,\nu}-\lambda_{\om,\nu}^{(1)}\bigr)
\phi_{\om,\nu}^{(1)}
=0$ yields
\begin{equation}\label{eq:partial-lam2}
\bigl(\mathcal L_{\om,\nu}-\lambda_{\om,\nu}^{(1)}\bigr)
\partial_\nu\phi_{\om,\nu}^{(1)}
=-
\Big(
\frac{2\nu}{r^2}-2\om
-\partial_\nu\lambda_{\om,\nu}^{(1)}
\Big)\phi_{\om,\nu}^{(1)}.
\end{equation}
Since \(\{\phi_{\Omega,\nu}^{(m)}\}_{m\geq1}\) is an orthonormal
basis of \(L^2(\mathbb R_+,r\,dr)\) and
\( \braketr{\phi_{\Omega,\nu}^{(1)}}{
\partial_\nu\phi_{\Omega,\nu}^{(1)}}=0\) due to \( \braketr{\phi_{\Omega,\nu}^{(1)}}{
\phi_{\Omega,\nu}^{(1)}}=1\), we have
\[
\partial_\nu\phi_{\Omega,\nu}^{(1)}
=\sum_{m\geq2}
\braketr{
\phi_{\Omega,\nu}^{(m)}}{
\partial_\nu\phi_{\Omega,\nu}^{(1)}
}
\phi_{\Omega,\nu}^{(m)}
\quad\text{in }L^2(\mathbb R_+,r\,dr).
\]
Taking the inner product of \eqref{eq:partial-lam2} with
\(\phi_{\Omega,\nu}^{(m)}\) for each \(m\geq2\) determines
these coefficients and yields
\begin{equation}\label{eq:partial-lam3}
\partial_\nu\phi_{\om,\nu}^{(1)}
=-\sum_{m\geq2}
\frac{1}{
\lambda_{\om,\nu}^{(m)}-\lambda_{\om,\nu}^{(1)}
}
\braketr{
\phi_{\om,\nu}^{(m)}}{
\Big(\frac{2\nu}{r^2}-2\om\Big)
\phi_{\om,\nu}^{(1)}}
\phi_{\om,\nu}^{(m)}
\end{equation}
in \(L^2(\mathbb R_+,r\,dr)\).  Differentiating \eqref{eq:partial-lam1}
once more and using
\(\partial_{\nu\nu}\mathcal L_{\om,\nu}=2/r^2\) gives
\[
\begin{aligned}
\partial_{\nu\nu}\lambda_{\om,\nu}^{(1)}
=&
\braketr{
\phi_{\om,\nu}^{(1)}}{
\frac2{r^2}\phi_{\om,\nu}^{(1)}}
+2 \braketr{
\partial_\nu\phi_{\om,\nu}^{(1)}}{
\Big(\frac{2\nu}{r^2}-2\om\Big)
\phi_{\om,\nu}^{(1)}}.
\end{aligned}
\]
Substituting \eqref{eq:partial-lam3} therefore yields
\begin{equation}\label{eq:partial-lam4}
\partial_{\nu\nu}\lambda_{\om,\nu}^{(1)}
= \braketr{
\phi_{\om,\nu}^{(1)}}{
\frac2{r^2}\phi_{\om,\nu}^{(1)}
}
-2\sum_{m\ge2}\frac{1}
{\lambda_{\om,\nu}^{(m)}-\lambda_{\om,\nu}^{(1)}
} \lf|
 \braketr{\phi_{\om,\nu}^{(m)}}{
\Big(\frac{2\nu}{r^2}-2\om\Big)
\phi_{\om,\nu}^{(1)}
}
\ri|^2.
\end{equation}
We next analyze the two terms in \eqref{eq:partial-lam4} uniformly in the window
\(|\nu-\tau_\om|=o(\sqrt\om)\).  For the first term, the change of variables
\(r=\rho_\om(t)\) gives
\begin{equation*}
\braketr{
\phi_{\om,\nu}^{(1)}}{
\frac2{r^2}\phi_{\om,\nu}^{(1)}}
=2\int_{I_\om}
\frac{1}{\rho_\om(t)} \lf|\psi_{\om,\nu}^{(1)}(t)\ri|^2\,dt.
\end{equation*}
For all sufficiently large \(\om\),
\(\{\rho_\om<1/2\}\subset\{t<-R\}\), and hence
\[
I_\om
=
[-R,R]
\,\cup\,
\bigl\{|t|>R,\ \rho_\om\geq1/2\bigr\}
\,\cup\,
\bigl\{\rho_\om<1/2\bigr\}.
\]
On \([-R,R]\), \(\rho_\om^{-1}\to1\) uniformly and
\(\psi_{\om,\nu}^{(1)}\to\chi_0\) in \(L^2(-R,R)\) as \(\om\to\infty\).  Therefore
\[
\int_{-R}^R
\frac{1}{\rho_\om} \lf|\psi_{\om,\nu}^{(1)} \ri|^2\,dt
\longrightarrow
\int_{-R}^R|\chi_0|^2\,dt\quad\hbox{uniformly for}\,\ |\nu-\tau_\om|\leq L_\om
\]
as \(\om\to\infty\).  On the second region,
\(\rho_\om^{-1}\leq4\rho_\om\), so \eqref{eq:bounded-q-low-energy} in 
Lemma~\ref{lem:bounded-q-radial-control}  gives
\[
\int_{\substack{|t|>R,\,\rho_\om\geq1/2}}
\frac{1}{\rho_\om} \lf|\psi_{\om,\nu}^{(1)} \ri|^2\,dt
\leq\frac{4}{R^2}
\int_{I_\om}t^2|\psi_{\om,\nu}^{(1)}|^2\rho_\om\,dt
\leq\frac{C}{R^2}.
\]
On the third region, \eqref{eq:proof-origin-weighted-estimate} gives
\[
\int_{\rho_\Omega<1/2}
\frac{1}{\rho_\om} \lf|\psi_{\om,\nu}^{(1)} \ri|^2\,dt
\leq\frac14\int_{\rho_\Omega<1/2}
\frac{1}{\rho^3_\om} \lf|\psi_{\om,\nu}^{(1)} \ri|^2\,dt
\leq\frac C{\Omega^2}.
\]
Letting first \(\om\to\infty\)
and then \(R\to\infty\), we obtain
\begin{equation}\label{eq:proof-r-inverse-limit}
\braketr{
\phi_{\om,\nu}^{(1)}}{
\frac2{r^2}\phi_{\om,\nu}^{(1)}
}
=2+o(1)\quad \hbox{uniformly for}\,\ |\nu-\tau_\om|\leq L_\om\,\ \hbox{as}\,\ \om\to\infty.
\end{equation}
For the second term, Lemma~\ref{lem:FH-spectral-sum-strict} gives
\[
\sum_{m\ge2}\frac{1}
{\lambda_{\om,\nu}^{(m)}
-\lambda_{\om,\nu}^{(1)}
} \lf|
 \braketr{\phi_{\om,\nu}^{(m)}}{
\Big(\frac{2\nu}{r^2}-2\om\Big)
\phi_{\om,\nu}^{(1)}} \ri|^2=\frac89+o(1)\quad\hbox{as}\,\ \om\to\infty.
\] 
Substituting Lemma~\ref{lem:FH-spectral-sum-strict} and
\eqref{eq:proof-r-inverse-limit} into \eqref{eq:partial-lam4} gives
\begin{equation}\label{eq:2ndpart-lam1}
\partial_{\nu\nu}\lambda_{\om,\nu}^{(1)}
=2-2\cdot\frac89+o(1)=\frac29+o(1)
\quad\hbox{as}\,\ \om\to\infty,
\end{equation}
which proves \eqref{eq:uniform-curvature-new}.

In the following, we prove
\begin{equation}\label{Sep25-2}
|\alpha_\om|\leq1\quad\hbox{for all sufficiently
large}\,\  \om.
\end{equation} 
Set \(F_\om(s)=\lambda_{\om,\tau_\om+s}^{(1)}\).  Since the function \(\nu\mapsto\lambda_{\Omega,\nu}^{(1)}\)
attains its minimum at \(\nu=\tau_\om\),
\(F_\om'(0)=0\).  Put
\[
B_\om
:=|\tau_\om-\om|+|n_\om-\om|
+\sqrt{\varepsilon_\om\om}+\om^{1/4},
\]
where \(\varepsilon_\om\) is the sequence in
\eqref{eq:proof-coarse-fiber-bound}.  By \eqref{eq:E-size-new1},
\(B_\om=o(\sqrt\om)\).  Choose \(K_\om\) such that
\begin{equation}\label{eq:adaptive-angular-window}
B_\om=o(K_\om),
\qquad K_\om=o(\sqrt\om).
\end{equation}
Then \(|\alpha_\om|=o(K_\om)\).  Applying
\eqref{eq:uniform-curvature-new} with \(K_\om\) in place of \(L_\om\),
we find that \(F_\om\) is strictly convex on
\([-K_\om,K_\om]\) for all sufficiently large \(\om\).  If
\(\alpha_\om>1\), then
\(-\alpha_\om<1-\alpha_\om<0\).  Since
\(F_\om'(0)=0\), the strict convexity implied by
\eqref{eq:2ndpart-lam1} gives \(F_\om'(x)<0\) for \( \mic{- K_\om \leq} x<0\), and hence
\[
\lambda_{\om,n_\om+1}^{(1)}
=F_\om(1-\alpha_\om)
<F_\om(-\alpha_\om)
=\lambda_{\om,n_\om}^{(1)},
\]
contrary to the minimizing property of \(n_\om\) over the integers.  The
competitor \(n_\om-1\) gives the analogous contradiction when
\(\alpha_\om<-1\).  This gives \eqref{Sep25-2}.

Based on \eqref{Sep25-2}, we next prove \eqref{eq:alpha-error}.  Taylor's formula and
\eqref{eq:2ndpart-lam1} give
\[
0\leq
\lambda_{\om,n_\om+1}^{(1)}
-\lambda_{\om,n_\om}^{(1)}
=F_\om(1-\alpha_\om)-F_\om(-\alpha_\om)
=\tfrac19(1-2\alpha_\om)+o(1)
\quad\hbox{if }\alpha_\om\geq0,
\]
and
\[
0\leq
\lambda_{\om,n_\om-1}^{(1)}
-\lambda_{\om,n_\om}^{(1)}
=F_\om(-1-\alpha_\om)-F_\om(-\alpha_\om)
=\tfrac19(1+2\alpha_\om)+o(1)
\quad\hbox{if }\alpha_\om<0.
\]
These two inequalities prove \eqref{eq:alpha-error}.

We now prove the uniform Taylor expansion
\eqref{eq:int-window-expansion}.  Define
\begin{equation*}
\kappa_\om
:=
\sup_{|\nu-\tau_\om|\leq L_\om}
\lf|
\partial_{\nu\nu}\lambda_{\om,\nu}^{(1)}-\tfrac29
\ri|.
\end{equation*}
Then \(\kappa_\om\to0\) as \(\om\to\infty\).   Taylor's formula gives, for
\(|s|\leq L_\om\),
\begin{equation}\label{eq:F-error}
\Big|
F_\om(s)-F_\om(0)-\tfrac19s^2
\Big|
\leq\tfrac{1}{2} \kappa_\om s^2.
\end{equation}
Since \(F_\om(0)=E_\om^{\rm cont}\) and
\(F_\om(-\alpha_\om)=E_\om^{\rm int}\), taking
\(s=-\alpha_\om\) yields
\begin{equation}\label{eq:E-erroe}
E_\om^{\rm int}-E_\om^{\rm cont}
=\tfrac19\alpha_\om^2+O(\kappa_\om).
\end{equation}
For \(|j-\alpha_\om|\leq L_\om\), we take
\(s=j-\alpha_\om\) in \eqref{eq:F-error}, subtract
\eqref{eq:E-erroe}, and use \eqref{eq:alpha-error}.  The resulting error
is bounded by
\(
C\kappa_\om\bigl(1+(j-\alpha_\om)^2\bigr).
\)
Since \(\kappa_\om\to0\), this proves
\eqref{eq:int-window-expansion}.

It remains to prove the global angular gap
\eqref{eq:angular-gap-new}.  Write \(d=j-\alpha_\om\).  Applying
\eqref{eq:int-window-expansion} with \(K_\om\) in place of
\(L_\om\), we obtain, uniformly for \(|d|\leq K_\om\),
\begin{equation*}
\lambda_{\om,n_\om+j}^{(1)}-E_\om^{\rm int}
=\tfrac19(d^2-\alpha_\om^2)
+o(1+d^2)
\end{equation*}
Since \(\alpha_\om\) is bounded,
this implies
\begin{equation}\label{eq:proof-inner-angular-gap}
\lambda_{\om,n_\om+j}^{(1)}-E_\om^{\rm int}
\geq c d^2-C
\qquad (|d|\leq K_\om).
\end{equation}
For the complementary region \(|d|> K_\om\), \eqref{eq:adaptive-angular-window} and
\(n_\om+j=\tau_\om+d\) give, uniformly for \(|d|>K_\om\),
\[
|n_\om+j-\om|
\geq |d|-|\tau_\om-\om|
\geq\tfrac12|d|
\]
for all sufficiently large \(\om\).  Since the definition of
\(\varepsilon_\om\) in \eqref{eq:def-varie} gives
\(E_\om^{\rm int}\leq\om(e_0+\varepsilon_\om)\), \eqref{eq:proof-coarse-fiber-bound} yields
\begin{equation*}
\lambda_{\om,n_\om+j}^{(1)}-E_\om^{\rm int}
\geq
c\min\{d^2,\om\}
-2\varepsilon_\om\om.
\end{equation*}
The definition of \(B_\om\) and
\eqref{eq:adaptive-angular-window} imply
\(\varepsilon_\om\om=o(K_\om^2)\).  Since
\(|d|>K_\om\) and \(K_\om=o(\sqrt\om)\),
\(\min\{d^2,\om\}\geq K_\om^2\),  the term
\(2\varepsilon_\om\om\) is absorbed into the positive term for all
sufficiently large \(\Omega\). Hence
\begin{equation}\label{eq:proof-exterior-angular-gap}
\lambda_{\Omega,n_\Omega+j}^{(1)}-E_\Omega^{\rm int}
\geq\frac{1}{2} c \min\{d^2,\Omega\}
\qquad (|d|>K_\Omega).
\end{equation}
Combining \eqref{eq:proof-exterior-angular-gap} and
\eqref{eq:proof-inner-angular-gap} proves
\eqref{eq:angular-gap-new}.  This completes the proof of
Proposition~\ref{prop:first-band-expansion}.
\end{proof}

\section{Reduction to the Effective Circle Problem}
\label{sec:uniform-circle-reduction}
In this section, we prove Theorem~\ref{thm:A}, which establishes
the asymptotic relation \eqref{eq:A-energy-expansion} between the minimum energies of the
two-dimensional GP and effective circle functionals,
together with the corresponding approximation \eqref{eq:A-L2-reduction} of two-dimensional
ground states.

\subsection{Preliminary estimates}
\label{sec:universal-estimates}
In this subsection, we establish some preliminary estimates
for the proof of Theorem~\ref{thm:A}.
\begin{lemma}
\label{lem:radial-profile-comparison}
Let \(L_\om\to\infty\) and \(L_\om=o(\sqrt\om)\) as
\(\om\to\infty\).  Uniformly, as $ \om \to \infty $, for
\(|j-\alpha_\om|\leq L_\om\),
\begin{align}
\lf\|\phi_{\om,n_\om+j}^{(1)}-\phi_{\om,\tau_\om}^{(1)} \ri\|_{L^2(\mathbb R_+,r\,dr)}
&=o(1),\label{eq:radial-profile-L2}\\
\lf\|\phi_{\om,n_\om+j}^{(1)}-\phi_{\om,\tau_\om}^{(1)} \ri\|_{L^4(\mathbb R_+,r\,dr)}
&=o(\om^{1/8}),\label{eq:radial-profile-approx-new}	\\
\int_0^\infty \lf|\phi_{\om,\tau_\om}^{(1)}(r) \ri|^4r\,dr
&=\sqrt\om\,\Gamma+o(\sqrt\om).
\label{eq:tau-profile-Gamma-new}
\end{align}
\end{lemma}

\begin{proof}[\normalfont\bfseries Proof]
Since
\(
|(n_\om+j)-\tau_\om|
=|j-\alpha_\om|\leq L_\om,
\)
\eqref{eq:low-mode-eigenfunction-conv} gives
\[
 \lf\|\psi_{\om,n_\om+j}^{(1)}
-\psi_{\om,\tau_\om}^{(1)}\ri\|_{\domain_\om}=o(1)\quad\hbox{uniformly for}\,\ |j-\alpha_\om|\leq L_\om,
\]
where $\domain_\om=L^2(I_\om,\rho_\om\,dt)$, $\rho_\om(t)=1+ t/\sqrt{\om}$.
By the unitary identification
\eqref{eq:radial-unitary-map},
\begin{align*}
	& \lf\|\phi_{\om,n_\om+j}^{(1)}
	-\phi_{\om,\tau_\om}^{(1)} \ri\|_
	{L^2(\mathbb R_+,r\,dr)}=
	\lf \|\psi_{\om,n_\om+j}^{(1)}
	-\psi_{\om,\tau_\om}^{(1)} \ri\|_
	{L^2(I_\om,\rho_\om\,dt)}
	=o(1).
\end{align*}
This proves \eqref{eq:radial-profile-L2}.

Set
\[
w_{\om,j}:=\psi_{\om,n_\om+j}^{(1)}
-\psi_{\om,\tau_\om}^{(1)}, \quad  J_\om:=\{t\in I_\om:\,\rho_\om(t)\geq1/2\}.
\]
We estimate the weighted \(L^4\) integral separately on
\(J_\Omega\) and \(I_\Omega\setminus J_\Omega:=\{t\in I_\om:\,0<\rho_\om(t)<1/2\}\).
We first prove 
\begin{equation}\label{eq:oct1}
	\int_{J_\om}|w_{\om,j}|^4\rho_\om(t)\,dt=o(1).
\end{equation}
By \eqref{eq:low-mode-eigenfunction-conv} and the definition of
\(w_{\om,j}\), and since \(\rho_\om\geq1/2\) on \(J_\om\),
\begin{equation}\label{eq:proof-profile-bulk-L2}
\lf\|w_{\om,j}\ri\|_{L^2(J_\om)}^2
\leq2\lf\|w_{\om,j}\ri\|_{\domain_\om}^2
\leq2\lf(
\lf\|\psi_{\om,n_\om+j}^{(1)}-\chi_0\ri\|_{\domain_\om}
+\lf\|\psi_{\om,\tau_\om}^{(1)}-\chi_0\ri\|_{\domain_\om}
\ri)^2=o(1).
\end{equation}
Moreover, \(|a-b|^2\leq2|a|^2+2|b|^2\) and
\(\rho_\om\geq1/2\) on \(J_\om\) imply
\begin{align}
&\quad\int_{J_\om}
\Big(|w_{\om,j}'|^2+t^2|w_{\om,j}|^2\Big)\,dt
\nonumber\\
&\leq4\int_{I_\om}
\lf( \lf|(\psi_{\om,n_\om+j}^{(1)})' \ri|^2
+t^2\lf|\psi_{\om,n_\om+j}^{(1)}\ri|^2 \ri)
\rho_\om\,dt
+4
\int_{I_\om}
\lf( \lf|(\psi_{\om,\tau_\om}^{(1)})' \ri|^2
+t^2\lf|\psi_{\om,\tau_\om}^{(1)} \ri|^2 \ri)
\rho_\om\,dt
\leq C,
\label{eq:proof-profile-bulk-energy}
\end{align}
where the last inequality follows from
\eqref{eq:bounded-q-low-energy}.  Hence, uniformly for
\(|j-\alpha_\om|\leq L_\om\),
\begin{equation}\label{eq:proof-profile-bulk-bounds}
\lf\|w_{\om,j}'\ri\|_{L^2(J_\om)}
+\lf\|tw_{\om,j}\ri\|_{L^2(J_\om)}\leq C.
\end{equation}
The one-dimensional Gagliardo--Nirenberg inequality on \(J_\om\)
and \eqref{eq:proof-profile-bulk-L2}--\eqref{eq:proof-profile-bulk-bounds}
give
\begin{equation}\label{eq:proof-profile-bulk-L4}
\int_{J_\om}|w_{\om,j}|^4\,dt=o(1),
\qquad
\lf\|w_{\om,j}\ri\|_{L^\infty(J_\om)}^2=o(1).
\end{equation}
Equations \eqref{eq:proof-profile-bulk-L2},
\eqref{eq:proof-profile-bulk-bounds}, and
\eqref{eq:proof-profile-bulk-L4} give
\begin{equation}\label{eq:proof-profile-bulk-weight}
\frac1{\sqrt\om}
\int_{J_\om}|t||w_{\om,j}|^4\,dt
\leq
\frac{1}{\sqrt\om} \lf\|w_{\om,j}\ri\|_{L^\infty(J_\om)}^2
\lf\|tw_{\om,j}\ri\|_{L^2(J_\om)}
\lf\|w_{\om,j}\ri\|_{L^2(J_\om)}=o(1).
\end{equation}
Combining \eqref{eq:proof-profile-bulk-L4} and \eqref{eq:proof-profile-bulk-weight} yields that
\eqref{eq:oct1} holds true.

We next prove
\begin{equation}\label{eq:proof-profile-endpoint-L4}
\int_{I_\Omega\setminus J_\Omega}
|w_{\Omega,j}|^4\rho_\Omega\,dt
\leq C\Omega^{-1/2}=o(1).
\end{equation}
Choose \(\chi\in C^\infty([0,\infty);[0,1])\) such that
\(\chi=1\) on \([0,1/2]\) and \(\chi=0\) on \([3/4,\infty)\). Applying \eqref{eq:critical-mass} to the radial function
\(x\mapsto\chi(|x|)\phi_{\Omega,\nu}^{(1)}(|x|)
\in H^1(\mathbb R^2)\), where \(\psi_{\Omega,\nu}^{(1)}(t)
=\Omega^{-1/4}\phi_{\Omega,\nu}^{(1)}(\rho_\Omega(t))\) is given by \eqref{eq:radial-unitary-map},
and using \eqref{eq:critical-mass} and
\eqref{eq:bounded-q-low-energy}, we obtain
\bml{\label{eq:oct2}
\int_{\rho_\Omega<1/2}
\lf|\psi_{\Omega,\nu}^{(1)} \ri|^4\rho_\Omega\,dt
=\Omega^{-1/2}
\int_0^{1/2} \lf|\phi_{\Omega,\nu}^{(1)} \ri|^4r\,dr \\
\leq C\Omega^{-1/2}
\left(\int_0^\infty \lf|\chi\phi_{\Omega,\nu}^{(1)} \ri|^2r\,dr\right)
\lf(\int_0^\infty
\lf| \lf(\chi\phi_{\Omega,\nu}^{(1)} \ri)' \ri|^2r\,dr \ri)\\
=C\Omega^{1/2}
\left(\int_{I_\Omega}
\lf|\chi(\rho_\Omega)\psi_{\Omega,\nu}^{(1)} \ri|^2\rho_\Omega\,dt\right)
\left(\int_{I_\Omega}
\left|\partial_t \lf(\chi(\rho_\Omega)
\psi_{\Omega,\nu}^{(1)} \ri)\right|^2\rho_\Omega\,dt\right)\\
\leq C\Omega^{-1/2}
\left(\int_{I_\Omega}
t^2\lf|\psi_{\Omega,\nu}^{(1)} \ri|^2\rho_\Omega\,dt\right)
\left(\int_{I_\Omega}
\left( \lf|(\psi_{\Omega,\nu}^{(1)})' \ri|^2
+\Omega^{-1} \lf|\psi_{\Omega,\nu}^{(1)} \ri|^2\right)
\rho_\Omega\,dt\right)\leq C\Omega^{-1/2}
}
uniformly for \(|\nu-\tau_\om|\leq L_\Omega\).
Applying \eqref{eq:oct2} with \(\nu=n_\Omega+j\)
and \(\nu=\tau_\om\), and using
\(|a-b|^4\leq8(|a|^4+|b|^4)\), we obtain
\eqref{eq:proof-profile-endpoint-L4}.

We conclude from
\eqref{eq:oct1}, \eqref{eq:proof-profile-bulk-weight} and
\eqref{eq:proof-profile-endpoint-L4}  that
\[
\|w_{\om,j}\|_{L^4(I_\om,\rho_\om\,dt)}=o(1)\quad\hbox{uniformly for}\,\ |j-\alpha_\om|\leq L_\om, 
\]
which, combined with the above scaling identity
\begin{equation*}
	\lf\|\phi_{\om,n_\om+j}^{(1)}
	-\phi_{\om,\tau_\om}^{(1)} \ri\|_{L^4(\mathbb R_+,r\,dr)}^4
	=\sqrt\om
	\lf\|w_{\om,j}\ri\|_{L^4(I_\om,\rho_\om\,dt)}^4,
\end{equation*}
gives \eqref{eq:radial-profile-approx-new}.

Finally, \eqref{eq:low-mode-eigenfunction-conv} applied with
\(\nu=\tau_\om\), gives 
\begin{equation*}
\lf\|\psi_{\om,\tau_\om}^{(1)}-\chi_0\ri\|_{L^2(I_\om,\rho_\om\,dt)}
\xrightarrow[\om \to \infty]{} 0.
\end{equation*}
Since
\(\|\chi_0'\|_{L^2(\mathbb R)}
+\|t\chi_0\|_{L^2(\mathbb R)}<\infty\),
the same argument of proving \eqref{eq:oct1} gives
\[
\int_{J_\Omega}
\lf|\psi_{\Omega,\tau_\om}^{(1)}-\chi_0\ri|^4
\rho_\Omega\,dt=o(1).
\]
On \(I_\Omega\setminus J_\Omega\),
\eqref{eq:oct2} with \(\nu=\tau_\om\), together with
the Gaussian decay of \(\chi_0\), gives
\[
\int_{I_\Omega\setminus J_\Omega}
|\psi_{\Omega,\tau_\om}^{(1)}-\chi_0|^4
\rho_\Omega\,dt=o(1).
\]
Combining these two estimates, we obtain
\[
\lf\|\psi_{\Omega,\tau_\om}^{(1)}-\chi_0\ri\|_
{L^4(I_\Omega,\rho_\Omega\,dt)}=o(1).
\]
Moreover,
\[
\int_{I_\om}|\chi_0(t)|^4\rho_\om(t)\,dt
\longrightarrow
\lf\|\chi_0\ri\|_{L^4(\mathbb R)}^4
=\int_{\mathbb R}|\chi_0(t)|^4\,dt=\Gamma.
\]
Therefore 
\begin{align*}
\int_0^\infty \lf|\phi_{\om,\tau_\om}^{(1)}(r)\ri|^4r\,dr
&=\sqrt\om
\int_{I_\om} \lf|\psi_{\om,\tau_\om}^{(1)}(t) \ri|^4
\rho_\om(t)\,dt
=\sqrt\om\,\Gamma+o(\sqrt\om),
\end{align*}
which proves \eqref{eq:tau-profile-Gamma-new}.
\end{proof}

We introduce the notation used in the following estimates. For \(u\in L^2(\mathbb R^2)\), the spectral expansion
\eqref{eq:full-spectral-decomp} in the eigenfunctions of
\(\mathcal L_{\Omega,n}\) reads
\begin{equation}\label{eq:decom}
u(r,\theta)=\frac1{\sqrt{2\pi}}
\sum_{n\in\mathbb Z}\sum_{m \in \N_+}
a_{n,m}\phi_{\Omega,n}^{(m)}e^{in\theta}
\quad\hbox{in }L^2(\mathbb R^2).
\end{equation}
For \(A\subset\mathbb Z\), assume that $u$ satisfies the above decomposition. We define  the following orthogonal projection
in \(L^2(\mathbb R^2)\): 
\begin{equation}\label{eq:first-band-set-projection}
	P_\Omega^A u
	:=
	\frac1{\sqrt{2\pi}}
	\sum_{j\in A}
	a_{n_\Omega+j,1}
	\phi_{\Omega,n_\Omega+j}^{(1)}(r)e^{i(n_\Omega+j)\theta}.
\end{equation}
Set
\(P_\om:=P_\om^{\mathbb Z}\). Assume \(\gsom\in\domain\) satisfies \eqref{eq:decom}, we define
\begin{equation}\label{eq:first-band-coefficients}
	\eta_\om:=(1-P_\om)u_\om
	=\frac1{\sqrt{2\pi}}
	\sum_{n\in\mathbb Z}\sum_{m\geq2}
	a_{n,m}\phi_{\om,n}^{(m)}(r)e^{in\theta}.
\end{equation}
For $\|u_\om\|_{L^2(\R^2)}=1$ satisfying \eqref{eq:decom}, we also define the corresponding linear energy excess by
\begin{equation}\label{eq:X-Omega-def}
	X_\om(u_\om)
	:=\mathcal Q_\om(u_\om)-E_\om^{\rm int}
	=
	\sum_{n\in\mathbb Z}\sum_{m\geq1}
	\bigl(\lambda_{\om,n}^{(m)}-E_\om^{\rm int}\bigr)|a_{n,m}|^2,
\end{equation}
where $\mathcal Q_\om(u_\om)$ is given by \eqref{eq:linear-form}.

The following lemma gives the preliminary bounds needed to prove
Theorem~\ref{thm:A}.

\begin{lemma}
\label{lem:rough-estimates}
Assume that \(\sigma_\Omega\) satisfies \eqref{eq: weak int attraction}.
Then,
	\begin{equation}\label{eq:X-improved-new}
		X_\om(u_\om)=O(1+\sigma_\om^2).
	\end{equation}
Moreover, with \(\eta_\Omega\) defined by
\eqref{eq:first-band-coefficients}, we have
\begin{equation}\label{eq:eta-improved-new}
	\|\eta_\om\|_2^2
	\leq
	C\frac{1+\sigma_\om^2}{\om}
	=o(1),
	\qquad
	\mathcal Q_\om(\eta_\om)
	\leq
	C(1+\sigma_\om^2),
\end{equation}
	and
	\begin{equation}\label{eq:angular-improved-new}
		\sum_{n\in\mathbb Z}
		\min\{(n-\tau_\om)^2,\Omega\}|a_{n,1}|^2
		\leq
		C(1+\sigma_\Omega^2).
	\end{equation}
\end{lemma}

\begin{proof}[\normalfont\bfseries Proof]
We first prove the preliminary estimate
\begin{equation}\label{eq:X-preliminary-new}
	X_\om(u_\om)=O(1+\sigma_\om\sqrt\om)=o(\om).
\end{equation}
The normalized linear ground state
\((2\pi)^{-1/2}\phi_{\om,n_\om}^{(1)}(r)e^{in_\om\theta}\)
has linear energy \(E_\om^{\rm int}\).  Since the interaction is attractive,
\(\gse\leq E_\om^{\rm int}\), and hence
\begin{equation}\label{eq:minimizer-upper-new}
	\enom(\gsom)
= \gseom =\mathcal Q_\om(u_\om)-\frac{a_\om}{2}\int_{\R^2}|u_\om|^4dx
\leq E_\Omega^{\rm int}\leq C\om.
\end{equation}
For \(A\in L^2_{\rm loc}(\mathbb R^d;\mathbb R^d)\) and
\(u\in L^2(\mathbb R^d)\) satisfying
\((\nabla-iA)u\in L^2(\mathbb R^d)\), the diamagnetic inequality
\cite[Theorem~7.21]{LiebLoss2001} gives
\(|u|\in H^1(\mathbb R^d)\) and
\begin{equation}\label{eq:diamagnetic}
|\nabla |u|(x)|\leq |(\nabla-iA)u(x)|
\quad\hbox{for almost every }x\in\mathbb R^d.
\end{equation}
Applying \eqref{eq:diamagnetic} with \(A=\om x^\perp\) and using the
two-dimensional Gagliardo--Nirenberg inequality yields
\begin{equation}\label{eq:GN-diamagnetic-new}
\lf\|u\ri\|_4^4
\leq C\lf\|u\ri\|_2^2\mathcal Q_\om(u).
\end{equation}
Since \(\lf\|\gsom\ri\|_2=1\), it follows that
\[
	\enom(\gsom)
\geq
(1-Ca_\om)\mathcal Q_\om(u_\om)
=
(1-Ca_\om)(E_\om^{\rm int}+X_\om(u_\om)).
\]
Together with \eqref{eq:minimizer-upper-new}, this gives
\[
X_\Omega(\gs)
\leq
\frac{Ca_\Omega E_\Omega^{\rm int}}{1-Ca_\Omega}\leq \frac{Ca_\Omega \om}{1-Ca_\Omega}.
\]
Moreover,
since \(a_\om=\sigma_\om/\sqrt\om\), we have
\(a_\om E_\om^{\rm int}
=O(\sigma_\om\sqrt\om)\), thus
\eqref{eq:X-preliminary-new} holds true.

Based on \eqref{eq:X-preliminary-new}, we next prove the $L^4$-estimate
\begin{equation}\label{eq:rough-L4-control-new}
\lf\|u_\om\ri\|_4^4
\leq
C\sqrt\om \lf(1+\sqrt{X_\om(u_\om)} \ri)
+
C\om^{-1} (1+X_\om(u_\om))^2.
\end{equation}
For brevity, write
\begin{equation}\label{eq:oct3}
	c_{\Omega,j}:=a_{n_\Omega+j,1},\qquad j\in\mathbb Z.
\end{equation}
Then we have
\begin{equation}\label{eq:first-band-decomp}
u_\om
=P_\om u_\om+\eta_\om
=\frac1{\sqrt{2\pi}}
\sum_{j\in\mathbb Z}
c_{\om,j}\phi_{\om,n_\om+j}^{(1)}(r)
e^{i(n_\om+j)\theta}
+\eta_\om.
\end{equation}
Denote
\begin{equation}\label{eq:rough-angular-window}
	\mathcal J_\om
	:=\big\{j\in\mathbb Z:\,|j-\alpha_\om|\leq\sqrt\om\big\}.
\end{equation}
Using this window, we decompose \(\gsom\) into three
orthogonal spectral components
\[
u_\om
=P_\om u_\om
+\eta_\om=
P_\om^{\mathcal J_\om}u_\om
+P_\om^{\mathcal J_\om^c}u_\om
+\eta_\om,
\]
where the two projections for the first band  correspond to
\(j\in\mathcal J_\om\) and \(j\notin\mathcal J_\om\), respectively.
The definitions of \(X_\Omega\) and \(c_{\Omega,j}\) in
\eqref{eq:X-Omega-def} and \eqref{eq:oct3}, respectively, give
\begin{equation}\label{eq:X-decompo}
X_\om(u_\om)
=
\sum_{j\in\mathbb Z}
\bigl(\lambda_{\om,n_\om+j}^{(1)}-E_\om^{\rm int}\bigr)
|c_{\om,j}|^2
+
\sum_{n\in\mathbb Z}\sum_{m\geq2}
\bigl(\lambda_{\om,n}^{(m)}-E_\om^{\rm int}\bigr)|a_{n,m}|^2.
\end{equation}
Following \eqref{eq:E-int-def}, \eqref{eq:radial-gap-new} and \eqref{eq:X-decompo}, we have
\begin{align}
\lf\|\eta_\om\ri\|_2^2=\sum_{n\in\mathbb Z}\sum_{m\geq2}
|a_{n,m}|^2
\leq \frac{C}{\om} \lf( 1+X_\om(u_\om) \ri).
\label{eq:eta-L2-in-terms-X-new}
\end{align}
We also derive from \eqref{eq:X-decompo}, \eqref{eq:angular-gap-new} and \eqref{eq:radial-gap-new} that
\begin{equation}\label{eq:angular-control-X-new}
	\sum_{j\in\mathbb Z}
	\min\{(j-\alpha_\om)^2,\om\}|c_{\om,j}|^2
	\leq C(1+X_\om(u_\om)).
\end{equation}

For the \(L^4\) estimate, group the last two components by setting
\[
w_\om:= \lf(1-P_\om^{\mathcal J_\om}\ri)u_\om
=P_\om^{\mathcal J_\om^c}u_\om+\eta_\om,
\qquad
u_\om=P_\om^{\mathcal J_\om}u_\om+w_\om.
\]
Then we derive from \eqref{eq:rough-angular-window}, \eqref{eq:eta-L2-in-terms-X-new} and \eqref{eq:angular-control-X-new} that
\begin{align}
\sum_{j\in\mathcal J_\om}
(j-\alpha_\om)^2|c_{\om,j}|^2
\leq C(1+X_\om(u_\om)),\qquad
\lf\|P_\om^{\mathcal J_\om^c}u_\om \ri\|_2^2
\leq \frac{C}{\om} \lf( 1+X_\om(u_\om) \ri).\label{eq:inner-angular-control-new}
\end{align}
Consequently, orthogonality of the spectral components yields
\begin{equation}\label{eq:oct4}
\lf\|w_\om\ri\|_2^2=\lf\|P_\om^{\mathcal J_\om^c}u_\om\ri\|_2^2+\lf\|\eta_\om\ri\|_2^2
\leq \frac{C}{\om} \lf( 1+X_\om(u_\om) \ri).
\end{equation}
Following above estimate and \eqref{eq:X-decompo}, we also obtain 
\begin{equation}\label{eq:wom}
\mathcal Q_\om(w_\om)
=
E_\om^{\rm int}\|w_\om\|_2^2
+
\sum_{\{n\in\mathbb Z,m\geq 2\}\cup\{n - n_\om \not\in\mathcal J_\om,m=1\}}
\bigl(\lambda_{\om,n}^{(m)}-E_\om^{\rm int}\bigr)
|a_{n,m}|^2\leq C(1+X_\om(u_\om)).
\end{equation}
Applying \eqref{eq:oct4} and \eqref{eq:wom} to \eqref{eq:GN-diamagnetic-new} gives
\begin{equation}\label{eq:high-remainder-L4-new}
\|w_\om\|_4^4
\leq \frac{C}{\om}(1+X_\om(u_\om))^2.
\end{equation}
The same argument applied separately to the two spectral components gives
\begin{align}
\lf\|\eta_\om \ri\|_4^4
\leq  \frac{C}{\om}(1+X_\om(u_\om))^2,\qquad
\lf\|P_\om^{\mathcal J_\om^c}u_\om \ri\|_4^4
\leq  \frac{C}{\om}(1+X_\om(u_\om))^2.\label{eq:eta-L4-in-terms-X-new}
\end{align}

If \(j\in\mathcal J_\om\), then \eqref{eq:E-size-new1} gives
\[
|n_\om+j-\om|
\leq |j-\alpha_\om|+|\tau_\om-\om|
\leq2\sqrt\om
\]
for all sufficiently large \(\om\).  Applying \cite[Corollary 2.2]{Rougerie2011} with
\(D_\Omega=1/4\) and \(\omega=\Omega\), we obtain
\begin{equation}\label{eq:radial-L4-bound-new}
	\sup_{j\in\mathcal J_\om}\int_0^\infty
	\lf|\phi_{\om,n_\om+j}^{(1)}(r) \ri|^4r\,dr
	\leq C\sqrt\om.
\end{equation}

For \(A=\mathcal J_\Omega\) and \(d_j=c_{\Omega,j}\), the quantities
\(M\) and \(K\) in Lemma~\ref{lem:first-band-L4} satisfy
\[
\begin{aligned}
M=\sum_{j\in A}|c_{\Omega,j}|^2\leq1,\,\ 
K=\sum_{j\in A}(1+|j-\alpha_\Omega|^2)|c_{\Omega,j}|^2
\leq C(1+X_\Omega(\gs)),
\end{aligned}
\]
by the normalization of \(\gs\) and \eqref{eq:inner-angular-control-new}.
Applying Lemma~\ref{lem:first-band-L4} in Appendix~\ref{app:first-band-L4} gives
\begin{equation}\label{eq:P-in-L4-control-new}
	\lf\|P_\om^{\mathcal J_\om}u_\om \ri\|_4^4
	\leq C\sqrt\om \lf(1+ \sqrt{X_\om(u_\om)} \ri).
\end{equation}
Since \(u_\om=P_\om^{\mathcal J_\om}u_\om+w_\om\),
\eqref{eq:high-remainder-L4-new} and
\eqref{eq:P-in-L4-control-new} imply
\eqref{eq:rough-L4-control-new}.

Based on \eqref{eq:X-preliminary-new} and \eqref{eq:rough-L4-control-new}, we finally prove \eqref{eq:X-improved-new}.
By \eqref{eq:minimizer-upper-new},
\[
X_\Omega(\gs)=\mathcal Q_\om(u_\om)-E_\om^{\rm int}
\leq
\frac{a_\Omega}{2}
\int_{\mathbb R^2}|\gs|^4\,dx.
\]
Using \(a_\om=\sigma_\om/\sqrt\om\) and
\eqref{eq:rough-L4-control-new}, we obtain
\begin{equation}\label{eq:X-feedback-new}
X_\Omega(\gs)
\leq
C \lf[ \sigma_\Omega\lf(1+ \sqrt{X_\om(u_\om)} \ri)
+
\sigma_\Omega\Omega^{-3/2} \lf(1+X_\Omega(\gs) \ri)^2 \ri].
\end{equation}
Then \eqref{eq:X-preliminary-new} and
\(\sigma_\om=o(\sqrt\om)\) imply
\[
\sigma_\Omega\Omega^{-3/2} \lf(1+X_\Omega(\gs) \ri)^2
=
\OO \Big(
\sigma_\om\om^{-3/2}(1+\sigma_\om^2\om)
\Big)
=o(1+\sigma_\om^2).
\]
Consequently,
\begin{equation}\label{eq:Xom}
X_\om(u_\om)
\leq C \lf(1+\sigma_\om^2
+\sigma_\om \sqrt{X_\om(u_\om)} \ri),
\end{equation}
which further implies \eqref{eq:X-improved-new}.  Since \(\eta_\om\) and
\(P_\om^{\mathcal J_\om^c}u_\om\) are spectrally orthogonal,   we obtain from \eqref{eq:X-improved-new}  and 
\eqref{eq:wom} that
\begin{equation*} 
\mathcal Q_\om(\eta_\om)\leq \mathcal Q_\om(\eta_\om)+\mathcal Q_\om\lf(P_\om^{\mathcal J_\om^c}u_\om\ri)=\mathcal Q_\om(w_\om)\leq C(1+\sigma_\om^2),
\end{equation*}
Combining \eqref{eq:X-improved-new} with
\eqref{eq:eta-L2-in-terms-X-new},
\eqref{eq:angular-control-X-new}, and the above estimate on
\(\mathcal Q_\Omega(\eta_\Omega)\), we obtain
\eqref{eq:eta-improved-new} and \eqref{eq:angular-improved-new}. 
\end{proof}

\subsection{Proof of Theorem~\ref{thm:A}}
\label{sec:unified-angular-window}
In this subsection, we first approximate each minimizer of \( \enom \)
by its first radial band projection restricted to angular modes
near \(n_\Omega\).
We then replace the radial eigenfunctions
\(\phi_{\Omega,n_\Omega+j}^{(1)}\) in this projection by
the common profile \(\phi_{\Omega,\tau_\om}^{(1)}\)
and estimate the resulting error in the quartic interaction.
Using these estimates, we prove the energy expansion
and profile approximation in Theorem~\ref{thm:A}.

We first introduce some symbols needed below. For \(h\in L^2(S^1)\), write its Fourier expansion as
\begin{equation}\label{eq:oct5}
h(\theta)=\frac1{\sqrt{2\pi}}
\sum_{j\in\mathbb Z}  h_j e^{ij\theta}
\quad\hbox{in }L^2(S^1).
\end{equation}
For \(A\subset\mathbb Z\) and \(h\in L^2(S^1)\) satisfying \eqref{eq:oct5}, we define the following projection operator by
\begin{equation}\label{eq:unified-window-projection}
  \lf( \Pi^A h \ri)(\theta)
:=\frac1{\sqrt{2\pi}}
\sum_{j\in A}  h_j e^{ij\theta}.
\end{equation}
For \(\gs\) with spectral coefficients \(a_{n,m}\) in
\eqref{eq:decom}, define  the angular profile associated with \(\gsom\) by
\begin{equation}\label{eq:g-unified-def}
[g_\Omega(u_\om)](\theta)
:=\frac1{\sqrt{2\pi}}
\sum_{j\in\mathbb Z}a_{n_\Omega+j,1}e^{ij\theta}.
\end{equation}
For simplicity, we suppress the dependence on \(\gsom\) and write \(g_\Omega\) below.
Choose \(M_\Omega\) such that as $ \Omega\to\infty $
\begin{equation}\label{eq:unified-diagonal-window}
M_\Omega\to\infty,
\quad
M_\Omega(1+\sigma_\Omega)=o(\sqrt{\Omega})
,
\end{equation}
and define
\begin{equation}\label{eq:unified-effective-window}
\mathcal W_\om
:=\bigl\{j\in\mathbb Z:
|j-\alpha_\om|\leq M_\om(1+\sigma_\om)\bigr\},
\qquad
\mathcal T_\om:=\mathcal J_\om\setminus\mathcal W_\om.
\end{equation}
Then \(\mathcal W_\om\subset\mathcal J_\om\) for all sufficiently large
\(\om\).  The angular projection satisfies
\begin{equation}\label{eq:gW-def-unified}
\lf(  \Pi^{\mathcal W_\Omega}g_\Omega \ri)(\theta)
=\frac1{\sqrt{2\pi}}
\sum_{j\in\mathcal W_\Omega}a_{n_\Omega+j,1}e^{ij\theta}.
\end{equation}
The corresponding component in the main window is
\begin{equation}\label{eq:PW-def-unified}
P_\Omega^{\mathcal W_\Omega}\gsom
=
\frac1{\sqrt{2\pi}}
\sum_{j\in\mathcal W_\Omega}
a_{n_\Omega+j,1}\phi_{\Omega,n_\Omega+j}^{(1)}(r)
e^{i(n_\Omega+j)\theta},
\end{equation}
where the projection operator $P_\om^{A}$ is given by \eqref{eq:first-band-set-projection}.

By \eqref{eq:rough-angular-window},
\(|j-\alpha_\om|^2\leq\om\) for \(j\in\mathcal J_\om\).
We claim that
\begin{equation}\label{eq:gJ-unified-H1-bound}
	\lf\|  \Pi^{\mathcal J_\om}g_\om \ri\|_{L^2(S^1)}\leq C,
	\qquad
	\lf\|D_{\alpha_\om}  \Pi^{\mathcal J_\om}g_\om \ri\|_{L^2(S^1)}^2
	\leq
	C(1+\sigma_\om^2),
\end{equation}
where
\(D_{\alpha_\om}:=-i\partial_\theta-\alpha_\om .\) Indeed,
\[
D_{\alpha_\om} \Pi^{\mathcal J_\om}g_\om
=
\frac1{\sqrt{2\pi}}
\sum_{j\in\mathcal J_\om}
(j-\alpha_\om)a_{n_\Omega+j,1}e^{ij\theta},
\]
and hence
\begin{equation}\label{eq:gJ-angular-moment-identity}
\lf\|D_{\alpha_\om} \Pi^{\mathcal J_\om}g_\om \ri\|_{L^2(S^1)}^2
=
\sum_{j\in\mathcal J_\om}
(j-\alpha_\om)^2|a_{n_\Omega+j,1}|^2.
\end{equation}
Combining this identity with \eqref{eq:angular-improved-new} gives
\begin{equation}\label{eq:gJ-angular-moment-bound}
\lf\|D_{\alpha_\om} \Pi^{\mathcal J_\om}g_\om\ri\|_{L^2(S^1)}^2
\leq
\sum_{j\in\mathbb Z}
\min\{(j-\alpha_\om)^2,\om\}|a_{n_\Omega+j,1}|^2
\leq
C(1+\sigma_\om^2).
\end{equation}
This proves the claim \eqref{eq:gJ-unified-H1-bound}.

We begin with the following lemma.
\begin{lemma}
\label{lem:nonlinear-reduction}
Under the same assumptions of
Lemma~\ref{lem:rough-estimates}, we have, as  $ \om \to \infty $, 
\begin{align}
\lf\|  \Pi^{\mathcal W_\Omega}g_\Omega \ri\|_{L^2(S^1)}^2&=1-o(1),
\label{eq:unified-window-mass}\\
\lf\|u_\om
-e^{in_\om\theta}\phi_{\om,\tau_\om}^{(1)}  \Pi^{\mathcal W_\Omega}g_\Omega
\ri\|_2&=o(1),	
\label{eq:unified-window-L2-profile}	\\
a_\om\int_{\mathbb R^2}|u_\om|^4\,dx
& =
\Gamma\sigma_\om
\int_{S^1} \lf|(\Pi^{\mathcal W_\Omega}g_\Omega)(\theta) \ri|^4\,d\theta
+
o(1+\sigma_\om^2).\label{eq:quartic-reduction}
\end{align}
\end{lemma}

\begin{proof}[\normalfont\bfseries Proof]
We first prove \eqref{eq:unified-window-mass} and
\eqref{eq:unified-window-L2-profile}.
We claim that
\begin{equation}\label{eq:unified-tail-L2}
\lf\| \Pi^{\mathcal T_\om}g_\om \ri\|_{L^2(S^1)}^2=\sum_{j\in \mathcal T_\om}
|a_{n_\Omega+j,1}|^2
\leq\frac{C}{M_\om^2},\quad\hbox{where}\,\ \mathcal T_\om=\mathcal J_\om\setminus\mathcal W_\om.
\end{equation}
For \(j\in \mathcal T_\om\), the definition
of \(\mathcal W_\om\) gives
\(
|j-\alpha_\om|>M_\om(1+\sigma_\om).
\)
Consequently, \eqref{eq:gJ-unified-H1-bound} yields
\[
\begin{aligned}
\lf\|  \Pi^{\mathcal T_\om}g_\om \ri\|_{L^2(S^1)}^2
=
\sum_{j\in \mathcal T_\om}
|a_{n_\Omega+j,1}|^2                                 &\leq
\frac{1}{M_\om^2(1+\sigma_\om)^2}
\sum_{j\in \mathcal T_\om}
(j-\alpha_\om)^2|a_{n_\Omega+j,1}|^2                    \\
&\leq
\frac{C(1+\sigma_\om^2)}
{M_\om^2(1+\sigma_\om)^2}
\leq
\frac{C}{M_\om^2}.
\end{aligned}
\]
This proves the claim \eqref{eq:unified-tail-L2}.

By 
\eqref{eq:eta-improved-new}, \eqref{eq:angular-improved-new}, and
\eqref{eq:unified-tail-L2},  we obtain 
\begin{equation}\label{eq:window-L2-truncation}
\lf\| (1 -P_\om^{\mathcal W_\om})u_\om \ri\|_2^2
=
\lf\|\eta_\om\ri\|_2^2
+
\sum_{j\notin\mathcal J_\om}|a_{n_\Omega+j,1}|^2
+
\sum_{j\in \mathcal T_\om}
|a_{n_\Omega+j,1}|^2
\leq
C\frac{1+\sigma_\om^2}{\om}
+\frac{C}{M_\om^2}
=o(1).
\end{equation}
By orthogonality and the normalization of \(\gsom\),
\[
\begin{aligned}
\lf\|\Pi^{\mathcal W_\Omega}g_\Omega \ri\|_{L^2(S^1)}^2
&=\sum_{j\in\mathcal W_\Omega}|a_{n_\Omega+j,1}|^2
=\lf\| P_\Omega^{\mathcal W_\Omega}\gsom \ri\|_2^2\\
&=1- \lf\|(1 - P_\Omega^{\mathcal W_\Omega})\gsom \ri\|_2^2
=1-o(1),
\end{aligned}
\]
which proves \eqref{eq:unified-window-mass}.

Recall from 
\eqref{eq:unified-diagonal-window} that \(\mathcal W_\om\) satisfies the assumption  of
Lemma~\ref{lem:radial-profile-comparison}. Thus
\begin{align}
&\Big\|P_\om^{\mathcal W_\om}u_\om
-e^{in_\om\theta}\phi_{\om,\tau_\om}^{(1)}
\widehat \Pi^{\mathcal W_\Omega}g_\Omega\Big\|_2^2=
\sum_{j\in\mathcal W_\om}|a_{n_\Omega+j,1}|^2
\Big\|\phi_{\om,n_\om+j}^{(1)}
-\phi_{\om,\tau_\om}^{(1)}\Big\|_
{L^2(\mathbb R_+,r\,dr)}^2
=o(1).\label{eq:window-radial-L2}
\end{align}
Combining \eqref{eq:window-L2-truncation} and
\eqref{eq:window-radial-L2} proves
\eqref{eq:unified-window-L2-profile}.

Finally, we prove the nonlinear reduction
\eqref{eq:quartic-reduction}.
By \eqref{eq:alpha-asymptotic-bound}, \(\alpha_\Omega\) is bounded. The
one-dimensional Sobolev inequality  gives 
\bml{\label{eq:GNs1}
	\lf\| h \ri\|_{L^\infty(S^1)}^2\leq \tfrac{1}{2\pi} \lf\| h \ri\|_{L^2(S^1)}^2
+2 \lf\| h \ri\|_{L^2(S^1)} \lf\|h' \ri\|_{L^2(S^1)}
\\
\leq C\lf\|h \ri\|_{L^2(S^1)}
\lf( \lf\|D_{\alpha_\Omega}h \ri\|_{L^2(S^1)}+\lf\| h \ri\|_{L^2(S^1)} \ri),
}
where $D_{\alpha_\Omega}h:=-ih'-\alpha_\Omega h$.
Therefore,
for every \(h\in H^1(S^1)\), 
\begin{equation}\label{eq:unified-Sobolev-magnetic}
\begin{aligned}
\lf\| h \ri\|_{L^4(S^1)}^4
&\leq \lf\| h \ri\|_{L^\infty(S^1)}^2 \lf\|  h \ri\|_{L^2(S^1)}^2
\leq C \lf\| h \ri\|_{L^2(S^1)}^3
\lf( \lf\|D_{\alpha_\Omega}h \ri\|_{L^2(S^1)}+\lf\|  h \ri\|_{L^2(S^1)} \ri)
\end{aligned}
\end{equation}
uniformly in \(\Omega\).
Since \(\mathcal W_\Omega\subset\mathcal J_\Omega\), orthogonality and
\eqref{eq:gJ-unified-H1-bound} give
\[
\lf\| \Pi^{\mathcal W_\Omega}g_\Omega \ri\|_{L^2(S^1)}\leq1,
\qquad
\lf\| D_{\alpha_\Omega} \Pi^{\mathcal W_\Omega}g_\Omega \ri\|_{L^2(S^1)}^2
\leq \lf\| D_{\alpha_\Omega} \Pi^{\mathcal J_\Omega}g_\Omega \ri\|_{L^2(S^1)}^2
\leq C(1+\sigma_\Omega^2).
\]
Substituting these bounds into \eqref{eq:unified-Sobolev-magnetic} gives
\begin{equation}\label{eq:int-angular-tail-L4}
\lf\| \Pi^{\mathcal W_\Omega}g_\Omega \ri\|_{L^4(S^1)}^4\leq C(1+\sigma_\Omega).
\end{equation}
We claim that the following
three estimates hold (recall \eqref{eq:first-band-set-projection} and that $ P_{\om} = P_{\om}^{\mathbb Z} $):
\begin{align}
a_\om
\lf|
\int_{\mathbb R^2}|u_\om|^4\,dx
-
\int_{\mathbb R^2}|P_\om u_\om|^4\,dx
\ri|
& =o(1+\sigma_\om^2),
\label{eq:unified-high-radial-quartic}	\\
a_\om
\lf|
\int_{\mathbb R^2}|P_\om u_\om|^4\,dx
-
\int_{\mathbb R^2} \lf| P_\om^{\mathcal W_\om} u_\om \ri|^4\,dx
\ri|
& =o(1+\sigma_\om^2). \label{eq:first-band-tail-quartic}	\\
\lf\|
P_\om^{\mathcal W_\om}u_\om
-
e^{in_\om\theta}\phi_{\om,\tau_\om}^{(1)}(r)
\lf( \Pi^{\mathcal W_\Omega}g_\Omega \ri)(\theta)
\ri\|_{L^4(\mathbb R^2)}
&=
o\big(\om^{1/8}(1+\sigma_\om)^{1/4}\big).\label{eq:radial-profile-replacement}
\end{align}
If \eqref{eq:unified-high-radial-quartic}--%
\eqref{eq:radial-profile-replacement} hold, then we can prove
\eqref{eq:quartic-reduction} as follows.  Indeed, by
\eqref{eq:tau-profile-Gamma-new} and
\eqref{eq:int-angular-tail-L4},
\[
 \lf\|
e^{in_\om\theta}\phi_{\om,\tau_\om}^{(1)}
\Pi^{\mathcal W_\Omega}g_\Omega
 \ri\|_4^4
\leq
C\sqrt\om(1+\sigma_\om).
\]
The same bound holds for \(P_\om^{\mathcal W_\om}u_\om\) by
\eqref{eq:radial-profile-replacement}.  Therefore, 
\eqref{eq:radial-profile-replacement} implies
\bmln{
	a_\om
	\lf|
	\int_{\mathbb R^2}
	\lf(
	\lf|P_\om^{\mathcal W_\om}u_\om \ri|^4
	-
	\lf|
	\phi_{\om,\tau_\om}^{(1)}\Pi^{\mathcal W_\Omega}g_\Omega
	\ri|^4
	\ri)\,dx
	\ri| \\
	\leq
	Ca_\om
	\lf(
	\lf\|P_\om^{\mathcal W_\om}u_\om \ri\|_4^3
	+
	\lf\|
	\phi_{\om,\tau_\om}^{(1)}\widehat \Pi^{\mathcal W_\Omega}g_\Omega
	\ri\|_4^3
	\ri)
	\lf\|
	P_\om^{\mathcal W_\om}u_\om
	-
	e^{in_\om\theta}\phi_{\om,\tau_\om}^{(1)}
	\Pi^{\mathcal W_\Omega}g_\Omega
	\ri\|_4\\
	=
	o\bigl(\sigma_\om(1+\sigma_\om)\bigr)
	=o(1+\sigma_\om^2).
}
Moreover, \eqref{eq:tau-profile-Gamma-new} and
\eqref{eq:int-angular-tail-L4} give
\[
a_\om
\int_{\mathbb R^2}
\Big|
\phi_{\om,\tau_\om}^{(1)}\widehat \Pi^{\mathcal W_\Omega}g_\Omega
\Big|^4\,dx
=
\sigma_\om\Gamma
\int_{S^1}| \Pi^{\mathcal W_\Omega}g_\Omega|^4\,d\theta
+
o(1+\sigma_\om^2).
\]
Combining the last two estimates with
\eqref{eq:unified-high-radial-quartic} and
\eqref{eq:first-band-tail-quartic} proves
\eqref{eq:quartic-reduction}.

It remains to prove
\eqref{eq:unified-high-radial-quartic}--%
\eqref{eq:radial-profile-replacement}.  We first prove
\eqref{eq:unified-high-radial-quartic}.
The identity
\(u_\om=P_\om u_\om+\eta_\om\), together with
\eqref{eq:rough-L4-control-new},
\eqref{eq:eta-L4-in-terms-X-new}, and
\eqref{eq:X-improved-new}, gives
\begin{equation}\label{eq:first-band-high-radial-L4}
\lf\|P_\om u_\om\ri\|_4^4\leq C\sqrt\om(1+\sigma_\om),
\qquad
\lf\|\eta_\om\ri\|_4^4
\leq \frac{C}{\om}(1+\sigma_\om^2)^2.
\end{equation}
By  H\"older's inequality,
\[
a_\om
\lf|\int |u_\om|^4-\int |P_\om u_\om|^4
\ri|\leq Ca_\om
\lf(
\lf\|P_\om u_\om \ri\|_4^3 \lf\|\eta_\om \ri\|_4
+ \lf\|\eta_\om \ri\|_4^4
\ri).
\]
Since $a_\om=\sigma_\om/\sqrt{\om}$, the first term satisfies
\[
a_\om
\lf\|P_\om u_\om \ri\|_4^3 \lf\|\eta_\om \ri\|_4
\leq
C\om^{-3/8}\sigma_\om(1+\sigma_\om)^{3/4}(1+\sigma_\om^2)^{1/2}
=
o(1+\sigma_\om^2),
\]
whereas
\[
a_\om \lf\|\eta_\om \ri\|_4^4
\leq
 C \om^{-3/2} \sigma_\om(1+\sigma_\om^2)^2
=o(1+\sigma_\om^2).
\]
This proves
\eqref{eq:unified-high-radial-quartic}.

We next prove \eqref{eq:first-band-tail-quartic}.  We first establish
the auxiliary estimate
\begin{equation}\label{eq:unified-first-band-tail-L4}
\lf\|P_\om^{\mathcal W_\om^c}u_\om \ri\|_4^4
\leq
C \lf( \sqrt\om M_\om^{-3} \lf(1+\sigma_\om \ri) + \om^{-1}
(1+\sigma_\om^2)^2 \ri).
\end{equation}
Since
\(\mathcal W_\om\subset\mathcal J_\om\), we decompose
\(
P_\om^{\mathcal W_\om^c}u_\om
=P_\om^{\mathcal T_\om}u_\om
+P_\om^{\mathcal J_\om^c}u_\om.
\)
We claim that
\begin{equation}\label{eq:first-band-inner-tail}
\lf\|P_\om^{\mathcal T_\om}u_\om \ri\|_4^4
\leq
C\sqrt\om M_\om^{-3} \lf(1+\sigma_\om \ri).
\end{equation}
Indeed, by \eqref{eq:unified-tail-L2},
\(
M_A:=\sum_{j\in\mathcal T_\om}|a_{n_\Omega+j,1}|^2
=\| \Pi^{\mathcal T_\om}g_\om\|_{L^2(S^1)}^2
\leq C/M_\om^2.
\)
Moreover, the fact that \(\|u_\om\|_2^2=1\) and
\eqref{eq:angular-improved-new} give
\[
K_A:=
\sum_{j\in \mathcal T_\om}
(1+|j-\alpha_\om|^2)|a_{n_\Omega+j,1}|^2
\leq \sum_{j\in \mathcal T_\om}
|a_{n_\Omega+j,1}|^2+\sum_{j\in \mathcal T_\om}
|j-\alpha_\om|^2|a_{n_\Omega+j,1}|^2
\leq C(1+\sigma_\om^2).
\]
Applying the \(L^4\) estimate for the first band 
\eqref{eq:P-in-precoeff-new} to \(\mathcal T_\om\), we obtain
\[
\lf\|P_\om^{\mathcal T_\om}u_\om \ri\|_4^4
\leq
C\sqrt\om\,M_A^{3/2}K_A^{1/2}.
\]
Substitution of the bounds for \(M_A\) and \(K_A\) into
\eqref{eq:P-in-precoeff-new} gives
\eqref{eq:first-band-inner-tail}.
For \(P_\om^{\mathcal J_\om^c}u_\om\), combining
\eqref{eq:eta-L4-in-terms-X-new} with
\eqref{eq:X-improved-new} gives 
\[
\lf\|P_\om^{\mathcal J_\om^c}u_\om\ri\|_4^4
\leq
\frac{C}{\om}(1+\sigma_\om^2)^2.
\]  
This together with
\eqref{eq:first-band-inner-tail} proves
\eqref{eq:unified-first-band-tail-L4}.

Based on \eqref{eq:unified-first-band-tail-L4}, we now prove \eqref{eq:first-band-tail-quartic} as follows. Combining \eqref{eq:unified-first-band-tail-L4} with
\(\|P_\om u_\om\|_4^4\leq
C\sqrt\om(1+\sigma_\om)\) and
\(\sigma_\om=o(\sqrt\om)\) gives
\[
\lf\|P_\om^{\mathcal W_\om}u_\om\ri\|_4^4
\leq
C\sqrt\om(1+\sigma_\om).
\]
On the other hand, H\"older's inequality gives
\[
a_\om
\lf|
\int |P_\om u_\om|^4
-\int \lf|P_\om^{\mathcal W_\om}u_\om \ri|^4
\ri| \leq
Ca_\om
\lf(
\lf\|P_\om u_\om \ri\|_4^3
+
\lf\|P_\om^{\mathcal W_\om}u_\om \ri\|_4^3
\ri)
\lf\|P_\om^{\mathcal W_\om^c}u_\om \ri\|_4.
\]
Using \eqref{eq:unified-first-band-tail-L4} and
\(a_\om=\sigma_\om/\sqrt\om\), the right-hand side is bounded by
\bmln{
C\sigma_\om\om^{-1/8}(1+\sigma_\om)^{3/4}
\left(
\om^{1/8}M_\om^{-3/4} (1+\sigma_\om)^{1/4} 
+(1+\sigma_\om^2)^{1/2}\om^{-1/4}
\right)
\\
\leq
C\sigma_\om \lf( M_\om^{-3/4}(1+\sigma_\om)
+\om^{-3/8} (1+\sigma_\om)^{3/4}
(1+\sigma_\om^2)^{1/2} \ri)
=o(1+\sigma_\om^2),
}
because \(M_\om\to\infty\) and
\(\sigma_\om=o(\sqrt\om)\).
This proves \eqref{eq:first-band-tail-quartic}.

Finally, we prove \eqref{eq:radial-profile-replacement}. Note from \eqref{eq:PW-def-unified} that
\begin{equation}\label{eq:oct7}
P_\Omega^{\mathcal W_\Omega}\gs
-e^{in_\Omega\theta}\phi_{\Omega,\tau_\om}^{(1)}  \Pi^{\mathcal W_\Omega}g_\Omega=\frac1{\sqrt{2\pi}}
\sum_{j\in\mathcal W_\Omega}
a_{n_\Omega+j,1}
\Big(
\phi_{\Omega,n_\Omega+j}^{(1)}
-\phi_{\Omega,\tau_\om}^{(1)}
\Big) e^{i(n_\Omega+j)\theta}.
\end{equation}
On
\(\mathcal W_\om\), Lemma~\ref{lem:radial-profile-comparison} and
\eqref{eq:unified-diagonal-window} give
\begin{equation}\label{eq:oct6}
\sup_{j\in\mathcal W_\om}
\Big\|
\phi_{\om,n_\om+j}^{(1)}
-
\phi_{\om,\tau_\om}^{(1)}
\Big\|_{L^4(\mathbb R_+,r\,dr)}
=o(\om^{1/8}).
\end{equation}
Since \(\mathcal W_\om\subset\mathcal J_\om\), the normalization
of \(u_\om\) and \eqref{eq:angular-improved-new} give
\[
\sum_{j\in\mathcal W_\om}|a_{n_\Omega+j,1}|^2\leq1,
\qquad
\sum_{j\in\mathcal W_\om}
\bigl(1+|j-\alpha_\om|^2\bigr)|a_{n_\Omega+j,1}|^2
\leq C(1+\sigma_\om^2).
\]
Since $\phi_{\om,n_\om+j}^{(1)}
-
\phi_{\om,\tau_\om}^{(1)}$ satisfies \eqref{eq:oct6},
the argument in the proof of Lemma~\ref{lem:first-band-L4} in
Appendix~\ref{app:first-band-L4} applies with
\(\phi_{\Omega,n_\Omega+j}^{(1)}\) replaced by
\(\phi_{\Omega,n_\Omega+j}^{(1)}
-\phi_{\Omega,\tau_\om}^{(1)}\).
Using the two coefficient bounds above, we obtain
\bml{\label{eq:oct8}
\Big\| 
\sum_{j\in\mathcal W_\Omega}
a_{n_\Omega+j,1}
\Big(
\phi_{\Omega,n_\Omega+j}^{(1)}
-\phi_{\Omega,\tau_\om}^{(1)}
\Big)
e^{i(n_\Omega+j)\theta}
\Big\|_4^4\\
\leq C(1+\sigma_\Omega)
\sup_{j\in\mathcal W_\Omega}
\lf\|
\phi_{\Omega,n_\Omega+j}^{(1)}
-\phi_{\Omega,\tau_\om}^{(1)}
\ri\|_{L^4(\mathbb R_+,r\,dr)}^4=o(\Omega^{1/2})(1+\sigma_\Omega).
}
Combining \eqref{eq:oct7} and \eqref{eq:oct8} gives \eqref{eq:radial-profile-replacement}.
Thus  the proof of Lemma~\ref{lem:nonlinear-reduction} is  complete.
\end{proof}

We next establish a Fourier truncation estimate for the circle functional.
\begin{lemma}
\label{lem:unified-1D-truncation}
Let \(h_\om(\theta)\in H^1(S^1)\) satisfy \(\|h_\om\|_{L^2(S^1)}=1\) and 
\(
\mathcal E_{\alpha_\om,\sigma_\om}(h_\om)
\leq C(1+\sigma_\om^2),
\)
then
\begin{equation}\label{eq:h-near-minimizer-bounds}
\int_{S^1}|D_{\alpha_\om}h_\om|^2\,d\theta
\leq C(1+\sigma_\om^2),
\qquad
\sigma_\om\int_{S^1}|h_\om|^4\,d\theta
\leq C(1+\sigma_\om^2),
\end{equation}
where $D_{\alpha_\om}=-i\partial_\theta-\alpha_\om$.
Moreover, let \(
\widetilde h_\om
:=
\frac{1}{\lf\| \Pi^{\mathcal W_\om} h_\om \ri\|_{2}} \Pi^{\mathcal W_\om} h_\om\),
where \( \Pi^{\mathcal W_\om}\) is defined in
\eqref{eq:unified-window-projection}, 
then we have
\begin{equation}\label{eq:1D-truncation-energy}
\big\|\widetilde h_\om \big\|_{L^2(S^1)}=1, \qquad \mathcal E_{\alpha_\om,\sigma_\om}(\widetilde h_\om)
\leq
\mathcal E_{\alpha_\om,\sigma_\om}(h_\om)
+
o(1+\sigma_\om^2).
\end{equation}
\end{lemma}

\begin{proof}[\normalfont\bfseries Proof]
We first prove \eqref{eq:h-near-minimizer-bounds}.    For \(\|h\|_{L^2(S^1)}=1\), we derive from \eqref{eq:unified-Sobolev-magnetic} that
\[
\int_{S^1}|h|^4\,d\theta
\leq
C \lf(1+ \lf\|D_{\alpha_\om}h \ri\|_{L^2(S^1)} \ri).
\]
Hence, there exist some constants \(c,C>0\) independent of \(\om\) such that
\begin{equation}\label{eq:coercive-E}
\mathcal E_{\alpha_\om,\sigma_\om}(h)
\geq
c \lf\|D_{\alpha_\om}h \ri\|_{L^2(S^1)}^2
-
C\sigma_\om \lf\|D_{\alpha_\om}h \ri\|_{L^2(S^1)}
-C(1+\sigma_\om)
\geq
\tfrac c2 \lf\|D_{\alpha_\om}h \ri\|_{L^2(S^1)}^2-C(1+\sigma_\om^2).
\end{equation}
Applying \eqref{eq:coercive-E} to \(h_\om\) and using the hypothesis
\(
\mathcal E_{\alpha_\om,\sigma_\om}(h_\om)
\leq C(1+\sigma_\om^2)
\)
gives the first estimate in
\eqref{eq:h-near-minimizer-bounds}.  The second one follows again
by \eqref{eq:GNs1}.

We claim that
\begin{equation}\label{eq:oct11}
	\big\| D_{\alpha_\om}\widetilde h_\om \big\|_{L^2(S^1)}^2
	-
	\lf\|D_{\alpha_\om}h_\om \ri\|_{L^2(S^1)}^2
	\leq
	CM_\om^{-2}
	\lf\| D_{\alpha_\om}h_\om \ri\|_{L^2(S^1)}^2
	=
	o(1+\sigma_\om^2).
\end{equation}
Indeed, write
\(
h_\om(\theta)
=\frac1{\sqrt{2\pi}}
\sum_{j\in\mathbb Z}h_{\om,j}e^{ij\theta}
\), then $\widetilde h_\om
=
\frac{ \Pi^{\mathcal W_\om} h_\om}
{\lf\| \Pi^{\mathcal W_\om} h_\om \ri\|_{2}}$, 
where
\[
\Pi^{\mathcal W_\om} h_\om
=
\frac1{\sqrt{2\pi}}
\sum_{j\in\mathcal W_\om}h_{\om,j}e^{ij\theta},
\qquad
\Pi^{\mathcal W_\om^c}h_\om
=
\frac1{\sqrt{2\pi}}
\sum_{j\notin\mathcal W_\om}h_{\om,j}e^{ij\theta}.
\]
Following Parseval's identity and the definition
\eqref{eq:unified-effective-window}, we obtain from
\eqref{eq:h-near-minimizer-bounds} that
\bml{\label{eq:oct9}
\lf\| \Pi^{\mathcal W_\om^c}h_\om \ri\|_{L^2(S^1)}^2
=
\sum_{j\notin\mathcal W_\om}|h_{\om,j}|^2\leq
\frac{1}{M_\om^2(1+\sigma_\om)^2}
\sum_{j\notin\mathcal W_\om}
(j-\alpha_\om)^2|h_{\om,j}|^2\\
\leq
\frac{1}
{M_\om^2(1+\sigma_\om)^2} \lf\|D_{\alpha_\om}h_\om \ri\|_{L^2(S^1)}^2
\leq
\frac{C}{M_\om^2}.
}
Since \( \Pi^{\mathcal W_\om}\) is an orthogonal Fourier projection commuting with
\(D_{\alpha_\om}\), we also have
\[
\lf\|\Pi^{\mathcal W_\om} h_\om \ri\|_{L^2(S^1)}\leq1,
\qquad
\lf\|D_{\alpha_\om} \Pi^{\mathcal W_\om} h_\om \ri\|_{L^2(S^1)}^2
+
\lf\|D_{\alpha_\om} \Pi^{\mathcal W_\om^c}h_\om \ri\|_{L^2(S^1)}^2
=
\lf\|D_{\alpha_\om}h_\om \ri\|_{L^2(S^1)}^2
\leq
C(1+\sigma_\om^2).
\]
We obtain from \eqref{eq:oct9} that
\begin{equation}\label{eq:oct13}
\lf\|\Pi^{\mathcal W_\om} h_\om \ri\|_{L^2(S^1)}^2
=
1- \lf\|  \Pi^{\mathcal W_\om^c}h_\om \ri\|_{L^2(S^1)}^2
=
1+\OO(M_\om^{-2}),
\qquad
\widetilde h_\om
=\bigl(1+\OO(M_\om^{-2})\bigr)  \Pi^{\mathcal W_\om} h_\om,
\end{equation}
and \( \Pi^{\mathcal W_\om}\) commutes with
\(D_{\alpha_\om}\), we have
\begin{equation*}
	\big\|D_{\alpha_\om}\widetilde h_\om \big\|_{L^2(S^1)}^2
	-
	\lf\|D_{\alpha_\om}h_\om \ri\|_{L^2(S^1)}^2
	\leq
	CM_\om^{-2}
	\lf\| D_{\alpha_\om}h_\om \ri\|_{L^2(S^1)}^2
	=
	o(1+\sigma_\om^2).
\end{equation*}
This proves the claim \eqref{eq:oct11}.

On the other hand, applying \eqref{eq:unified-Sobolev-magnetic} to 
\( \Pi^{\mathcal W_\om}h_\om\)  and
\( \Pi^{\mathcal W_\om^c}h_\om\), respectively, and using
\eqref{eq:h-near-minimizer-bounds} together with  \eqref{eq:oct9} gives
\begin{equation}\label{eq:oct10}
\lf\|  \Pi^{\mathcal W_\om} h_\om \ri\|_{L^4(S^1)}^4
\leq
C(1+\sigma_\om),
\qquad
\lf\|  \Pi^{\mathcal W_\om^c}h_\om \ri\|_{L^4(S^1)}^4
\leq
CM_\om^{-3} \lf(1+\sigma_\om \ri).
\end{equation}
It follows from \eqref{eq:oct9} and \eqref{eq:oct10} that
\bml{\label{eq:erroe-L4}
\sigma_\om
\lf|
\int_{S^1} |h_\om|^4-
\int_{S^1}\lf| \Pi^{\mathcal W_\om} h_\om \ri|^4
\ri| \leq
C\sigma_\om
\lf(
\lf\|h_\om \ri\|_{L^4(S^1)}^3
+
\lf\|  \Pi^{\mathcal W_\om} h_\om \ri\|_{L^4(S^1)}^3
\ri)
\lf\|  \Pi^{\mathcal W_\om^c}h_\om \ri\|_{L^4(S^1)} \\ \leq
C\frac{\sigma_\om(1+\sigma_\om)}
{M_\om^{3/4}}
=
o(1+\sigma_\om^2).
}
Moreover, \eqref{eq:oct13} gives
\(\|  \Pi^{\mathcal W_\Omega}h_\Omega\|_{L^2(S^1)}^{-4}-1
= \OO(M_\Omega^{-2})\).
By the definition of \(\widetilde h_\Omega\),
\eqref{eq:erroe-L4}, and \eqref{eq:oct10}, we obtain
\bml{ \label{eq:oct12}
 \sigma_\Omega
 \lf|\int_{S^1}|\widetilde h_\Omega|^4
-\int_{S^1}|h_\Omega|^4
\ri|\\
\leq \sigma_\Omega
\lf|
\int_{S^1} \lf|  \Pi^{\mathcal W_\Omega}h_\Omega \ri|^4
-\int_{S^1}|h_\Omega|^4
\ri|+\sigma_\Omega
\lf( \lf\|  \Pi^{\mathcal W_\Omega}h_\Omega \ri\|_{L^2(S^1)}^{-4}-1 \ri)
\lf\|  \Pi^{\mathcal W_\Omega}h_\Omega \ri\|_{L^4(S^1)}^4\\
\leq o(1+\sigma_\Omega^2)
+CM_\Omega^{-2}\sigma_\Omega
\|\widehat \Pi^{\mathcal W_\Omega}h_\Omega\|_{L^4(S^1)}^4= o(1+\sigma_\Omega^2).
}
Combining \eqref{eq:oct11} and \eqref{eq:oct12} gives \eqref{eq:1D-truncation-energy}, and the proof is completed. 
\end{proof}

\begin{proof}[\normalfont\bfseries Proof of Theorem~\ref{thm:A}]
We first
prove the following upper bound:
\begin{equation}\label{eq:upper}
\gseom-E_\om^{\rm int} \leq e_{\alpha_\om,\sigma_\om}+o(1+\sigma_\om^2).
\end{equation} 
Equation \eqref{eq:coercive-E} gives
\(e_{\alpha_\om,\sigma_\om}>-\infty\).  Hence, we may choose
\(\varepsilon_\om\searrow0\) and
\(h_\om\in H^1(S^1)\), with \(\|h_\om\|_{L^2(S^1)}=1\), such that
\begin{equation}\label{eq:upper-1D-near-minimizer}
	\mathcal E_{\alpha_\om,\sigma_\om}(h_\om)
	\le
	e_{\alpha_\om,\sigma_\om}
	+
	\varepsilon_\om(1+\sigma_\om^2).
\end{equation}
For the normalized constant trial state
\(
h_{0}(\theta):=(2\pi)^{-1/2},
\)
\[
e_{\alpha_\om,\sigma_\om}
\leq
\mathcal E_{\alpha_\om,\sigma_\om}(h_{0})
=
-\frac{\Gamma\sigma_\Omega}{4\pi}
\leq0.
\]
Consequently, \eqref{eq:upper-1D-near-minimizer} gives
\(
\mathcal E_{\alpha_\om,\sigma_\om}(h_\om)
\leq
\varepsilon_\om(1+\sigma_\om^2),
\)
so Lemma~\ref{lem:unified-1D-truncation} applies.  Define
\[
\widetilde h_\om
:=
\frac{  \Pi^{\mathcal W_\om} h_\om}
{\lf\|  \Pi^{\mathcal W_\om} h_\om \ri\|_{L^2(S^1)}}.
\]
By Lemma~\ref{lem:unified-1D-truncation}, for all sufficiently large
\(\om\),
\[
\widetilde h_\om(\theta)
=
\frac1{\sqrt{2\pi}}
\sum_{j\in\mathcal W_\om}\widetilde h_{\om,j}e^{ij\theta},
\qquad
\sum_{j\in\mathcal W_\om} \big|\widetilde h_{\om,j} \big|^2
= 
\big\|\widetilde h_\om\big\|_{L^2(S^1)}^2
=1.
\]
Moreover, \eqref{eq:upper-1D-near-minimizer} and
\eqref{eq:1D-truncation-energy} give
\begin{equation}\label{eq:truncated-near-minimizer}
	\mathcal E_{\alpha_\om,\sigma_\om}(\widetilde h_\om)
	\le
	e_{\alpha_\om,\sigma_\om}
	+
	o(1+\sigma_\om^2).
\end{equation}

Next, we define
\begin{equation}\label{eq:unified-first-band-trial}
	u_{\om,\widetilde h_\om}(r,\theta)
	:=
	\frac1{\sqrt{2\pi}}
	\sum_{j\in\mathcal W_\om}
	\widetilde h_{\om,j}
	\phi_{\om,n_\om+j}^{(1)}(r)
	e^{i(n_\om+j)\theta},
\end{equation}
then we have
\[
\big\|u_{\om,\widetilde h_\om}\big\|_{L^2(\mathbb R^2)}^2
=
\sum_{j\in\mathcal W_\om} \big|\widetilde h_{\om,j} \big|^2
=
\big\|\widetilde h_\om \big\|_{L^2(S^1)}^2
=1.
\]
Thus \(u_{\om,\widetilde h_\om}\) is an admissible
two-dimensional trial state.
For its linear energy, since \(E_\om^{\rm int}=\lambda_{\om,n_\om}^{(1)}\),  orthogonality and
Proposition~\ref{prop:first-band-expansion} give
\bmln{
\mathcal Q_\om(u_{\om,\widetilde h_\om})
-E_\om^{\rm int} =
\sum_{j\in\mathcal W_\om}
\lf(\lambda_{\om,n_\om+j}^{(1)}-E_\om^{\rm int} \ri)
\big|\widetilde h_{\om,j} \big|^2  \\
 =
\frac19\int_{S^1} \big|D_{\alpha_\om}\widetilde h_\om \big|^2\,d\theta
-
\frac19\alpha_\om^2 \big\|\widetilde h_\om \big\|_{L^2(S^1)}^2
+
o(1+\sigma_\om^2).
}
Since \(e_{\alpha_\om,\sigma_\om}\leq0\),
\eqref{eq:truncated-near-minimizer} shows that
\(\widetilde h_\om\) satisfies the hypothesis of
Lemma~\ref{lem:unified-1D-truncation}.  Hence
\eqref{eq:h-near-minimizer-bounds} and its Fourier expansion give
\[
\sum_{j\in\mathcal W_\om}
\bigl(1+(j-\alpha_\om)^2\bigr)|\widetilde h_{\om,j}|^2
\leq
C(1+\sigma_\om^2).
\]
Following above estimate and \(\|\widetilde h_\om\|_{L^2(S^1)}^2
=1\), we can repeat the same argument used in proving
\eqref{eq:radial-profile-replacement}   
with \(a_{n_\Omega+j,1}\) replaced by \(\widetilde h_{\om,j}\), to obtain
\[
\Big\|
u_{\om,\widetilde h_\om}
-
e^{in_\om\theta}\phi_{\om,\tau_\om}^{(1)}
\widetilde h_\om
\Big\|_{L^4(\mathbb R^2)}
=
o(\om^{1/8}(1+\sigma_\om)^{1/4}).
\]
Recall from \eqref{eq:h-near-minimizer-bounds} that
\(\|\widetilde h_\om\|_{L^4(S^1)}^4\leq C(1+\sigma_\om)\).  
The preceding \(L^4\) approximation and \eqref{eq:tau-profile-Gamma-new} give
\[
a_\om
\int_{\mathbb R^2} \big|u_{\om,\widetilde h_\om} \big|^4\,dx
=
\Gamma\sigma_\om
\int_{S^1} |\widetilde h_\om(\theta)|^4\,d\theta
+
o(1+\sigma_\om^2).
\]
Combining the linear and nonlinear estimates with
\eqref{eq:truncated-near-minimizer}, we conclude that
\[
\begin{aligned}
\gseom -E_\om^{\rm int}
&\leq
\enom(u_{\om,\widetilde h_\om})
-E_\om^{\rm int}=
\mathcal E_{\alpha_\om,\sigma_\om}(\widetilde h_\om)
+o(1+\sigma_\om^2)\leq
e_{\alpha_\om,\sigma_\om}
+o(1+\sigma_\om^2).
\end{aligned}
\]
This proves \eqref{eq:upper}.

We next prove the lower bound:
\begin{align}
\enom(u_\om)-E_\om^{\rm int}
\geq e_{\alpha_\om,\sigma_\om}
+o(1+\sigma_\om^2).
\label{eq:lowerbound}
\end{align}
Let \(g_\Omega\) be the angular profile associated with this
\(\gsom\) in \eqref{eq:g-unified-def}.
Recall from \eqref{eq:decom} and
\eqref{eq:first-band-coefficients} that
\[
u_\om
=
\frac1{\sqrt{2\pi}}
\sum_{j\in\mathbb Z}
a_{n_\Omega+j,1}\phi_{\om,n_\om+j}^{(1)}(r)
e^{i(n_\om+j)\theta}
+
\eta_\om,
\]
where \(\eta_\om=(1-P_\om)u_\om\) contains the higher radial
bands.  
Lemma~\ref{lem:nonlinear-reduction} gives
\(\|\Pi^{\mathcal W_\Omega}g_\Omega\|_{L^2(S^1)}^2=1-o(1)\) and
\eqref{eq:quartic-reduction}.  Hence, for all sufficiently
large \(\om\), we may define
\begin{equation}\label{eq:unified-normalized-profile}
f_\om 
:=
\frac{\Pi^{\mathcal W_\Omega}g_\Omega}{\lf\| \Pi^{\mathcal W_\Omega}g_\Omega \ri\|_{L^2(S^1)}},\quad\hbox{where $  \Pi^{\mathcal W_\Omega}g_\Omega$ is given by \eqref{eq:gW-def-unified}}.
\end{equation}
The bounds \eqref{eq:gJ-unified-H1-bound},
\eqref{eq:int-angular-tail-L4}, and
\eqref{eq:unified-window-mass} imply
\bml{\label{eq:normalization-energy}
\frac19 \lf\|D_{\alpha_\om}  \Pi^{\mathcal W_\Omega}g_\Omega \ri\|_{L^2(S^1)}^2
-\frac{\alpha_\om^2}{9} \lf \|  \Pi^{\mathcal W_\Omega}g_\Omega \ri\|_{L^2(S^1)}^2
-\frac{\Gamma\sigma_\om}{2}
\int_{S^1} \lf|  \Pi^{\mathcal W_\Omega}g_\Omega \ri|^4\,d\theta\\
 =
\mathcal E_{\alpha_\om,\sigma_\om}
( f_\om)
+
o(1+\sigma_\om^2). 
}
By \eqref{eq:energy-excess-spectral} and \eqref{eq:E-int-def},
\bml{
\enom(u_\om)-E_\om^{\rm int}
 =
\sum_{j\in\mathcal W_\om}
\lf(\lambda_{\om,n_\om+j}^{(1)}
-E_\om^{\rm int} \ri)|a_{n_\Omega+j,1}|^2
+\sum_{j\notin\mathcal W_\om}
\lf(\lambda_{\om,n_\om+j}^{(1)}
-E_\om^{\rm int} \ri)|a_{n_\Omega+j,1}|^2
\\
+
\sum_{n\in\mathbb Z}\sum_{m\geq2}
\lf(\lambda_{\om,n}^{(m)}
-E_\om^{\rm int} \ri)|a_{n,m}|^2
-\frac{a_\om}{2}
\int_{\mathbb R^2}|u_\om|^4\,dx
 \\
\geq
\sum_{j\in\mathcal W_\om}
\lf(\lambda_{\om,n_\om+j}^{(1)}
-E_\om^{\rm int} \ri)|a_{n_\Omega+j,1}|^2
-\frac{a_\om}{2}
\int_{\mathbb R^2}|u_\om|^4\,dx.
\label{eq:spectral-truncation-lb}
}
Recall that \(E_\om^{\rm int}=\lambda_{\om,n_\om}^{(1)}\).
On \(\mathcal W_\om\), Proposition~\ref{prop:first-band-expansion}
and the Fourier representation \eqref{eq:gW-def-unified} give
\bml{\label{eq:oct14}
\sum_{j\in\mathcal W_\om}
\lf(\lambda_{\om,n_\om+j}^{(1)}
-E_\om^{\rm int} \ri)|a_{n_\Omega+j,1}|^2=
\frac19
\sum_{j\in\mathcal W_\om}
(j-\alpha_\om)^2|a_{n_\Omega+j,1}|^2
-\frac{\alpha_\om^2}{9}
\sum_{j\in\mathcal W_\om}|a_{n_\Omega+j,1}|^2
+o(1+\sigma_\om^2)\\
=
\frac19 \lf\|D_{\alpha_\om}  \Pi^{\mathcal W_\Omega}g_\Omega \ri\|_{L^2(S^1)}^2
-\frac{\alpha_\om^2}{9} \lf\|  \Pi^{\mathcal W_\Omega}g_\Omega \ri\|_{L^2(S^1)}^2
+o(1+\sigma_\om^2).
}
We conclude from \eqref{eq:normalization-energy}, \eqref{eq:oct14} and \eqref{eq:quartic-reduction} that
\begin{equation}\label{eq:oct15}
\sum_{j\in\mathcal W_\om}
\lf(\lambda_{\om,n_\om+j}^{(1)}
-E_\om^{\rm int} \ri)|a_{n_\Omega+j,1}|^2
-\frac{a_\om}{2}
\int_{\mathbb R^2}|u_\om|^4\,dx=
\mathcal E_{\alpha_\om,\sigma_\om}
(f_\om)
+
o(1+\sigma_\om^2).
\end{equation}
Combining \eqref{eq:oct15} with
\eqref{eq:spectral-truncation-lb} gives
\begin{align}
\enom(u_\om)-E_\om^{\rm int}
\geq
\mathcal E_{\alpha_\om,\sigma_\om}
(f_\om)
+o(1+\sigma_\om^2)
\geq
e_{\alpha_\om,\sigma_\om}
+o(1+\sigma_\om^2)\label{eq:unified-lower-bound}.
\end{align}
This proves \eqref{eq:lowerbound}.
Since \( \enom(\gsom)= \gseom \),
\eqref{eq:upper} and \eqref{eq:lowerbound} prove
\eqref{eq:A-energy-expansion}.

The function \(f_\Omega \) defined in
\eqref{eq:unified-normalized-profile} belongs to \(H^1(S^1)\)
and satisfies \(\|f_\Omega \|_{L^2(S^1)}=1\).
By \eqref{eq:e-circle}, \eqref{eq:unified-lower-bound}, and
\eqref{eq:A-energy-expansion},
\[
e_{\alpha_\Omega,\sigma_\Omega}
\leq \mathcal E_{\alpha_\Omega,\sigma_\Omega}(f_\Omega)
\leq e_{\alpha_\Omega,\sigma_\Omega}+o(1+\sigma_\Omega^2).
\]
Thus \(  f_\Omega \) satisfies \eqref{eq:A-angular-almost-minimizer}. Moreover,
\eqref{eq:unified-window-mass} and
\eqref{eq:unified-window-L2-profile}, together with the normalization in
\eqref{eq:unified-normalized-profile}, give
\eqref{eq:A-L2-reduction}. This completes the proof of Theorem~\ref{thm:A}.
\end{proof}

\section{Analysis of the Effective Circle Problem}
\label{sec:circle-consequences}
In this section, we prove Theorems~\ref{thm:B} and~\ref{thm:C}
using the circle reduction established in Theorem~\ref{thm:A}.
More precisely, in Subsection~\ref{sec:homogeneous-circle}, we consider the case where
\(\alpha_\Omega\to\alpha\in(-1/2,1/2)\) and
\(\sigma_\Omega\to\sigma\in[0,\sigma_{\rm c}(\alpha))\),
and prove Theorem~\ref{thm:B}.
In Subsection~\ref{sec:strong-effective-coupling}, we treat
\(\sigma_\Omega\to\infty\) with
\(\sigma_\Omega=o(\sqrt{\Omega})\), and prove
Theorem~\ref{thm:C}.

\subsection{The effective circle problem}
\label{sec:homogeneous-circle}
Assume
\begin{equation}\label{eq:bounded-sigma}
	\sigma_\om\to\sigma\in[0,\infty),
	\qquad
	\alpha_\om\to\alpha\in[-1/2,1/2].
\end{equation}
Then \(1+\sigma_\om^2=\OO(1)\), so Theorem~\ref{thm:A}
gives
\begin{equation}\label{eq:bounded-energy}
	\gseom
	=
	E_\om^{\rm int}+
	e_{\alpha_\om,\sigma_\om}
	+o(1).
\end{equation}
Let \( f_\Omega \) be the angular profile constructed in
the proof of Theorem~\ref{thm:A}, as defined in
\eqref{eq:unified-normalized-profile}. Then it is bounded in
\(H^1(S^1)\).  By the compact embedding
\(H^1(S^1)\hookrightarrow L^q(S^1)\) $(q>1)$, there exists
\(f\in H^1(S^1)\) such that, up to a subsequence,
\[
  f_\om \xrightarrow[\om \to \infty]{\mathrm{w}-H^1(S^1)} f,	\qquad
f_\om \xrightarrow[\om \to \infty]{L^4(S^1)} f,
\]
and \( \lf\| g \ri\|_{L^2(S^1)}=1\).  Passing to the limit in the circle functional gives that
\(g\) minimizes \(\mathcal E_{\alpha,\sigma}\).  Therefore, to  
prove Theorem~\ref{thm:B}, it suffices to prove
the constant profile
\begin{equation}\label{eq:homogeneous-profile}
	h_0(\theta)=\frac{1}{\sqrt{2\pi}}
\end{equation}
is the unique minimizer of \(\mathcal E_{\alpha,\sigma}\)
up to a constant phase.

Motivated by \cite{Bonheure2021,Dolbeault2018}, we have the following result.
\begin{lemma}
\label{lem:sharp-circle-interpolation}
For every \(\alpha\in(-1/2,1/2)\) and every \(h \in H^1(S^1)\) with
\( \lf\| h \ri\|_{L^2(S^1)}=1\), one has
\begin{equation}\label{eq:scis}
	\int_{S^1}|h|^4\,d\theta-\frac1{2\pi}
	\le
	\frac{2}{\pi(1-4\alpha^2)}
	\lf(
	\int_{S^1}|(-i\partial_\theta-\alpha)h|^2\,d\theta-\alpha^2
	\ri).
\end{equation}
Moreover, the constant \(2/[\pi(1-4\alpha^2)]\) in \eqref{eq:scis}
is sharp.
\end{lemma}

The proof of Lemma~\ref{lem:sharp-circle-interpolation} is
given in Appendix~\ref{app:sharp-circle-interpolation}.

\begin{proposition}
\label{prop:homogeneous-circle}
Let \(e_{\alpha,\sigma}\) be the circle minimization problem defined by \eqref{eq:e-circle} with \(|\alpha|<1/2\).  Then, the following  holds:
\begin{enumerate}
\item[(i)] If
\begin{equation}\label{eq:homogeneous-small-threshold}
	0\le \Gamma\sigma
	\le
	\tfrac{\pi}{9}(1-4\alpha^2),
\end{equation}
then \(h_0\) is a global minimizer of
\(\mathcal E_{\alpha,\sigma}\) under the constraint \(\lf\|h\ri\|_{L^2(S^1)}=1\).
In particular, if $0\le \Gamma\sigma
<
\tfrac{\pi}{9}(1-4\alpha^2)$, then
the minimizer is unique up to multiplication by a constant phase, that can be chosen in such a way that the minimizer equals \(1/\sqrt{2\pi}\).

\item[(ii)] The threshold is sharp for every \(|\alpha|<1/2\): if
\begin{equation}\label{eq:homogeneous-sharp-threshold}
	\Gamma\sigma>\tfrac{\pi}{9}(1-4\alpha^2),
\end{equation}
then \(h_0\) is not a minimizer.
\end{enumerate}
\end{proposition}

\begin{proof}[\normalfont\bfseries Proof]
(i). For \(\|h\|_{L^2(S^1)}=1\), set
\[
K_\alpha(h):=
\int_{S^1}|(-i\partial_\theta-\alpha)h|^2\,d\theta-\alpha^2,
\qquad
D(h):=\int_{S^1}|h|^4\,d\theta-\frac1{2\pi}.
\]
Lemma~\ref{lem:sharp-circle-interpolation} gives
\begin{equation}\label{eq:quartic-gap-control}
	D(h)\leq\frac{2}{\pi(1-4\alpha^2)}K_\alpha(h).
\end{equation}
Since
$\mathcal E_{\alpha,\sigma}(h_0)=-\frac{\Gamma\sigma}{4\pi},
$,
we obtain
\begin{equation}\label{eq:homogeneous-energy-gap}
\mathcal E_{\alpha,\sigma}(h)
-\mathcal E_{\alpha,\sigma}(h_0)
=\frac19K_\alpha(h)-\frac{\Gamma\sigma}{2}D(h)\\
\geq
\lf[
	\frac19-\frac{\Gamma\sigma}{\pi(1-4\alpha^2)}
\ri]K_\alpha(h).
\end{equation}

For every \(h\in H^1(S^1)\) with \(\|h\|_{L^2(S^1)}=1\), we have
\begin{equation}\label{eq:homogeneous-gap-H1}
K_\alpha(h)\geq
(1-2|\alpha|)\int_{S^1}|h'|^2\,d\theta\geq 0.
\end{equation}
Indeed, writing
\(h=\sum_{n\in\mathbb Z}c_n e^{in\theta}/\sqrt{2\pi}\), we obtain
\[
K_\alpha(h)
=\sum_{n\in\mathbb Z}(n^2-2\alpha n)|c_n|^2
\geq (1-2|\alpha|)\sum_{n\in\mathbb Z}n^2|c_n|^2
=(1-2|\alpha|)\int_{S^1}|h'|^2\,d\theta.
\]
Since \(|\alpha|<1/2\), this bound and the normalization imply that
\(K_\alpha(h)=0\) if and only if
\(h(\theta)=e^{i\gamma}/\sqrt{2\pi}\) for some \(\gamma\in\mathbb R\).
Under \eqref{eq:homogeneous-small-threshold}, the pre-factor in
\eqref{eq:homogeneous-energy-gap} is nonnegative, so \(h_0\) is a
global minimizer. If the threshold inequality is strict and \(h\)
is a minimizer, the energy gap vanishes and the positive coefficient
forces \(K_\alpha(h)=0\). Thus \(h=e^{i\gamma}h_0\) for some
\(\gamma\in\mathbb R\). This proves (i).

%

(ii). To prove sharpness for arbitrary \(|\alpha|<1/2\), set
\[
\eta_\alpha(\theta):=
\frac{(1+2\alpha)e^{i\theta}+(1-2\alpha)e^{-i\theta}}
{2\sqrt{2\pi}},
\qquad
h_\varepsilon:=
\frac{h_0+\varepsilon\eta_\alpha}
{\sqrt{1+\varepsilon^2\|\eta_\alpha\|_{L^2(S^1)}^2}} .
\]
Then \( \lf\|h_\varepsilon \ri\|_{L^2(S^1)}=1\), and a direct expansion gives
\[
K_\alpha(h_\varepsilon)
=\tfrac{1-4\alpha^2}{2}\varepsilon^2+O(\varepsilon^4),
\qquad
D(h_\varepsilon)
=\tfrac1\pi\varepsilon^2+O(\varepsilon^4).
\]
Consequently,
\[
\mathcal E_{\alpha,\sigma}(h_\varepsilon)
-\mathcal E_{\alpha,\sigma}(h_0)
=
\frac{\varepsilon^2}{2\pi}
\Big[
	\frac{\pi}{9}(1-4\alpha^2)-\Gamma\sigma
\Big]
+O(\varepsilon^4).
\]
Under \eqref{eq:homogeneous-sharp-threshold}, this quantity is negative for
all sufficiently small \(\varepsilon>0\).  Hence \(h_0\) is not a minimizer.
\end{proof}

We first establish two profile convergence results used in the proof
of Theorem~\ref{thm:B}.

\begin{lemma}
\label{lem:giant-angular-rigidity}
Let the assumptions of Theorem~\ref{thm:B} be satisfied. Then, for any
$ f_\om $ as in \eqref{eq:unified-normalized-profile},
there exist constant phases
\(\gamma_\om\in\mathbb R\) such that
\begin{equation}\label{eq:giant-angular-H1}
 e^{-i\gamma_\om} f_\om \xrightarrow[\om \to \infty]{H^1(S^1)}  h_0 = \tfrac{1}{\sqrt{2\pi}}.
\end{equation}
\end{lemma}

\begin{proof}[\normalfont\bfseries Proof]
Applying \eqref{eq:homogeneous-energy-gap} with
\((\alpha,\sigma)=(\alpha_\om,\sigma_\om)\), and then using
\eqref{eq:homogeneous-gap-H1}, gives
\[
\mathcal E_{\alpha_\om,\sigma_\om}(h)
-\mathcal E_{\alpha_\om,\sigma_\om}(h_0)\geq
\lf[
\frac19-
\frac{\Gamma\sigma_\om}{\pi(1-4\alpha_\om^2)}
\ri]
(1-2|\alpha_\om|) \lf\|h' \ri\|_{L^2(S^1)}^2.
\]
The strict inequality in \eqref{eq:bounded-regime} and
\(\alpha\in(-1/2,1/2)\) imply that both factors on the right-hand side
have positive lower bounds independent of all sufficiently large
\(\om\).  Hence there exists \(c>0\) such that
\begin{equation}\label{eq:giant-circle-coercivity}
 \mathcal E_{\alpha_\om,\sigma_\om}(h)
 -\mathcal E_{\alpha_\om,\sigma_\om}(h_0)
 \geq c \lf\|h' \ri\|_{L^2(S^1)}^2,
 \qquad \lf\|h \ri\|_{L^2(S^1)}=1.
\end{equation}
Since \(u_\om\) is a minimizer and \(\sigma_\om=\OO(1)\),
\eqref{eq:A-energy-expansion} gives
\[
\enom(\gsom)-E_\om^{\rm int}
=e_{\alpha_\om,\sigma_\om}+o(1).
\]
Combining this identity with \eqref{eq:unified-lower-bound} gives
\[
e_{\alpha_\om,\sigma_\om}\leq\mathcal E_{\alpha_\om,\sigma_\om}
(f_\om)
\leq e_{\alpha_\om,\sigma_\om}+o(1).
\]
Hence
\(
\mathcal E_{\alpha_\om,\sigma_\om}
(f_\om)
=e_{\alpha_\om,\sigma_\om}+o(1).
\)
Proposition~\ref{prop:homogeneous-circle} and the strict
inequality in \eqref{eq:bounded-regime} give
\(
e_{\alpha_\om,\sigma_\om}
=\mathcal E_{\alpha_\om,\sigma_\om}(h_0)
\)
for all sufficiently large \(\om\).  Consequently,
\[
\mathcal E_{\alpha_\om,\sigma_\om}
(f_\om)
-\mathcal E_{\alpha_\om,\sigma_\om}(h_0)
=o(1).
\]
Combining this identity with \eqref{eq:giant-circle-coercivity} gives
\(\lf\| f_\om' \ri\|_{L^2(S^1)}=o(1)\).  Set
\[
\tilde{f}_\om
:=
\frac1{2\pi}\int_{S^1}  f_\om (\theta)\,d\theta.
\]
Poincar\'e's inequality gives
\bdm
	\lf\|  f_\om -  \tilde{f}_\om \ri \|_{L^2(S^1)}\leq \lf\| f_\Omega ' \ri\|_{L^2(S^1)}=o(1).
\edm
Since
\(\|  f_\om \|_{L^2(S^1)}=1\), it follows that
\(\sqrt{2\pi}| \tilde{f}_\om|\to1\).  Choosing
\(\gamma_\om=\arg \tilde f_\om\) for all sufficiently large
\(\om\), we obtain \eqref{eq:giant-angular-H1}.
\end{proof}

\begin{lemma}
\label{lem:giant-local-profile}
Under the same  assumptions of Lemma~\ref{lem:giant-angular-rigidity}, let \(\gamma_\om\) be
the phases obtained in Lemma~\ref{lem:giant-angular-rigidity}.  Then, for
every fixed \(R>0\),
\[
\sup_{\substack{|t|\leq R,\,\theta\in S^1}}
\lf|
\tfrac{1}{\om^{1/4}} e^{-i\gamma_\om-i n_\om\theta}
u_\om \lf(1+ \tfrac{t}{\sqrt\om},\theta \ri)
-\tfrac{1}{\sqrt{2\pi}} \chi_0(t)
 \ri| \xrightarrow[\om\to\infty]{} 0.
\]
\end{lemma}
\begin{proof}[\normalfont\bfseries Proof]
Let \(R>0\) be fixed and recall that
\[
w_\om=\lf(1-P_\om^{\mathcal J_\om}\ri)u_\om,
\qquad
u_\om=P_\om^{\mathcal J_\om}u_\om+w_\om,
\]
where $P_\om^{\mathcal J_\om}u_\om$ is given by \eqref{eq:first-band-set-projection} with $A=\mathcal J_\om$ and $u=u_\om$.
To prove Lemma~\ref{lem:giant-local-profile}, it is enough to prove the following two estimates:
\begin{align}
	\lf\|w_\om \ri\|_{L^\infty(\{\sqrt\om\,||x|-1|\leq R\})}
	&=o(1), \label{eq:giant-outer-Linfty-small}	\\
	\tfrac{1}{\om^{1/4}} e^{-i n_\om\theta}
\lf(P_\om^{\mathcal J_\om} u_\om \ri)\lf(1+\tfrac{t}{\sqrt\om},\theta\ri)
& =\tfrac{1}{\sqrt{2\pi}} e^{i\gamma_\om}\chi_0(t)+o(1) \label{eq:giant-low-uniform-profile}
\end{align}
uniformly for \(|t|\leq R\) and \(\theta\in S^1\).
Indeed, we have
\bmln{
\tfrac{1}{\om^{1/4}} e^{-i\gamma_\om-i n_\om\theta}
u_\om\lf(1+\tfrac{t}{\sqrt\om},\theta\ri)
-\tfrac{1}{\sqrt{2\pi}}\chi_0(t)
\\
= e^{-i\gamma_\om}\left[
\tfrac{1}{\om^{1/4}} e^{-i n_\om\theta}
\lf(P_\om^{\mathcal J_\om}u_\om \ri)\lf(1+\tfrac{t}{\sqrt\om},\theta\ri)
-\tfrac{1}{\sqrt{2\pi}}e^{i\gamma_\om}\chi_0(t)
\right]
+ \tfrac{1}{\om^{1/4}} e^{-i\gamma_\om-i n_\om\theta}
w_\om\lf(1+\tfrac{t}{\sqrt\om},\theta\ri).
}
The first term on the right is \(o(1)\) by
\eqref{eq:giant-low-uniform-profile}, while the second is
\(o(\om^{-1/4})=o(1)\) by
\eqref{eq:giant-outer-Linfty-small}.  It therefore remains to prove these
two estimates.

We first prove \eqref{eq:giant-outer-Linfty-small}.
We claim that
\begin{equation}\label{eq:giant-outer-form-small}
	\|w_\om\|_2^2=o(\om^{-1}),
	\qquad \mathcal Q_\om(w_\om)=o(1).
\end{equation}
For \(A\subset\mathbb Z\), define
\begin{equation}\label{eq:giant-discarded-energy}
\begin{split}
\mathcal R_\Omega(A):=\sum_{j\notin A}
\lf(\lambda_{\Omega,n_\Omega+j}^{(1)}
-E_\Omega^{\rm int} \ri)|a_{n_\Omega+j,1}|^2
+\sum_{n\in\mathbb Z}\sum_{m\geq2}
\lf(\lambda_{\Omega,n}^{(m)}
-E_\Omega^{\rm int} \ri)|a_{n,m}|^2\geq0,
\end{split}
\end{equation}
where \(a_{n,m}\), \(n\in\mathbb Z\), \(m\in \N_+ \), are the spectral
coefficients of \(\gsom\) in \eqref{eq:decom}.
In particular, \(\mathcal R_\Omega(\mathcal J_\Omega)
=\mathcal Q_\Omega(w_\Omega)-E_\Omega^{\rm int}\|w_\Omega\|_2^2\).
We establish the following intermediate estimate:
\begin{equation}\label{eq:giant-discarded-small}
0\leq\mathcal R_\Omega(\mathcal J_\Omega)
\leq\mathcal R_\Omega(\mathcal W_\Omega)=o(1),
\end{equation}
where \(\mathcal W_\Omega\) is defined in \eqref{eq:unified-effective-window}.
Since \(\lambda_{\Omega,n_\Omega+j}^{(1)}-E_\Omega^{\rm int}\geq0\),
\(\lambda_{\Omega,n}^{(m)}-E_\Omega^{\rm int}\geq0\), and
\(\mathcal W_\Omega\subset\mathcal J_\Omega\), we have
\(0\leq\mathcal R_\Omega(\mathcal J_\Omega)
\leq\mathcal R_\Omega(\mathcal W_\Omega)\).
Consequently, it suffices to show that
\(\mathcal R_\Omega(\mathcal W_\Omega)=o(1)\).
Since \(\sigma_\Omega=O(1)\) by \eqref{eq:bounded-regime}, the errors in
\eqref{eq:A-energy-expansion} and \eqref{eq:oct15} are \(o(1)\).
Combining these estimates with the spectral identity in
\eqref{eq:spectral-truncation-lb}, we obtain
\bml{ \label{eq:giant-retained-lower-bound}
 e_{\alpha_\Omega,\sigma_\Omega}+o(1)
= \enom(\gsom)-E_\Omega^{\rm int}\\
=\mathcal E_{\alpha_\Omega,\sigma_\Omega}(f_\Omega)
+\mathcal R_\Omega(\mathcal W_\Omega)+o(1)\geq e_{\alpha_\Omega,\sigma_\Omega}
+\mathcal R_\Omega(\mathcal W_\Omega)+o(1).
}
The last inequality follows from
\(\| f_\Omega \|_{L^2(S^1)}=1\) and \eqref{eq:e-circle}.
Since \(\mathcal R_\Omega(\mathcal W_\Omega)\geq0\),
\eqref{eq:giant-retained-lower-bound} yields
\(\mathcal R_\Omega(\mathcal W_\Omega)=o(1)\), proving
\eqref{eq:giant-discarded-small}.

Based on \eqref{eq:giant-discarded-small}, we can prove \eqref{eq:giant-outer-form-small} as follows. Recall from
\eqref{eq:oct3}  and
\eqref{eq:first-band-decomp} that
\begin{equation}\label{eq:fouerie-form}
c_{\om,j}=a_{n_\om+j,1},
\qquad
u_\om
=
\frac1{\sqrt{2\pi}}
\sum_{j\in\mathbb Z}
c_{\om,j}\phi_{\om,n_\om+j}^{(1)}(r)
e^{i(n_\om+j)\theta}
+
\eta_\om,
\end{equation}
where \(\eta_\om=(1-P_\om)u_\om\) contains the higher radial
bands.    
Consequently,
\[
\begin{aligned}
w_\om
&=\lf(1-P_\om^{\mathcal J_\om} \ri)u_\om=
\frac1{\sqrt{2\pi}}
\sum_{j\notin\mathcal J_\om}
c_{\om,j}\phi_{\om,n_\om+j}^{(1)}(r)
e^{i(n_\om+j)\theta}
+\eta_\om.
\end{aligned}
\]
For \(j\notin\mathcal J_\Omega\), we have
\(|j-\alpha_\Omega|>\sqrt{\Omega}\) by \eqref{eq:rough-angular-window}.
Hence the angular gap
\eqref{eq:angular-gap-new} and the radial gap
\eqref{eq:radial-gap-new} give
\[
(c\Omega-C)\|w_\Omega\|_2^2
\leq\mathcal R_\Omega(\mathcal J_\Omega)
\leq\mathcal R_\Omega(\mathcal W_\Omega)
=o(1),
\]
where the last equality follows from \eqref{eq:giant-discarded-small}.
It follows that \(\|w_\Omega\|_2^2=o(\Omega^{-1})\).
Since \(E_\Omega^{\rm int}=O(\Omega)\) by \eqref{eq:E-size-new},
\[
\mathcal Q_\Omega(w_\Omega)
=\mathcal R_\Omega(\mathcal J_\Omega)
+E_\Omega^{\rm int}\|w_\Omega\|_2^2=o(1).
\]
This proves \eqref{eq:giant-outer-form-small}.

In the following, we use \eqref{eq:giant-outer-form-small} and elliptic regularity theory to prove \eqref{eq:giant-outer-Linfty-small}. Let  
\begin{equation}\label{eq:giant-magnetic-operator}
H_\om:=(-i\nabla-\om x^\perp)^2+\tfrac{1}{8}\om^2(|x|^2-1)^2.
\end{equation}
Equivalently, in polar coordinates,
\begin{equation}\label{eq:Hom}
H_\om=-\frac1r\partial_r(r\partial_r)
+\left(\frac{-i\partial_\theta}{r}-\om r\right)^2
+\frac{\om^2}{8}(r^2-1)^2.
\end{equation}
The minimizers $u_\om$ satisfy
\begin{equation}\label{eq:giant-EL}
 H_\om u_\om-a_\om |u_\om|^2u_\om
 =\mu_\om u_\om,
 \qquad |\mu_\om|\leq C\om.
\end{equation}
Indeed, by \eqref{eq:X-Omega-def}, \eqref{eq:X-improved-new}, and
\eqref{eq:E-size-new}, we have
\(\mathcal Q_\Omega(\gs)=O(\Omega)\).
Testing the equation with \(\gs\) then yields  \(|\mu_\om|\leq C\om\). Moreover,
\eqref{eq:diamagnetic} and the Gagliardo--Nirenberg inequality give
\begin{equation}\label{eq:giant-L6-rough}
 \lf\|u_\om\ri\|_6^3
 \leq C \lf\|\nabla|u_\om| \ri\|_2^2 \lf\|u_\om \ri\|_2
 \leq C\om.
\end{equation}
Using \eqref{eq:fouerie-form} and the polar coordinate form of \(H_\om\) in \eqref{eq:Hom},
we obtain
\[
\lf(1-P_\om^{\mathcal J_\om} \ri)H_\om u_\om
=H_\om \lf(1-P_\om^{\mathcal J_\om} \ri)u_\om.
\]
Therefore, we can apply \(1-P_\om^{\mathcal J_\om}\) to
\eqref{eq:giant-EL}   give 
\begin{equation}\label{eq:giant-w-equation}
H_\om w_\om
-a_\om \lf(1-P_\om^{\mathcal J_\om} \ri) \lf(|u_\om|^2u_\om \ri)
=\mu_\om w_\om.
\end{equation}
Choose \(\zeta\in C_c^\infty((-2R,2R))\) such that $\zeta\equiv 1$ on
\([-R,R]\), and set
\begin{equation}\label{eq:def-zom}
\xi_\om(x):=\zeta\bigl(\sqrt\om(|x|-1)\bigr),
\qquad
z_\om:=\xi_\om w_\om.
\end{equation}
On the support of \(\xi_\om\),
\begin{equation}\label{eq:giant-cutoff-derivatives}
\tfrac{1}{8}\Omega^2(|x|^2-1)^2=\OO_R(\Omega),
\qquad
\|\nabla\xi_\om\|_\infty= \OO_R(\sqrt\om),
\qquad
\|\Delta\xi_\om\|_\infty= \OO_R(\om).
\end{equation}
Equation \eqref{eq:giant-outer-form-small} gives
\[
\lf\|(-i\nabla-\om x^\perp)w_\om \ri\|_2=o(1),
\qquad
\lf\|w_\om\ri\|_2=o(\om^{-1/2}).
\]
Using this identity,
\[
(-i\nabla-\om x^\perp)z_\om
=\xi_\om(-i\nabla-\om x^\perp)w_\om
-i(\nabla\xi_\om)w_\om,
\]
together with \eqref{eq:def-zom}, \eqref{eq:diamagnetic}, \eqref{eq:giant-outer-form-small} and
\eqref{eq:giant-cutoff-derivatives}, we obtain
\begin{equation}\label{eq:giant-cutoff-bounds}
\big\||z_\om|\big\|_{H^1(\mathbb R^2)}=o(1).
\end{equation}

Writing \eqref{eq:giant-w-equation} in terms of the magnetic Laplacian
and applying  Leibniz's rule gives 
\begin{equation}\label{eq:giant-cutoff-equation}
-(-i\nabla-\om x^\perp)^2z_\om=\zeta_\om,
\end{equation}
where
\bmln{
\zeta_\om
:=\xi_\Omega\lf(\tfrac{1}{8}\Omega^2(|x|^2-1)^2-\mu_\om \ri)w_\Omega
-a_\om\xi_\om \lf(1-P_\om^{\mathcal J_\om}\ri)
\bigl(|u_\om|^2u_\om\bigr)\\
+2i\nabla\xi_\om\cdot
(-i\nabla-\om x^\perp)w_\om
+(\Delta\xi_\om)w_\om.
}
The projection \(1-P_\om^{\mathcal J_\om}\) is a contraction.  Hence
\eqref{eq:giant-L6-rough} and
\(a_\om=\OO(\om^{-1/2})\) give
\[
a_\om \lf\| \lf(1-P_\om^{\mathcal J_\om} \ri)
\bigl(|u_\om|^2u_\om\bigr) \ri\|_2
\leq Ca_\om \lf\|u_\om \ri\|_6^3
=\OO(\sqrt\om).
\]
The remaining terms in \(\zeta_\om\) are \(o(\sqrt\om)\) by
\eqref{eq:giant-cutoff-derivatives},
\(|\mu_\om|=O(\om)\), and
\eqref{eq:giant-outer-form-small}.  Therefore
\begin{equation}\label{eq:giant-forcing-bound}
\lf\| \zeta_\om \ri\|_2=O(\sqrt\om).
\end{equation}
We next apply the diamagnetic resolvent inequality
\cite[(A.23)]{Hupfer2001}.  For every \(s>0\) and \(G\in L^2(\mathbb R^2)\),
it gives the pointwise estimate
\[
\Big|\big((-i\nabla-\om x^\perp)^2+s\big)^{-1}G\Big|
\leq
(-\Delta+s)^{-1}|G|.
\]
The Fourier representation of the free resolvent and the Cauchy--Schwarz
inequality give
\[
\begin{aligned}
\lf\|(-\Delta+s)^{-1}G \ri\|_\infty
&\leq
C \lf\|\frac{1}{|\xi|^2+s} \ri\|_{L^2(\mathbb R^2)} \lf\|G\ri\|_2\leq Cs^{-1/2} \lf\|G\ri\|_2.
\end{aligned}
\]
By \eqref{eq:giant-cutoff-bounds}, \(z_\om\) belongs to the quadratic-form
domain of \((-i\nabla-\om x^\perp)^2\).  Equations
\eqref{eq:giant-cutoff-equation} and \eqref{eq:giant-forcing-bound} then
imply that \(z_\om\) belongs to the operator domain.  Thus, for every
\(s>0\),
\[
z_\om
=
\big((-i\nabla-\om x^\perp)^2+s\big)^{-1}
(sz_\om-\zeta_\om),
\]
and consequently
\[
\lf\|z_\om\ri\|_\infty
\leq
C\Big(s^{1/2} \lf\|z_\om\ri\|_2
+s^{-1/2} \lf\|\zeta_\om\ri\|_2\Big).
\]
If \(z_\om\neq0\), choose
\[
s=1+\frac{\lf\|\zeta_\om\ri\|_2}{\lf\|z_\om\ri\|_2};
\]
the case \(z_\om=0\) is immediate.
Using \(\lf\|z_\om\ri\|_2\leq C\lf\|w_\om\ri\|_2\),
\eqref{eq:giant-outer-form-small}, and
\eqref{eq:giant-forcing-bound}, we obtain
\[
\lf\|z_\om\ri\|_\infty
\leq
C\lf( \lf\|z_\om \ri\|_2
+ \lf\|z_\om \ri\|_2^{1/2} \lf\|\zeta_\om \ri\|_2^{1/2} \ri)
=o(1).
\]
Since \(z_\om=w_\om\) when
\(\sqrt\om\,||x|-1|\leq R\), this proves \eqref{eq:giant-outer-Linfty-small}.

We next prove \eqref{eq:giant-low-uniform-profile}.
By
\eqref{eq:first-band-set-projection}, the corresponding rough window
first band projection satisfies
\begin{equation}\label{eq:PJ-def-giant}
P_\om^{\mathcal J_\om} u_\om
=
\frac1{\sqrt{2\pi}}
\sum_{j\in\mathcal J_\om}
c_{\om,j}\phi_{\om,n_\om+j}^{(1)}(r)
e^{i(n_\om+j)\theta},
\end{equation}
where \(\mathcal J_\om\) is given by \eqref{eq:rough-angular-window}.
Recall from
\eqref{eq:rescaled-eigenpairs} that
\(\psi_{\om,j}:=\psi_{\om,n_\om+j}^{(1)}
=U_\om\phi_{\om,n_\om+j}^{(1)}\).  The definition of
\(P_\om^{\mathcal J_\om}\) in \eqref{eq:PJ-def-giant} and
\eqref{eq:radial-unitary-map} give
\[
\tfrac{1}{\om^{1/4}} e^{-in_\om\theta}
\lf(P_\om^{\mathcal J_\om}u_\om \ri) \lf(1+\tfrac{t}{\sqrt\om},\theta \ri)
=
\frac1{\sqrt{2\pi}}
\sum_{j\in\mathcal J_\om}
c_{\om,j}\psi_{\om,j}(t)e^{ij\theta}.
\]
The difference between the sum on the right-hand side and
\(e^{i\gamma_\om}\chi_0(t)\) equals
\begin{equation}\label{eq:oct16}
\begin{aligned}
&\sum_{j\in\mathcal W_\om}
\bigl(c_{\om,j}-e^{i\gamma_\om}\delta_{j0}\bigr)
\psi_{\om,j}(t)e^{ij\theta}
+e^{i\gamma_\om}\bigl(\psi_{\om,0}(t)-\chi_0(t)\bigr)+
\sum_{j\in \mathcal T_\om}
c_{\om,j}\psi_{\om,j}(t)e^{ij\theta},
\end{aligned}
\end{equation}
where $\delta_{j0}=1$ for $j=0$, and $\delta_{j0}=0$ for $j\neq 0$.

We estimate these three terms in \eqref{eq:oct16} in order.  First, for
\(j\in\mathcal J_\om\), \eqref{eq:E-size-new1} and
\eqref{eq:rough-angular-window} give
\[
|q_{\om,n_\om+j}|
\leq
\frac{|n_\om-\om|+|j-\alpha_\om|+|\alpha_\om|}
{\sqrt\om}
\leq 2\quad\hbox{for all sufficiently large}\,\ \om,
\]
where \(q_{\om,n_\om+j}\) is defined by \eqref{eq:def-qom}.  Moreover,
\(\rho_\om\geq1/2\) on \([-R,R]\).  Hence
Lemma~\ref{lem:bounded-q-radial-control} gives a uniform
\(H^1(-R,R)\) bound, and the one-dimensional Sobolev inequality yields
\begin{equation}\label{eq:suppsi}
\sup_{j\in\mathcal J_\om}\sup_{|t|\leq R}
|\psi_{\om,j}(t)|\leq C_R,
\end{equation}
By \eqref{eq:unified-normalized-profile},
\eqref{eq:unified-window-mass}, and \eqref{eq:giant-angular-H1},
\[
e^{-i\gamma_\Omega}  \Pi^{\mathcal W_\Omega}g_\Omega
=
\lf\|\Pi^{\mathcal W_\Omega}g_\Omega \ri\|_{L^2(S^1)}
e^{-i\gamma_\Omega}  f_\Omega 
\xrightarrow[\om \to \infty]{H^1(S^1)} h_0.
\]
Using the Fourier representation \eqref{eq:gW-def-unified} of
\(  \Pi^{\mathcal W_\Omega}g_\Omega\), Parseval's identity
for the function and its derivative gives
\begin{equation}\label{eq:giant-central-coefficients}
 \sum_{j\in\mathcal W_\om}(1+j^2)
 \lf|c_{\om,j}-e^{i\gamma_\om}\delta_{j0} \ri|^2
 \longrightarrow0.
\end{equation}
Consequently, the first term is bounded uniformly for \(|t|\leq R\) and
\(\theta\in S^1\) by
\[
C_R
\bigg(\sum_{j\in\mathcal W_\om}(1+j^2)
\lf|c_{\om,j}-e^{i\gamma_\om}\delta_{j0}\ri|^2 \bigg)^{1/2}
\bigg(\sum_{j\in\mathbb Z}(1+j^2)^{-1} \bigg)^{1/2}
=o(1).
\]
For the second term in \eqref{eq:oct16}, Corollary~\ref{lem:uniform-two-mode-HO}, applied at
\(\nu=n_\om\), gives
\(\psi_{\om,0}\to\chi_0\) in \(H^1(-R,R)\), hence uniformly on
\([-R,R]\).  Therefore,
\[
\sup_{|t|\leq R}
\lf|e^{i\gamma_\om}
\big(\psi_{\om,0}(t)-\chi_0(t)\big) \ri|=o(1).
\]
For the third term in \eqref{eq:oct16}, \eqref{eq:gJ-angular-moment-identity} and
\eqref{eq:gJ-angular-moment-bound}, together with the boundedness of
\(\sigma_\Omega\), imply
\begin{equation}\label{eq:giant-middle-l1-tail}
 \sum_{j\in \mathcal T_\om}|c_{\om,j}|
 \leq
 \Big(\sum_{j\in\mathcal J_\om}
 |j-\alpha_\om|^2|c_{\om,j}|^2\Big)^{1/2}
 \Big(\sum_{|j-\alpha_\om|>M_\om}
 |j-\alpha_\om|^{-2}\Big)^{1/2}\\
 \leq C M_\om^{-1/2}=o(1).
\end{equation}
Together with \eqref{eq:suppsi}, this gives
\[
\sum_{j\in \mathcal T_\om}
c_{\om,j}\psi_{\om,j}(t)e^{ij\theta}=o(1)\quad\hbox{uniformly for}\,\ |t|\leq R,\,\ \theta\in S^1.
\]
The three estimates prove \eqref{eq:giant-low-uniform-profile} and complete
the proof.
\end{proof}

\begin{proof}[\normalfont\bfseries Proof of Theorem~\ref{thm:B}]
Let \(u_\om\) be any family of minimizers along the sequence stated in Theorem~\ref{thm:B}.  Combining  Lemmas~\ref{lem:giant-angular-rigidity}
and
\ref{lem:giant-local-profile} proves
\eqref{eq:giant-local-profile}.

We next prove \eqref{eq:giant-momentum-limit}.  The first band
projection \eqref{eq:PJ-def-giant} gives the orthogonal decomposition
\[
\begin{aligned}
\gs(r,\theta)
=P_\Omega^{\mathcal J_\Omega}\gs(r,\theta)+w_\Omega(r,\theta)=\frac1{\sqrt{2\pi}}\sum_{j\in\mathcal J_\Omega}
c_{\Omega,j}\phi_{\Omega,n_\Omega+j}^{(1)}(r)
e^{i(n_\Omega+j)\theta}+w_\Omega(r,\theta).
\end{aligned}
\]
The remainder contains the first radial modes outside
\(\mathcal J_\Omega\) and all higher radial modes.  Orthogonality,
the radial normalization, and \( \lf\|\gs\ri\|_2=1\) give
\[
1= \lf\|P_\Omega^{\mathcal J_\Omega}\gsom \ri\|_2^2+ \lf\|w_\Omega\ri\|_2^2
=\sum_{j\in\mathcal J_\Omega}|c_{\Omega,j}|^2+ \lf\|w_\Omega\ri\|_2^2.
\]
Since \(L_z=-i\partial_\theta\),
\[
L_zP_\Omega^{\mathcal J_\Omega}\gs
=\frac1{\sqrt{2\pi}}\sum_{j\in\mathcal J_\Omega}
(n_\Omega+j)c_{\Omega,j}\phi_{\Omega,n_\Omega+j}^{(1)}(r)
e^{i(n_\Omega+j)\theta}.
\]
Thus the  inner
product of \(L_zP_\Omega^{\mathcal J_\Omega}\gs\) and \(w_\Omega\) vanishes, and
\[
\meanlrlr{\gsom}{L_z}{ \gsom}
=\sum_{j\in\mathcal J_\Omega}(n_\Omega+j)|c_{\Omega,j}|^2
+ \meanlrlr{w_\Omega}{L_z}{ w_\Omega}.
\]
Subtracting \(n_\Omega\) times the mass identity and using
\(\mathcal T_\Omega=\mathcal J_\Omega\setminus\mathcal W_\Omega\)
from \eqref{eq:unified-effective-window}, we obtain
\begin{equation}\label{eq:giant-momentum-decomposition}
\begin{aligned}
 \meanlrlr{\gsom}{L_z}{\gsom} -n_\Omega
&=\sum_{j\in\mathcal J_\Omega}j|c_{\Omega,j}|^2
+ \meanlrlr{w_\Omega}{L_z-n_\Omega}{w_\Omega}\\
&=\sum_{j\in\mathcal W_\Omega}j|c_{\Omega,j}|^2
+\sum_{j\in\mathcal T_\Omega}j|c_{\Omega,j}|^2+ \meanlrlr{w_\Omega}{L_z-n_\Omega}{w_\Omega}.
\end{aligned}
\end{equation}
For the first term in
\eqref{eq:giant-momentum-decomposition}, recall from
\eqref{eq:gW-def-unified} and \eqref{eq:unified-normalized-profile} that
\[
\lf(\Pi^{\mathcal W_\Omega}g_\Omega \ri)(\theta)
=\frac1{\sqrt{2\pi}}\sum_{j\in\mathcal W_\Omega}
c_{\Omega,j}e^{ij\theta},
\qquad
f_\Omega =\frac{\Pi^{\mathcal W_\Omega} g_\Omega}{\lf\|  \Pi^{\mathcal W_\Omega} g_\Omega \ri\|_{L^2(S^1)}}.
\]
By \eqref{eq:giant-angular-H1},
\(e^{-i\gamma_\Omega} f_\Omega \to h_0=(2\pi)^{-1/2}\)
in \(H^1(S^1)\).  Parseval's identity gives
\bml{\label{eq:sep-1}
\bigg| \sum_{j\in\mathcal W_\Omega}j|c_{\Omega,j}|^2 \bigg| \leq\sum_{j\in\mathcal W_\Omega}j^2|c_{\Omega,j}|^2
=\lf\| \lf(\Pi^{\mathcal W_\Omega}g_\Omega\ri)' \ri\|_{L^2(S^1)}^2
\\
= \lf\|  \Pi^{\mathcal W_\Omega}g_\Omega \ri\|_{L^2(S^1)}^2 \lf\|f_\Omega' \ri\|_{L^2(S^1)}^2=o(1).
}
We next consider the second term in
\eqref{eq:giant-momentum-decomposition}.
For \(j\in\mathcal T_\Omega=\mathcal J_\Omega\setminus\mathcal W_\Omega\),
the definitions \eqref{eq:rough-angular-window} and
\eqref{eq:unified-effective-window} give
\[
M_\Omega(1+\sigma_\Omega)<|j-\alpha_\Omega|\leq\sqrt{\Omega}.
\]
Since \(M_\Omega\to\infty\) and \(\alpha_\Omega=O(1)\), the angular
gap \eqref{eq:angular-gap-new} implies, uniformly on \(\mathcal T_\Omega\),
\[
\lambda_{\Omega,n_\Omega+j}^{(1)}-E_\Omega^{\rm int}
\geq c(j-\alpha_\Omega)^2-C\geq c_0j^2
\]
for some \(c_0>0\) and all sufficiently large \(\Omega\).
Since \(\mathcal T_\Omega\subset\mathbb Z\setminus\mathcal W_\Omega\) and all coefficients in \eqref{eq:giant-discarded-energy} are nonnegative, we have
\[
\mathcal R_\Omega(\mathcal W_\Omega)
\geq\sum_{j\in\mathcal T_\Omega}
\bigl(\lambda_{\Omega,n_\Omega+j}^{(1)}-E_\Omega^{\rm int}\bigr)
|c_{\Omega,j}|^2
\geq c_0\sum_{j\in\mathcal T_\Omega}j^2|c_{\Omega,j}|^2.
\]
Combining this inequality with \eqref{eq:giant-discarded-small}, we obtain
\begin{equation}\label{eq:sep-2}
\sum_{j\in\mathcal T_\Omega}j^2|c_{\Omega,j}|^2
\leq c_0^{-1}\mathcal R_\Omega(\mathcal W_\Omega)=o(1).
\end{equation}
For the last term in \eqref{eq:giant-momentum-decomposition}, 
\eqref{eq:giant-outer-form-small} gives
\(
\|w_\Omega\|_2^2=o(\Omega^{-1})\) and
\(\mathcal Q_\Omega(w_\Omega)=o(1)
\).
The potential term in \eqref{eq:linear-form} and the Cauchy--Schwarz
inequality imply
\bmln{
\lf\||x|w_\Omega\ri\|_2^2
\leq\lf\|w_\Omega \ri\|_2^2
+\left(\int_{\mathbb R^2}(|x|^2-1)^2|w_\Omega|^2\,dx\right)^{1/2}
\lf\|w_\Omega \ri\|_2\\
\leq\lf\|w_\Omega \ri\|_2^2
+\frac{\sqrt{8\mathcal Q_\Omega(w_\Omega)}}{\Omega}\lf\|w_\Omega \ri\|_2
=o(\Omega^{-1}).
}
Since \(x^\perp\cdot\nabla=\partial_\theta\),
\begin{equation}\label{eq:giant-momentum-identity}
\meanlrlr{w_\Omega}{L_z}{w_\Omega}
=\Omega \lf\||x|w_\Omega \ri\|_2^2
+\operatorname{Re}\int_{\mathbb R^2}\overline{w_\Omega}
x^\perp\cdot(-i\nabla-\Omega x^\perp)w_\Omega\,dx.
\end{equation}
Consequently,
\[
 \lf| \meanlrlr{w_\Omega}{L_z}{w_\Omega} \ri|
\leq\Omega \lf\||x|w_\Omega \ri\|_2^2
+\lf\||x|w_\Omega \ri\|_2\mathcal Q_\Omega(w_\Omega)^{1/2}=o(1).
\]
Since \(n_\Omega=\OO(\Omega)\) by \eqref{eq:E-size-new1}, we also have
\begin{equation}\label{eq:giant-remainder-momentum}
\meanlrlr{w_\Omega}{L_z-n_\Omega}{w_\Omega}=o(1).
\end{equation}
We conclude from  \eqref{eq:sep-1}, \eqref{eq:sep-2} and
\eqref{eq:giant-remainder-momentum}  that all three terms in 
\eqref{eq:giant-momentum-decomposition} are \(o(1)\).
Finally, \eqref{eq:E-size-new1} gives
\(n_\Omega=\Omega+o(\sqrt{\Omega})\), which proves
\eqref{eq:giant-momentum-limit}.
\end{proof}

\subsection{Intermediate coupling and point concentration}\label{sec:strong-effective-coupling}
In this subsection, we prove Theorem~\ref{thm:C} on the energy
asymptotics and angular concentration of ground states in the regime
\begin{equation}\label{eq:growing-sigma}
	\sigma_\Omega\to\infty,
	\qquad
	\sigma_\Omega=o(\sqrt{\Omega}).
\end{equation}

The minimizers of the circle functional $\mathcal E_{\alpha_\om,\sigma_\om}(h)$ concentrate on the angular scale \(\sigma_\Omega^{-1}\), with limiting rescaled profiles determined by the following variational problem on \(\mathbb R\).
\begin{lemma}
\label{lem:line-compactness}
We have
\begin{equation}\label{eq:line-problem-new}
I_{\mathbb R}
:=
\inf_{ w\in H^1(\mathbb R), \lf\|w\ri\|_{2}=1}
 \lf\{
\frac19\int_{\mathbb R}|w'|^2\,ds
-
\frac{\Gamma}{2}\int_{\mathbb R}|w|^4\,ds
 \ri\}
=
-\frac{3\Gamma^2}{16}.
\end{equation}
Its minimizers are precisely
\[
e^{i\beta}H(\,\cdot-y),
\qquad \beta,y\in\mathbb R,
\]
where \(H\) is defined in \eqref{eq:line-profile-main}.
Moreover, let \(\{w_n\}_{n \in \N} \subset H^1(\mathbb R)\) be a minimizing sequence
for \(I_{\mathbb R}\) with \(\|w_n\|_{L^2(\R)}=1\).
Then, after passing to a subsequence, there exist \(y_n\in\mathbb R\) and
\(\beta_n\in\mathbb R\) such that
\beq
e^{-i\beta_n}w_n(y_n+\cdot) \xrightarrow[n \to \infty]{H^1(\mathbb R)} H.
\eeq
\end{lemma}

\begin{proof}[\normalfont\bfseries Proof]
The sharp one-dimensional Gagliardo--Nirenberg inequality \cite[Proposition~3.1]{Oh2022} gives, for
\(\|w\|_{L^2(\R)}=1\),
\[
\int_{\mathbb R}|w|^4\,ds
\leq\frac1{\sqrt3}\|w'\|_{L^2(\R)},
\]
and hence, with \(x=\|w'\|_{L^2(\R)}\),
\[
\frac19\int_{\mathbb R}|w'|^2\,ds
-\frac{\Gamma}{2}\int_{\mathbb R}|w|^4\,ds
\geq
\frac{x^2}{9}-\frac{\Gamma x}{2\sqrt3}
\geq-\frac{3\Gamma^2}{16}.
\]
The normalized function \(H\) in \eqref{eq:line-profile-main} attains the
last value, which proves \eqref{eq:line-problem-new}.

We next identify all minimizers.  If \(W\) is a minimizer,
\eqref{eq:diamagnetic} shows that \(Q:=|W|\) is also a minimizer.  The corresponding
Euler--Lagrange equation and the strong maximum principle give \(Q>0\) and
\begin{equation}\label{eq:line-minimizer-EL}
-\frac19Q''-\Gamma Q^3=\lambda Q
\quad\hbox{on }\mathbb R
\end{equation}
for some \(\lambda\in\mathbb R\).  Taking the inner product with \(Q\) and
using \(\|Q\|_{L^2(\R)}=1\) gives
\[
\lambda
=I_{\mathbb R}-\frac{\Gamma}{2}\int_{\mathbb R}Q^4\,ds<0.
\]
The classification of the optimizers
\cite[Remark~3.2]{Oh2022}, together with
\(\|Q\|_{L^2(\mathbb R)}=1\), therefore gives
\(Q=H(\,\cdot-y)\) for some \(y\in\mathbb R\).
Substitution into \eqref{eq:line-minimizer-EL} yields
\(\lambda=-9\Gamma^2/16\).

It remains to prove the compactness of minimizing sequences.
The sequence is bounded in \(H^1(\mathbb R)\) by the one-dimensional
Gagliardo--Nirenberg inequality.  The concentration--compactness principle of
Lions~\cite{Lions1984a,Lions1984b} applies.  Vanishing is excluded by
\(I_{\mathbb R}<0\).  If \(I_{\mathbb R}(\mu)\) denotes the infimum under the
constraint \(\|w\|_{L^2(\R)}^2=\mu\), then scaling gives
\(I_{\mathbb R}(\mu)=\mu^3I_{\mathbb R}\); hence strict subadditivity excludes
dichotomy.  Compactness modulo translations follows, and convergence of the
energy upgrades the resulting convergence to strong \(H^1(\mathbb R)\)
convergence.  The preceding classification identifies the limit with \(H\)
up to a constant phase.  
\end{proof}

\begin{proposition}
\label{prop:strong-circle-limit-new}
Assume that \(\sigma_\om\) satisfies
\eqref{eq:growing-sigma} and that
\(\sup_\om|\alpha_\om|<\infty\).  Then
\begin{equation}\label{eq:strong-circle-energy-new}
\inf_{\lf\|h\ri\|_{L^2(S^1)}=1}\mathcal E_{\alpha_\om,\sigma_\om}(h)
=
-\frac{3\Gamma^2}{16}\sigma_\om^2+o(\sigma_\om^2).
\end{equation}
Moreover, suppose that \(\|h_\om\|_{L^2(S^1)}=1\) and
\(\mathcal E_{\alpha_\om,\sigma_\om}(h_\om)=
\inf_{\|h\|_{L^2(S^1)}=1}\mathcal E_{\alpha_\om,\sigma_\om}(h)
+o(\sigma_\om^2)\).
Up to a subsequence, there are \(\theta_\om\to\theta_*\in S^1\)
and \(\beta_\om\in\mathbb R\) such that
\begin{equation}\label{eq:strong-circle-profile-new}
e^{-i\beta_\om}\sigma_\om^{-1/2}h_\om
 \lf(\theta_\om+\tfrac{s}{\sigma_\om} \ri)
\xrightarrow[\om \to \infty]{L^2(\mathbb R)} H(s).
\end{equation}
The rescaled function is read on
\(s\in(-\pi\sigma_\om,\pi\sigma_\om]\) and set equal to zero outside that
interval for the stated \(L^2(\mathbb R)\) convergence.  
\end{proposition}

\begin{proof}[\normalfont\bfseries Proof]
We first prove \eqref{eq:strong-circle-energy-new}.
Fix \(\theta_0\in S^1\) and choose a real-valued
\(\chi\in C_c^\infty((-\pi,\pi))\) with \(\chi\equiv1\) on $(-1/2,1/2)$.
Define the \(2\pi\)-periodic trial function by
\[
h_\Omega^{\rm tr}(\theta_0+\theta)
=c_\Omega\sigma_\Omega^{1/2}\chi(\theta)
H(\sigma_\Omega\theta),
\qquad -\pi<\theta\leq\pi,
\]
where \(c_\Omega>0\) is chosen so that
\(\|h_\Omega^{\rm tr}\|_{L^2(S^1)}=1\). 
The decay of \(H\)
and \(\alpha_\Omega=O(1)\) give
\begin{equation}\label{eq:strong-circle-upper-proof}
e_{\alpha_\om,\sigma_\om}
\leq \sigma_\om^2 I_{\mathbb R}+o(\sigma_\om^2).
\end{equation}

For \(L>0\), let \(\mathbb T_L:=\mathbb R/(L\mathbb Z)\) denote the
circle of length \(L\).  We identify functions on \(\mathbb T_L\) with
\(L\)-periodic functions on \(\mathbb R\).  In particular,
\[
H^1(\mathbb T_L):=\Big\{v\in H^1((-L/2,L/2);\, \mathbb C):\quad v(-L/2)=v(L/2)\Big\}.
\]  
Integrals and norms on \(\mathbb T_L\) are taken over one period.
The circle distance is
\[
d_L(s,b):=\min_{k\in\mathbb Z}|s-b+kL|,
\qquad s,b\in\mathbb R.
\]
Set \(L_\Omega:=2\pi\sigma_\Omega\).  For \(h\in H^1(S^1)\) with
\(\|h\|_{L^2(S^1)}=1\), use its \(2\pi\)-periodic angular
representative to define
\begin{equation}\label{eq:rescale-h}
v(s)=\sigma_\Omega^{-1/2}h(s/\sigma_\Omega),
\qquad s\in\mathbb R.
\end{equation}
Then \(v\in H^1_{\mathrm{loc}}(\mathbb R)\) is \(L_\Omega\)-periodic,
\(\|v\|_{L^2(\mathbb T_{L_\Omega})}=1\), and
\begin{equation}\label{eq:circle-rescaled-energy}
\begin{aligned}
\sigma_\om^{-2}\mathcal E_{\alpha_\om,\sigma_\om}(h)
={}&\frac19\int_{\mathbb T_{L_\om}}
\left|\left(-i\partial_s-\frac{\alpha_\om}{\sigma_\om}\right)v\right|^2\,ds
-\frac{\Gamma}{2}\int_{\mathbb T_{L_\om}}
|v|^4\,ds
-\frac{\alpha_\om^2}{9\sigma_\om^2}.
\end{aligned}
\end{equation}
After rescaling the circle inequalities in
\cite[Lemma~4.1(a)]{Lebowitz1988} and
\cite[Lemma~3.3(i)]{Oh2022}, for every \(\delta>0\) there exists
\(C_\delta>0\) such that
\begin{equation}\label{eq:expanding-circle-GN-proof}
\int_{\mathbb T_L}|v|^4\,ds
\leq \left(\tfrac1{\sqrt3}+\delta\right)
\lf\|v'\ri\|_{L^2(\mathbb T_L)}\lf\|v\ri\|_{L^2(\mathbb T_L)}^3
+\frac{C_\delta}{L}\lf\|v\ri\|_{L^2(\mathbb T_L)}^4,
\qquad L\geq1.
\end{equation}
Choose \(\delta=\delta(L_\om)\searrow0\) sufficiently slowly that
\(C_{\delta(L_\om)}/L_\om\to0\).  Since
\(\alpha_\Omega/\sigma_\Omega=o(1)\), the one-dimensional periodic form of
\eqref{eq:diamagnetic},
\eqref{eq:circle-rescaled-energy}, and
\eqref{eq:expanding-circle-GN-proof}, followed by completion of the square,
give, uniformly over normalized \(h\),
\[
\sigma_\om^{-2}\mathcal E_{\alpha_\om,\sigma_\om}(h)
\geq
\inf_{x\geq0} \lf(\frac{x^2}{9}
-\frac{\Gamma x}{2\sqrt3} \ri)-o(1)
=I_{\mathbb R}-o(1).
\]
Together with \eqref{eq:strong-circle-upper-proof}, this proves
\eqref{eq:strong-circle-energy-new}.

Let \(h_\om\) satisfy \(\|h_\om\|_{L^2(S^1)}=1\) and
\(\mathcal E_{\alpha_\om,\sigma_\om}(h_\om)=
\inf_{\|h\|_{L^2(S^1)}=1}\mathcal E_{\alpha_\om,\sigma_\om}(h)
+o(\sigma_\om^2)\), and let \(v_\om\) be its rescaling given by
\eqref{eq:rescale-h}.  The same   estimate, together with the
almost minimality of \(h_\om\) and \eqref{eq:strong-circle-energy-new}, gives
\[
\lf\| \lf(-i\partial_s-\frac{\alpha_\om}{\sigma_\om} \ri)
v_\om \ri\|_{L^2(\mathbb T_{L_\om})}=\OO(1),
\qquad \lf\| v_\om' \ri\|_{L^2(\mathbb T_{L_\om})}=\OO(1).
\]
Define
\[
\mathcal J_L(v):=\frac19\int_{\mathbb T_L}|v'|^2\,ds
-\frac{\Gamma}{2}\int_{\mathbb T_L}|v|^4\,ds.
\]
Since \(\alpha_\om/\sigma_\om=o(1)\), the magnetic kinetic term differs
from \(\|v_\om'\|_{L^2(\mathbb T_{L_\om})}^2\) by \(o(1)\).  Hence
\eqref{eq:circle-rescaled-energy} becomes
\begin{equation}\label{eq:strong-circle-defect-proof}
\sigma_\Omega^{-2}\mathcal E_{\alpha_\Omega,\sigma_\Omega}(h_\Omega)
=\mathcal J_{L_\Omega}(v_\Omega)+o(1).
\end{equation}
The almost minimality of \(h_\Omega\),
\eqref{eq:strong-circle-defect-proof}, and
\eqref{eq:strong-circle-energy-new} imply
\begin{equation}\label{eq:circle-minimizing-defect}
\mathcal J_{L_\Omega}(v_\Omega) \xrightarrow[\om \to 0]{} I_{\mathbb R}.
\end{equation}

The rescaled functions \(v_\Omega\) belong to
\(H^1(\mathbb T_{L_\Omega})\), whereas
Lemma~\ref{lem:line-compactness} applies to normalized minimizing
sequences in \(H^1(\mathbb R)\).
We now cut off \(v_\Omega\) on an arc of
\(\mathbb T_{L_\Omega}\) where the integrals of
\(|v_\Omega|^2\), \(|v_\Omega'|^2\), and \(|v_\Omega|^4\)
are \(o(1)\), and extend the truncated functions by zero to
\(\mathbb R\).
After normalization in \(L^2(\mathbb R)\), we obtain a minimizing
sequence for \eqref{eq:line-problem-new}.  
The preceding bounds and
\eqref{eq:expanding-circle-GN-proof} give
\[
\int_{\mathbb T_{L_\om}}
\bigl(|v_\om|^2+|v_\om'|^2+|v_\om|^4\bigr)\,ds\leq C.
\]
Choose \(R_\Omega\to\infty\) with \(R_\Omega=o(L_\Omega)\), so that
\(4R_\Omega<L_\Omega\) for all sufficiently large \(\Omega\).
Averaging over \(b\in\mathbb T_{L_\Omega}\) gives
\[
\begin{aligned}
&\frac1{L_\om}\int_{\mathbb T_{L_\om}}
\int_{\{d_{L_\om}(s,b)<2R_\om\}}
\bigl(|v_\om|^2+|v_\om'|^2+|v_\om|^4\bigr)\,ds\,db
\\
=&\frac{4R_\om}{L_\om}
\int_{\mathbb T_{L_\om}}
\bigl(|v_\om|^2+|v_\om'|^2+|v_\om|^4\bigr)\,ds
\geq
\inf_{b\in\mathbb T_{L_\om}}
\int_{\{d_{L_\om}(s,b)<2R_\om\}}
\bigl(|v_\om|^2+|v_\om'|^2+|v_\om|^4\bigr)\,ds.
\end{aligned}
\]
The first expression is \( \OO(R_\Omega/L_\Omega)=o(1)\).  Hence we may
choose a real representative \(b_\Omega\in(-L_\Omega/2,L_\Omega/2]\) for which the
last integral satisfies
\[
\int_{\{d_{L_\om}(s,b_{\om})<2R_\om\}}
\bigl(|v_\om|^2+|v_\om'|^2+|v_\om|^4\bigr)\,ds=O(R_\Omega/L_\Omega)=o(1).
\]
Set \(\tilde\theta_\Omega:=b_\Omega/\sigma_\Omega-\pi\).
We retain the notation \(v_\Omega\) for the translated profile:
\[
v_\Omega(s)
=\sigma_\Omega^{-1/2}
h_\Omega \lf(\tilde\theta_\Omega+\tfrac{s}{\sigma_\Omega} \ri),
\qquad s\in(-\pi\sigma_\Omega,\pi\sigma_\Omega].
\]
Then we have
\begin{equation}\label{eq:strong-circle-small-cut-arc}
\int_{\{L_\Omega/2-2R_\Omega<|s|\leq L_\Omega/2\}}
\bigl(|v_\Omega|^2+|v_\Omega'|^2+|v_\Omega|^4\bigr)\,ds=o(1).
\end{equation}
Choose \(\chi_\om\in C_c^\infty(\mathbb R)\) such that
\[
0\leq\chi_\om\leq1,
\qquad
\chi_\om=
\begin{cases}
	1	&	\hbox{if }|s|\leq L_\om/2-2R_\om,	\\
	0	&	\hbox{if }|s|\geq L_\om/2-R_\om,
\end{cases}
\qquad
|\chi_\om'|\leq \frac{C}{R_\om}.
\]
Equation \eqref{eq:strong-circle-small-cut-arc} gives
\[
\begin{aligned}
0\leq 1- \lf\|\chi_\Omega v_\Omega\ri\|_{L^2(\mathbb R)}^2
&=\int_{-L_\Omega/2}^{L_\Omega/2}
(1-\chi_\Omega^2)|v_\Omega|^2\,ds=o(1).
\end{aligned}
\]
Thus the cutoff retains mass \(1-o(1)\), and
\( \lf\|\chi_\Omega v_\Omega \ri\|_{L^2(\mathbb R)}>0\) for all sufficiently
large \(\Omega\).  Define
\[
\widetilde v_\Omega
:=\frac{\chi_\Omega v_\Omega}{\lf\|\chi_\Omega v_\Omega \ri\|_{L^2(\mathbb R)}}.
\]
Then \(\widetilde v_\om\in H^1(\mathbb R)\).  Equation
\eqref{eq:strong-circle-small-cut-arc}, the bound on \(\chi_\om'\), and the
limit in \eqref{eq:circle-minimizing-defect} give
\[
\frac19\int_{\mathbb R}|\widetilde v_\om'|^2\,ds
-\frac{\Gamma}{2}\int_{\mathbb R}|\widetilde v_\om|^4\,ds
=I_{\mathbb R}+o(1).
\]
Thus \(\widetilde v_\om\) is a minimizing sequence for
\eqref{eq:line-problem-new}.  Moreover,
\begin{equation}\label{eq:oct18}
\big\|v_\om-\widetilde v_\om \big\|_{L^2((-L_\om/2,L_\om/2])}
=o(1).
\end{equation}
Lemma~\ref{lem:line-compactness} gives, up to a subsequence,
\(y_\Omega,\beta_\Omega\in\mathbb R\) such that
\[
e^{-i\beta_\Omega}\widetilde v_\Omega(y_\Omega+\cdot)
\xrightarrow[\om \to \infty]{H^1(\mathbb R)} H.
\]
Since \(\operatorname{supp}\widetilde v_\Omega
\subset[-L_\Omega/2+R_\Omega,L_\Omega/2-R_\Omega]\) and
\( \lf\| H \ri\|_2 >0\), this convergence implies
\(L_\Omega/2-|y_\Omega|\geq R_\Omega-1\to\infty\).
Thus for every fixed $A>0$,
\begin{equation}\label{eq:oct19}
[y_\om-A,y_\om+A]\subset(-L_\Omega/2,L_\Omega/2)\quad\hbox{for sufficiently large}\,\ \om.
\end{equation}
For every fixed \(A>0\), \eqref{eq:oct19} gives
\bml{\label{eq:oct20}
	\lf\|e^{-i\beta_\Omega}v_\Omega(y_\Omega+\cdot)-H
 \ri\|_{L^2(-A,A)}
\\
\leq
\lf\|v_\Omega-\widetilde v_\Omega \ri\|_{L^2(-L_\Omega/2,L_\Omega/2)}
+
\lf\|e^{-i\beta_\Omega}\widetilde v_\Omega(y_\Omega+\cdot)-H
\ri\|_{L^2(\mathbb R)}
 \xrightarrow[\om \to \infty]{} 0.
}
Set
\[
\theta_\Omega
:=\tilde\theta_\Omega+\frac{y_\Omega}{\sigma_\Omega}
\pmod{2\pi}\in S^1.
\]
Restrict \(s\mapsto e^{-i\beta_\Omega}v_\Omega(y_\Omega+s)\)
to
\(
(-L_\Omega/2,L_\Omega/2]
=(-\pi\sigma_\Omega,\pi\sigma_\Omega]
\)
and extend it by zero outside this interval.
Thus, combining \eqref{eq:oct20} with the fact that \(\|H\|_{L^2(\mathbb R)}=\|v_\om\|_{L^2(\mathbb T_{L_\Omega)}}=1\), the convergence is strong:
\begin{equation}\label{eq:circle-torus-compactness}
	e^{-i\beta_\Omega}v_\Omega(y_\Omega+s)
	=e^{-i\beta_\Omega}\sigma_\Omega^{-1/2}h_\Omega
	\lf(\theta_\Omega+\tfrac{s}{\sigma_\Omega}\ri)
	\xrightarrow[\om \to \infty]{L^2(\mathbb R)} H(s).
\end{equation}
After passing to a further subsequence, assume that
\(\theta_\Omega\to\theta_*\in S^1\). Together with
\eqref{eq:circle-torus-compactness}, this proves
\eqref{eq:strong-circle-profile-new} and completes the proof.
\end{proof}

\begin{proof}[\normalfont\bfseries Proof of Theorem~\ref{thm:C}]
We first prove the energy expansion \eqref{eq:strong-main-energy}.
Theorem~\ref{thm:A} gives
\[
\gseom
=E_\om^{\rm int}+e_{\alpha_\om,\sigma_\om}
+o(\sigma_\om^2),
\]
while
Proposition~\ref{prop:strong-circle-limit-new} gives
\[
e_{\alpha_\om,\sigma_\om}
=-\tfrac{3\Gamma^2}{16}\sigma_\om^2+o(\sigma_\om^2).
\]
Combining these identities proves \eqref{eq:strong-main-energy}.

We next prove the profile convergence \eqref{eq:strong-rescaled-profile}.
Since \(\gsom\) is a minimizer, Theorem~\ref{thm:A} gives
a normalized function \( f_\Omega \in H^1(S^1)\) such that
\begin{equation}\label{eq:uniform2}
	\mathcal E_{\alpha_\Omega,\sigma_\Omega}(f_\Omega)
	=e_{\alpha_\om,\sigma_\om}+o(1+\sigma_\om^2)
\end{equation}
and
\begin{equation}\label{eq:uniform1}
	\Big\|u_\om
	-e^{in_\Omega\theta}\phi_{\Omega,\tau_\om}^{(1)}  f_\Omega 
	\Big\|_2=o(1).
\end{equation}
Since \(\sigma_\om\to\infty\), \eqref{eq:uniform2} shows that
\(f_\Omega \) satisfies the assumption \eqref{eq:strong-circle-energy-new}. Therefore, Proposition~\ref{prop:strong-circle-limit-new} gives
centers \(\theta_\om\to\theta_*\in S^1\) and phases
\(\gamma_\om\) such that
\begin{equation}\label{eq:Aug-4}
e^{-i\gamma_\Omega}\sigma_\Omega^{-1/2} f_\Omega
\lf(\theta_\om+\tfrac{s}{\sigma_\om} \ri)
\xrightarrow[\om \to \infty]{L^2(\mathbb R)} H(s).
\end{equation}
Here the function on the left is read on
\((-\pi\sigma_\om,\pi\sigma_\om]\) and extended by zero outside that
interval.

For \(t\in I_\om=(-\sqrt\om,\infty)\) and
\(s\in(-\pi\sigma_\om,\pi\sigma_\om]\), define
\[
R_\om(t):=
\tfrac{1}{\om^{1/4}} \lf(1+\tfrac{t}{\sqrt\om} \ri)^{1/2}
\phi_{\om,\tau_\om}^{(1)}
\lf(1+\tfrac{t}{\sqrt\om} \ri),
\qquad
A_\om(s):=
e^{-i\gamma_\Omega}\sigma_\Omega^{-1/2} f_\Omega
\lf( \theta_\om+\tfrac{s}{\sigma_\om} \ri).
\]
The angular argument is understood modulo \(2\pi\), so that \(s=0\)
corresponds to \(\theta_\om\).  We extend \(R_\om\) and \(A_\om\)
by zero outside the displayed intervals.  The radial and angular changes of
variables give
\(\|R_\om\|_{L^2(\mathbb R)}=\|A_\om\|_{L^2(\mathbb R)}=1\).
Moreover, under the change of variables and phase correction in
\eqref{eq:strong-rescaled-definition}, the rescaling of
\(e^{in_\Omega\theta}\phi_{\Omega,\tau_\om}^{(1)} f_\Omega \) is exactly
\(R_\om(t)A_\om(s)\).

We now establish convergence of the two factors.  The  angular
convergence \eqref{eq:Aug-4} gives \(A_\om\to H\) strongly in \(L^2(\mathbb R)\).  Since
\(R_\om=\rho_\om^{1/2}
U_\om\phi_{\om,\tau_\om}^{(1)}\) on \(I_\om\),
\eqref{eq:rescaled-eigenpairs} gives
\(\psi_{\om,\tau_\om}^{(1)}
=U_\om\phi_{\om,\tau_\om}^{(1)}\).  Since
\(|\tau_\om-\tau_\om|=0\leq L_\om\),
\eqref{eq:low-mode-eigenfunction-conv} gives
\[
\lf\|U_\om\phi_{\om,\tau_\om}^{(1)}-\chi_0 \ri\|_{\domain_\om}
\xrightarrow[\om \to \infty]{} 0.
\]
Together with \eqref{eq:radial-unitary-map}, this yields
\[
 \lf\|R_\om-\chi_0\ri\|_{L^2(\mathbb R)}
 \leq
\lf\|U_\om\phi_{\om,\tau_\om}^{(1)}-\chi_0 \ri\|_{\domain_\om}
+ \lf\|(\rho_\om^{1/2}-1)\chi_0 \ri\|_{L^2(I_\om)}
+ \lf\| \chi_0 \ri\|_{L^2(\mathbb R\setminus I_\om)}
\xrightarrow[\om \to \infty]{} 0,
\]
where the last two terms tend to zero by the Gaussian decay of \(\chi_0\).
Consequently,
\[
\|R_\om A_\om-\chi_0H\|_{L^2(\mathbb R^2)}
\leq
\|R_\om-\chi_0\|_{L^2(\mathbb R)}\|A_\om\|_{L^2(\mathbb R)}
+\|\chi_0\|_{L^2(\mathbb R)}\|A_\om-H\|_{L^2(\mathbb R)}
=o(1).
\]
A direct change of variables in \eqref{eq:strong-rescaled-definition} and
the preceding product identity give
\[
\lf\|V_\om-R_\om A_\om \ri\|_{L^2(\mathbb R^2)}
= \lf\|u_\om-e^{in_\om\theta}
\phi_{\Omega,\tau_\om}^{(1)} f_\Omega \ri\|_{L^2(\mathbb R^2)}
=o(1),
\]
where the last equality follows from \eqref{eq:uniform1}.  Hence
\[
\lf\|V_\om-\chi_0H\ri\|_{L^2(\mathbb R^2)}
\leq
\lf\|V_\om-R_\om A_\om \ri\|_{L^2(\mathbb R^2)}
+ \lf\|R_\om A_\om-\chi_0H \ri\|_{L^2(\mathbb R^2)}
=o(1),
\]
which proves \eqref{eq:strong-rescaled-profile}. This completes the proof of Theorem~\ref{thm:C}.
\end{proof}

\section{Strong Interaction}\label{subsec:endpoint}
In this section, we prove Theorem~\ref{thm:ep-concentration} for fixed \(a\in(0,a^*)\). We establish the energy asymptotics \eqref{eq:endpoint-main-energy} and the rescaled profile convergence \eqref{eq:endpoint-rescaled-profile}. The proof combines the variational properties of the limiting energy \(\mathfrak e_b\), defined in \eqref{eq:endpoint-lge}, with sharp upper and lower bounds for \( \gseom \) as \(\Omega\to\infty\).

Recall that 
\begin{equation}\label{eq:endpoint-regime}
\energy(u)
=\mathcal Q_\om(u)-\frac a2\int_{\mathbb R^2}|u|^4,
\qquad \|u\|_2=1,
\qquad 0<a<a^*\ \hbox{fixed}.
\end{equation}
Lemma~\ref{lem:standard-spectral-input} gives
\begin{equation}\label{eq:eldef}
\lim_{\om\to\infty}\frac{E_\om^{\rm int}}{\om}=e_0=\tfrac{3\sqrt2}{2},
\end{equation}
where \(E_\om^{\rm int}\) and \(e_0\) are defined in
\eqref{eq:E-int-def} and \eqref{eq:first-two-HO-modes}, respectively.
Since \(0<a<a^*\),
there exist \(\delta>0\) and \(C>0\), independent of \(\om\), such that for all sufficiently large \(\Omega\) and every
\(u\in\domain\) with \(\|u\|_2\leq1\),
\begin{equation}\label{eq:endpoint-stability-GN}
\frac a2\int_{\mathbb R^2}|u|^4
\le
(1-\delta)
\Big(\mathcal Q_\om(u)-E_\om^{\rm int}\|u\|_2^2\Big)
+
C\om\|u\|_2^2.
\end{equation}
Consequently, if
\begin{equation}\label{eq:endpoint-energy-ub}
	\energy(u_\om)\le E_\om^{\rm int}+C\om,
\end{equation}
then
\begin{equation}\label{eq:endpoint-coercivity}
	\frac{1}{\om}
	\Big(
	\mathcal Q_\om(u_\om)-E_\om^{\rm int}
	\Big)
	+
	\frac1\om\int_{\mathbb R^2}|u_\om|^4\,dx
	\le C .
\end{equation}
In particular, the mass is radially confined to the annulus \(|r-1|=O(\om^{-1/2})\):
for every fixed \(T>1\),
\begin{equation}\label{eq:endpoint-radial-mass-tail}
	\int_{\{\sqrt\om|r-1|>T\}}|u_\om|^2\,dx
	\le \frac{C}{T^2},
\end{equation}
and
\begin{equation}\label{eq:endpoint-radial-L4-tail}
	\lim_{T\to\infty}\limsup_{\om\to\infty}
	\frac1\om
	\int_{\{\sqrt\om|r-1|>T\}}|u_\om|^4\,dx
	=0 .
\end{equation}

\subsection{Effective minimization problem and its properties}
In this subsection, we mainly establish an upper bound of $\enom$ as $\om\to\infty$ and prove some qualitative properties of the associated limiting minimization problem \(\mathfrak e_b\) given by \eqref{eq:endpoint-lge}. 

Recall \(\mathfrak q\), \(\mathscr D(\mathfrak q)\) from
\eqref{eq:endpoint-local-form} and \eqref{eq:endpoint-form-domain}.  We equip the form
domain with the norm
\begin{equation}\label{eq:endpoint-form-domain-norm}
\lf\| v \ri\|_{\mathscr D(\mathfrak q)}^2
:=\int_{\mathbb R^2}\Big(|\partial_t v|^2
+|(-i\partial_s-2t)v|^2
+(1+t^2)|v|^2\Big)\,dt\,ds .
\end{equation}

Estimate~\eqref{eq:endpoint-radial-mass-tail} shows that the mass of minimizers concentrates radially near the unit circle on the scale \(\Omega^{-1/2}\). The following lemma provides the linear energy expansion for trial states localized radially on this scale, yielding an upper bound for \( \gse \).
\begin{lem}\label{lem:5.1}
Let \(J_\om=(\theta_\om,\theta_\om+L_\om)\pmod {2\pi}\), where
\(\theta_\om\in[0,2\pi)\) and \(0<L_\om<2\pi\), and set
\begin{equation}\label{eq:Aom}
A_{\om,J}:=\Big\{(r,\theta):
|r-1|<\tfrac{\log\om}{\sqrt{\om}},\ \theta\in J_\om\Big\}.
\end{equation}
Assume that \(u_\om\in C_c^\infty(A_{\om,J})\) satisfies
\( \lf\|u_\om \ri\|_2=1\) and \eqref{eq:endpoint-coercivity}.
For \(\widetilde\theta_\om\in J_\om\), define
\[
v_\om(t,s)
:=
\tfrac{1}{\om^{1/2}}
\lf(1+\tfrac{t}{\sqrt\om} \ri)^{1/2}
e^{-i n_\om(\widetilde\theta_\om+\tfrac{s}{\sqrt\om})}
u_\om \lf(
1+\tfrac{t}{\sqrt\om},\,
\widetilde\theta_\om+\tfrac{s}{\sqrt\om}
\ri),
\]
where \(n_\om\) is defined in \eqref{eq:E-int-def}.  Then, the zero extension
of \(v_\om\) belongs to \(\mathscr D(\mathfrak q)\),
\( \lf\|v_\om \ri\|_2=1\), and
\begin{equation}\label{limtf}
	\frac{\mathcal Q_\om(u_\om)-E_\om^{\rm int}}{\om}
	=
	\mathfrak q(v_\om)+o(1)
	\qquad\hbox{as}\,\ \om\to\infty,
\end{equation}
uniformly for \(J_\om\) and \(u_\om\).
\end{lem}
\begin{proof}[\normalfont\bfseries Proof]
Set \(t=\sqrt\om(r-1)\) and
\(s=\sqrt\om(\theta-\widetilde\theta_\om)\).  Then
\(\om\,d\theta\,dr=dt\,ds\), and the change of variables gives
\[
\int_{\mathbb R^2}|v_\om(t,s)|^2\,dt\,ds
=\int_0^\infty\int_0^{2\pi}|u_\om(r,\theta)|^2r\,d\theta\,dr=1.
\]
The compact support assumption also shows that the zero extension belongs to
\(\mathscr D(\mathfrak q)\).  Recall from \eqref{eq:linear-fpr} that
\begin{equation*}
\mathcal Q_\om(u)
=\int_0^\infty\int_0^{2\pi}
\Big\{|\partial_r u|^2
+\Big|\Big(\frac{-i\partial_\theta}{r}
-\om r\Big)u\Big|^2
+\frac{\om^2}{8}(r^2-1)^2|u|^2
\Big\}r\,d\theta\,dr .
\end{equation*}
Direct computations yield that
\[
\partial_r u_\om
=
\om^{1/2}r^{-1/2}e^{in_\om\theta}
\Big(
\sqrt\om\,\partial_t v_\om
-
\frac{1}{2r}v_\om
\Big),
\]
hence
\begin{equation}\label{eq:5.1}
\frac1\om
\int_0^\infty\int_0^{2\pi}|\partial_r u_\om|^2\,r\,dr\,d\theta
=\int_{\R^2}\Big|\partial_t v_\om-
\frac{1}{2r\sqrt\om}v_\om\Big|^2\,dt\,ds  \\
=\int_{\R^2}|\partial_t v_\om|^2\,dt\,ds+o(1),
\end{equation}
where we used \eqref{eq:endpoint-coercivity} in the last equality.
For the magnetic angular part,
\[
\Big(
\frac{-i\partial_\theta}{r}
-\om r
\Big)u_\om
=
\om^{1/2}r^{-1/2}e^{in_\om\theta}
\Big[
\Big(
\frac{n_\om-i\sqrt\om\,\partial_s}{r}
-\om r
\Big)v_\om
\Big].
\]
Therefore
\[
\begin{aligned}
\frac1\om\int_0^\infty\int_0^{2\pi}\Big|\Big(\frac{-i\partial_\theta}{r}
-\om r\Big)u_\om\Big|^2r\,dr\,d\theta 
=\int_{\R^2}\Big|\frac1{\sqrt\om}
\Big(\frac{n_\om-i\sqrt\om\,\partial_s}{r}
-\om r\Big)v_\om\Big|^2\,dt\,ds.
\end{aligned}
\]
By \eqref{eq:E-size-new1},
\((n_\om-\om)/\sqrt\om\to0\), while
\(\operatorname{supp}v_\om\subset\{|t|<\log\om\}\).  Therefore
\[
\frac1{\sqrt\om}
\Big(
\frac{n_\om-i\sqrt\om\,\partial_s}{r}
-\om r
\Big)
=
-i\partial_s-2t+o(1),
\]
and hence
\begin{equation}\label{eq:5.2}
\frac1\om\int_0^\infty\int_0^{2\pi}\Big|\Big(
\frac{-i\partial_\theta}{r}-\om r\Big)u_\om\Big|^2
r\,dr\,d\theta
=\int_{\R^2}|(-i\partial_s-2t)v_\om|^2\,dt\,ds+o(1).
\end{equation}
Finally, for the trapping potential,
since
\[
r^2-1
=
\lf(1+\tfrac t{\sqrt\om} \ri)^2-1
=
\tfrac{2t}{\sqrt\om}
+
\tfrac{t^2}{\om},
\]
we have
\[
\tfrac{1}{8}\om(r^2-1)^2=\tfrac12t^2+o(1)
\]
uniformly on the support of \(v_\om\).  Therefore
\begin{equation}\label{eq:5.3}
\frac1\om
\int_0^\infty\int_0^{2\pi}\frac{\om^2}{8}(r^2-1)^2|u_\om|^2
r\,dr\,d\theta
=\frac12\int_{\R^2}t^2|v_\om|^2\,dt\,ds+o(1).
\end{equation}
Combining \eqref{eq:5.1}--\eqref{eq:5.3} with \eqref{eq:eldef} proves
\eqref{limtf}.
\end{proof}

We next establish the properties of \(\mathfrak e_b\) needed for the energy lower bound and the concentration of minimizers.
\begin{lemma}\label{lem:endpoint-gbp}
For \(\mathfrak e_b\) defined in \eqref{eq:endpoint-lge}, the following
assertions hold for \(0\leq b<a^*\).
\begin{enumerate}
\item[(i)] $
\mathfrak e_0=0 $, and  $ -\infty<\mathfrak e_b<0 $ for any $ 0<b<a^* $;
\item[(ii)] \(b\mapsto\mathfrak e_b\) is continuous on
\([0,a^*)\), and, for every
\(0\le \mu\le1\), \(0\le b<a^*\),
\begin{equation}\label{eq:endpoint-concavity-useful}
	\mathfrak e_{b\mu}\ge \mu\mathfrak e_b;
\end{equation}
\item[(iii)] for every \(0<b<a^*\), \(\mathfrak e_b\) is attained in
\(\mathscr D(\mathfrak q)\).
\end{enumerate}
\end{lemma}

\begin{proof}[\normalfont\bfseries Proof]
(i). By \eqref{eq:diamagnetic} and the Gagliardo--Nirenberg inequality \eqref{eq:critical-mass}, the
functional
\[
v\mapsto
\mathfrak q(v)-\frac b2\int_{\mathbb R^2}|v|^4
\]
is bounded from below on \(\{v:\|v\|_2=1\}\) for every \(b<a^*\). Hence
\(\mathfrak e_b>-\infty\) for $0<b<a^*$.

We first prove \(\mathfrak e_0=0\). By Lemma~\ref{lem:5.1}, for every
normalized \(v\in C_c^\infty(\mathbb R^2)\), the corresponding trial state
\(u_\om\) satisfies
\[
\frac{\mathcal Q_\om(u_\om)-E_\om^{\rm int}}{\om}
=
\mathfrak q(v)+o_\om(1)
\quad\hbox{as}\,\ \om\to\infty.
\]
Since \(E_\om^{\rm int}\) is the bottom of the linear energy,
\[
\mathcal Q_\om(u_\om)-E_\om^{\rm int}\ge0.
\]
Letting \(\om\to\infty\), we obtain
\[
\mathfrak q(v)\ge0
\qquad
\hbox{for every normalized }v\in C_c^\infty(\mathbb R^2).
\]
By density and lower semicontinuity, \(\mathfrak q\ge0\) on its form domain.
Therefore
\[
\mathfrak e_0=\inf_{\|v\|_2=1}\mathfrak q(v)\ge0.
\]

It remains to prove the opposite inequality.  By the definitions of \(\beta\) and \(\chi_0\)
in \eqref{eq:limiting-HO}, respectively, we obtain
\begin{equation}\label{eq:oct21}
\Big(-\partial_t^2+\tfrac92 t^2\Big)\chi_0=\beta\chi_0=\tfrac{3\sqrt{2}}{2}\chi_0.
\end{equation}
Choose a real-valued function \(\phi\in C_c^\infty(\mathbb R)\) with
\(\|\phi\|_{L^2(\R)}=1\), and define, for \(L>1\),
\[
v_L(t,s):=L^{-1/2}\chi_0(t)\phi(s/L).
\]
Then \(\|v_L\|_2=1\). Since \(v_L\) is real-valued, the cross term in
\(|(-i\partial_s-2t)v_L|^2\) vanishes. \eqref{eq:oct21} gives
\[
\begin{aligned}
\mathfrak q(v_L)
&=
\int_{\mathbb R}
\lf(
|\chi_0'|^2+\tfrac92t^2|\chi_0|^2
-\tfrac{3\sqrt2}{2}|\chi_0|^2
\ri)\,dt
+
L^{-2}\int_{\mathbb R}|\phi'(y)|^2\,dy=
L^{-2}\int_{\mathbb R}|\phi'(y)|^2\,dy.
\end{aligned}
\]
Thus \(\mathfrak e_0\le \mathfrak q(v_L)\to0\) as \(L\to\infty\). Hence
$\mathfrak e_0=0$.

For \(b>0\), the same trial functions give
\[
\int_{\mathbb R^2}|v_L|^4\,dt\,ds
=
L^{-1}
\Big(\int_{\mathbb R}|\chi_0|^4\,dt\Big)
\Big(\int_{\mathbb R}|\phi|^4\,dy\Big)
=:C_0L^{-1},
\qquad C_0>0.
\]
Therefore
\[
\mathfrak q(v_L)-\frac b2\int_{\mathbb R^2}|v_L|^4
=
L^{-2}\int_{\mathbb R}|\phi'|^2\,dy
-
\frac b2 C_0L^{-1}.
\]
For \(L\) large enough, the right-hand side is negative. Hence
\[
\mathfrak e_b<0
\qquad
\hbox{for every }0<b<a^*.
\]
This proves (i).

(ii). If \(0\leq b_1<b_2<a^*\), the variational definition directly gives
\(\mathfrak e_{b_2}\leq\mathfrak e_{b_1}\).  For concavity, note that
for each fixed normalized \(v\), the energy $\mathfrak q(v)-\frac b2\int_{\mathbb R^2}|v|^4dtds$ depends linearly on the parameter $b$. Hence, taking
the infimum over \(v\), we get
\begin{equation}\label{eq:concave}
\mathfrak e_{\theta b_1+(1-\theta)b_2}
\ge
\theta\mathfrak e_{b_1}+(1-\theta)\mathfrak e_{b_2},
\qquad 0\le\theta\le1.
\end{equation}
Therefore \(b\mapsto\mathfrak e_b\) is concave. 
Since \(b\mapsto\mathfrak e_b\) is finite and concave,  for every compact interval \(K\subset(0,a^*)\), there exists \(C_K>0\) such that
\[
|\mathfrak e_{b_1}-\mathfrak e_{b_2}|
\leq C_K|b_1-b_2|,
\qquad b_1,b_2\in K.
\]
Thus $\mathfrak e_{b}$ is continuous on $(0,a^*)$. To verify continuity of $\mathfrak e_{b}$ at \(b=0\),   concavity and (i) give, for \(b_0\in(0,a^*)\),
\[
\frac{b}{b_0}\mathfrak e_{b_0}
\leq\mathfrak e_b\leq0,
\qquad 0\leq b\leq b_0.
\]
Thus \(\mathfrak e_b\to0=\mathfrak e_0\) as \(b\searrow0\), proving continuity on \([0,a^*)\).

By concavity and \(\mathfrak e_0=0\),
\[
\mathfrak e_{b\mu}
=
\mathfrak e_{\mu b+(1-\mu)0}
\ge
\mu\mathfrak e_b+(1-\mu)\mathfrak e_0
=
\mu\mathfrak e_b ,
\qquad 0\le\mu\le1.
\]
This proves  \eqref{eq:endpoint-concavity-useful}.

(iii). Let \(0<b<a^*\) and \(\{v_n\}_{n \in \N} \) be a
minimizing sequence for \(\mathfrak e_b\), with \(\|v_n\|_2=1\). Since \(b<a^*\), we can use \eqref{eq:diamagnetic} and the
Gagliardo--Nirenberg inequality \eqref{eq:critical-mass} to obtain that \(\{v_n\}\) is bounded
in \(\mathscr D(\mathfrak q)\) uniformly in \(n\).  Since the only translation invariance
is in the variable \(s\), the concentration-compactness principle gives, up to
translations in \(s\), a weak limit \(v\) in the form domain:
\[
v_n \xrightarrow[n \to \infty]{\mathrm{w}-D(\mathfrak q)} v.
\]
Vanishing cannot occur. Indeed, if vanishing occurred, then
\(\int_{\mathbb R^2}|v_n|^4\to0\) as \(n\to\infty\).
Since \(\mathfrak q\ge0\), this would imply
\[
\mathfrak e_b
=
\lim_{n\to\infty}
\lf(
\mathfrak q(v_n)-\frac b2\int_{\mathbb R^2}|v_n|^4
\ri)
\ge0,
\]
contradicting \(\mathfrak e_b<0\).

Set \(\mu:=\|v\|_2^2\).
Then \(\mu>0\). We claim that \(\mu=1\). Suppose instead that \(0<\mu<1\).
By the Brezis--Lieb decomposition and the lower semicontinuity of
\(\mathfrak q\), using \eqref{eq:concave} for \(v\) and for
the remainder \(v_n-v\), we obtain
\[
\mathfrak e_b
\ge
\mu\,\mathfrak e_{b\mu}
+
(1-\mu)\mathfrak e_{b(1-\mu)} .
\]
Using \eqref{eq:endpoint-concavity-useful}, this yields
\[
\mathfrak e_b
\ge
\bigl(\mu^2+(1-\mu)^2\bigr)\mathfrak e_b .
\]
Since \(0<\mu<1\), one has
\(\mu^2+(1-\mu)^2<1\).
Together with \(\mathfrak e_b<0\), this is impossible. Hence \(\mu=1\).

Thus  \(v_n\to v\) strongly in \(L^2(\mathbb R^2)\) as $n\to\infty$.
Using the Gagliardo--Nirenberg inequality \eqref{eq:critical-mass} again, we get
\[
\int_{\mathbb R^2}|v_n|^4 \xrightarrow[n \to \infty]{} \int_{\mathbb R^2}|v|^4.
\]
By lower semicontinuity of \(\mathfrak q\),
\[
\mathfrak q(v)-\frac b2\int_{\mathbb R^2}|v|^4
\le
\liminf_{n\to\infty}
\lf(
\mathfrak q(v_n)-\frac b2\int_{\mathbb R^2}|v_n|^4
\ri)
=
\mathfrak e_b .
\]
Since \(\|v\|_2=1\), the reverse inequality follows from the definition of
\(\mathfrak e_b\). Hence \(v\) is a minimizer.  
This proves (iii).
\end{proof}

Applying Lemma~\ref{lem:5.1} to localized trial states and using Lemma~\ref{lem:endpoint-gbp} (i), we obtain the following upper bound for \(\gse \).
\begin{lemma}\label{lem:endpoint-upper-bound}
Let \(0<a<a^*\).  Then,
\begin{equation}\label{eq:endpoint-upper-bound}
\limsup_{\om\to\infty}
\frac{ \gse -E_\om^{\rm int}}{\om}
\le
\mathfrak e_a.
\end{equation}
\end{lemma}

\begin{proof}[\normalfont\bfseries Proof]
Choose \(\theta_0\in S^1\), and take
\(v\in C_c^\infty(\mathbb R^2)\) with \( \lf\|v\ri\|_2=1\).  For
\((r,\theta)\in[0,\infty)\times[0,2\pi)\), define
\[
u^{\rm tr}_\om(r,\theta)
=
\om^{1/2}r^{-1/2}e^{in_\om\theta}
v\bigl(\sqrt\om(r-1),\sqrt\om(\theta-\theta_0)\bigr).
\]
For all sufficiently large \(\om\), this is a normalized trial state
supported in an angular coordinate chart about \(\theta_0\).  Lemma~\ref{lem:5.1}
then gives
\[
\frac{\mathcal Q_\om(u^{\rm tr}_\om)-E_\om^{\rm int}}{\om}
=
\mathfrak q(v)+o(1).
\]
Moreover, the same change of variables and the compact support of \(v\) give
\[
\frac1\om\int_{\mathbb R^2}|u^{\rm tr}_\om|^4\,dx
=
\int_{\mathbb R^2}|v|^4\,dt\,ds+o(1).
\]
Therefore
\[
\limsup_{\om\to\infty}
\frac{ \gse-E_\om^{\rm int}}{\om}
\le \limsup_{\om\to\infty}
\frac{\energy(u^{\rm tr}_\om)-E_\om^{\rm int}}{\om}
=
\mathfrak q(v)-\frac a2\int_{\mathbb R^2}|v|^4.
\]
By Lemma~\ref{lem:endpoint-gbp} (i), $\mathfrak e_a$ is finite for $a\in(0,a^*)$.   Taking the infimum over normalized \(v\in C_c^\infty(\mathbb R^2)\) and using the density of \(C_c^\infty(\mathbb R^2)\) in the form domain of \(\mathfrak q\), we obtain \eqref{eq:endpoint-upper-bound}.
\end{proof}

\subsection{Lower bound in the strong interaction regime}
In this subsection, we derive a lower bound for $ \gse$ as $\om\to\infty$. We begin with the following energy estimate.
\begin{lemma}\label{lem:llbp}
Let \(J_\om\) and \(A_{\om,J}\) be as in Lemma~\ref{lem:5.1}, and
let \(w_\om\in\mathscr D(\mathcal Q_\om)\) have compact support in
\(A_{\om,J}\).  Set
\[
0<\mu_\om:=\|w_\Omega\|_2^2\leq1.
\]
Assume that
\begin{equation}\label{eq:oct22}
\frac{\mathcal Q_\om(w_\om)-E_\om^{\rm int}\mu_\om}
{\om\mu_\om}
+\frac1{\om\mu_\om^2}
\int_{\mathbb R^2}|w_\om|^4\,dx
\leq C.
\end{equation}
Then
\begin{equation}\label{eq:endpoint-piece-lb}
\begin{aligned}
\frac1\om\lf[\mathcal Q_\om(w_\om)
-E_\om^{\rm int}\mu_\om
-\frac a2\int_{\R^2}|w_\om|^4\,dx \ri]
\ge\mu_\om \lf[\mathfrak e_{a\mu_\om}
-o(1) \ri],
\end{aligned}
\end{equation}
uniformly for \(J_\om\).
\end{lemma}
\begin{proof}[\normalfont\bfseries Proof]
By approximation in the form norm of \(\mathcal Q_\Omega\) and in \(L^4(\mathbb R^2)\), it suffices to consider \(w_\Omega\in C_c^\infty(A_{\Omega,J})\).
Choose
\(\hat\theta_\om\in J_\om\), and set
\[
t=\sqrt\om(r-1),
\qquad
s=\sqrt\om(\theta-\hat\theta_\om).
\]
Define
\[
v_\om(t,s)
:=
\tfrac{1}{\mu_\om^{1/2}\om^{1/2}}
 \lf(1+\tfrac t{\sqrt\om} \ri)^{1/2}
e^{-i n_\om(\hat\theta_\om+\tfrac{s}{\sqrt\om})}
w_\om \lf(
1+\tfrac t{\sqrt\om},
\hat\theta_\om+\tfrac s{\sqrt\om}
\ri).
\]
Note that $v_\Omega$ satisfies $\|v_\Omega\|_2=1$ and the assumption \eqref{eq:endpoint-coercivity} in Lemma~\ref{lem:5.1}. Since \(\mathfrak q(v_\Omega)\geq0\), we derive from the expansion \eqref{limtf} in Lemma~\ref{lem:5.1}  
\begin{equation}\label{eq:linear-log-comparison}
\frac{\mathcal Q_\om(w_\om)
-E_\om^{\rm int}\mu_\om
}{\om}
\ge\mu_\om
\Big[(1-\varepsilon_\om)\mathfrak q(v_\om)
-\varepsilon_\om
\Big].
\end{equation}
Here \(\varepsilon_\om\to0\) uniformly for \(J_\om\) and \(w_\om\).

We next prove \eqref{eq:nonlinear-log-comparison}.  Using the definition of \(v_\om\), and the fact that
\[
 \lf(1+\tfrac t{\sqrt\om} \ri)^{-1}
=1+o(1)\quad\hbox{on}\,\ |t|\leq\log\om,
\]
we obtain
\begin{equation}\label{eq:nonlinear-log-comparison}
	\frac1{\om\mu_\om^2}
	\int_{\mathbb R^2}|w_\om|^4\,dx
	\le
	\Big(1+\varepsilon_\om\Big)
	\int_{\mathbb R^2}|v_\om|^4\,dt\,ds.
\end{equation}

Combining \eqref{eq:linear-log-comparison} and
\eqref{eq:nonlinear-log-comparison}, we get
\bmln{
\frac1\om
\lf[
\mathcal Q_\om(w_\om)
-E_\om^{\rm int}\mu_\om
-\frac a2\int_{\mathbb R^2}|w_\om|^4\,dx
\ri]\\
 \ge\mu_\om
\lf[
(1-\varepsilon_\om)\mathfrak q(v_\om)
-\frac{a\mu_\om(1+\varepsilon_\om)}2
\int_{\mathbb R^2}|v_\om|^4\,dt\,ds
-\varepsilon_\om
\ri].
}
Since \(\|v_\om\|_2=1\), we may factor out \(1-\varepsilon_\om\) and use
the definition of \(\mathfrak e_b\):
\bmln{
 (1-\varepsilon_\om)\mathfrak q(v_\om)
-\frac{a\mu_\om(1+\varepsilon_\om)}2
\int_{\mathbb R^2}|v_\om|^4\,dt\,ds\\
 =(1-\varepsilon_\om)
\lf[
\mathfrak q(v_\om)
-\frac{a\mu_\om(1+\varepsilon_\om)}
{2(1-\varepsilon_\om)}
\int_{\mathbb R^2}|v_\om|^4\,dt\,ds
\ri]
 \ge(1-\varepsilon_\om)
\mathfrak e_{
\frac{a\mu_\om(1+\varepsilon_\om)}
{1-\varepsilon_\om}
}.
}
Since \(0<\mu_\om\leq1\) and
\(a<a^*\), the parameter
above remains in a compact subinterval of \([0,a^*)\) for \(\om\) large.
By the continuity of \(b\mapsto\mathfrak e_b\),
\[
(1-\varepsilon_\om)
\mathfrak e_{
	\frac{a\mu_\om(1+\varepsilon_\om)}
	{1-\varepsilon_\om}
}
\ge
\mathfrak e_{a\mu_\om}
-
o(1).
\]
Therefore
\[
\frac1\om
\lf[
\mathcal Q_\om(w_\om)
-E_\om^{\rm int}\mu_\om
-\frac a2\int_{\mathbb R^2}|w_\om|^4\,dx
\ri]
\ge\mu_\om
\lf[
\mathfrak e_{a\mu_\om}
-o(1)
\ri].
\]
This proves \eqref{eq:endpoint-piece-lb}.
\end{proof}

Lemma~\ref{lem:llbp} gives a lower bound for functions supported in a single angular interval. To apply it to a general state, we decompose the state into angularly localized pieces. Since this decomposition may introduce a loss in the attractive nonlinear term, we use a translated partition for which the loss can be made small by a suitable choice of the translation.

Fix \(N\ge3\) and \(\eta>0\) small enough, and set
\(\delta_N=2\pi/N\). For \(\tau\in[0,\delta_N)\) and
\(j=1,\cdots,N\), let
\begin{equation}\label{eq:shifted-angular-intervals}
I_{j,N}^\tau:=\bigl(\tau+(j-1-\eta)\delta_N,\,
\tau+(j+\eta)\delta_N\bigr)\,\ 
\mod\,2\pi .
\end{equation}
Choose \(\chi_1,\cdots,\chi_N\in C^\infty(S^1)\) with
\(0\leq\chi_j\leq1\) such that
\[
\chi_j(\theta)=
\begin{cases}
1,&\theta\in\bigl[(j-1+\eta)\delta_N,\,(j-\eta)\delta_N\bigr],\\
0,&\theta\notin I_{j,N}^0.
\end{cases}
\]
All angular intervals are understood modulo \(2\pi\).  Define
\(\chi_j^\tau(\theta):=\chi_j(\theta-\tau)\).
The functions can be chosen so that,
uniformly for \(\tau\in[0,\delta_N)\),
\begin{equation}\label{eq:shifted-angular-partition}
	\operatorname{supp}\chi_j^\tau\Subset I_{j,N}^\tau,
	\qquad
	\sum_{j=1}^N|\chi_j^\tau(\theta)|^2=1,
	\qquad
	|\partial_\theta\chi_j^\tau(\theta)|
	\le \frac{CN}{\eta}
\end{equation}
for every \(j=1,\cdots,N\) and \(\theta\in S^1\). Here $A\Subset B$ means that $\overline{A}$ is a compact subset of $B$.

\begin{lemma}\label{lem:fsa-partition}
Let \(N\geq3\), \(\delta_N=2\pi/N\), and let \(\eta>0\) be sufficiently
small.  For the partitions defined in
\eqref{eq:shifted-angular-intervals}, \eqref{eq:shifted-angular-partition} and
every \(u\in L^4(\mathbb R^2)\), there exists
\(\tau=\tau_u\in[0,\delta_N)\) such that
\begin{equation}\label{eq:qgs}
	\int_{\mathbb R^2}|u|^4\,dx
	\le
	\sum_{j=1}^N\int_{\mathbb R^2}|\chi_j^\tau u|^4\,dx
	+
	C\eta\int_{\mathbb R^2}|u|^4\,dx .
\end{equation}
\end{lemma}
\begin{proof}[\normalfont\bfseries Proof]
For \(\tau\in[0,\delta_N)\), set
\[
\Theta_\tau
:=
\lf\{
\theta\in S^1:\quad
\sum_{j=1}^N|\chi_j^\tau(\theta)|^4<1
\ri\}.
\]
Since the family \(\chi_j^\tau\) is obtained by translating a fixed
partition, the transition region satisfies
\[
\frac1{\delta_N}
\int_0^{\delta_N}
\mathds 1_{\Theta_\tau}(\theta)\,d\tau
\le C\eta
\qquad
\hbox{for every }\theta\in S^1 .
\]

For \(u\in L^4(\mathbb R^2)\), define
\[
F_u(\theta)
:=
\int_0^\infty |u(r,\theta)|^4r\,dr .
\]
Then
\(\int_{S^1}F_u(\theta)\,d\theta
=\int_{\mathbb R^2}|u|^4\,dx\).  By Fubini's theorem,
\[
	\frac1{\delta_N}
	\int_0^{\delta_N}
	\int_{\Theta_\tau}F_u(\theta)\,d\theta\,d\tau
	=
	\int_{S^1}
	F_u(\theta)
	\Big[
	\frac1{\delta_N}
	\int_0^{\delta_N}
	\mathds 1_{\Theta_\tau}(\theta)\,d\tau
	\Big]d\theta
	\le
	C\eta
	\int_{S^1}F_u(\theta)\,d\theta .
\]
Hence there exists \(\tau=\tau_u\in[0,\delta_N)\) such that
\[
\int_{\Theta_{\tau_u}}F_u(\theta)\,d\theta
\le
C\eta\int_{\mathbb R^2}|u|^4\,dx .
\]
Since
\(0\leq1-\sum_{j=1}^N|\chi_j^{\tau_u}(\theta)|^4
\leq\mathds 1_{\Theta_{\tau_u}}(\theta)\),
we get
\bmln{
 \int_{\mathbb R^2}|u|^4\,dx
-\sum_{j=1}^N\int_{\mathbb R^2}|\chi_j^{\tau_u} u|^4\,dx
=\int_{\mathbb R^2}
\Big(1-\sum_{j=1}^N|\chi_j^{\tau_u}(\theta)|^4\Big)|u|^4\,dx
\\
 \le
\int_{\Theta_{\tau_u}}
\int_0^\infty |u(r,\theta)|^4r\,dr\,d\theta
=
\int_{\Theta_{\tau_u}}F_u(\theta)\,d\theta
\le C\eta\int_{\mathbb R^2}|u|^4\,dx.
}
This proves \eqref{eq:qgs}.
\end{proof}

Fix \(\zeta,\xi\in C^\infty(\mathbb R)\) such that
\begin{equation}\label{eq:oct25}
\xi_1^2+\xi_2^2=1,\qquad0\leq\xi_j\leq1,\qquad\operatorname{supp} \xi_1 \subset(-1,1)\qquad \xi_1=1\quad\hbox{on}\quad[-1/2,1/2].
\end{equation}
Set
\[
\xi_{1,\om}(r):=\xi_1 \lf(\tfrac{\sqrt\om(r-1)}{\log\om} \ri),
\qquad
\xi_{2,\om}(r):=\xi_2 \lf(\tfrac{\sqrt\om(r-1)}{\log\om} \ri).
\]

\begin{lem}\label{lem:endpoint-fplb}
Fix \(a\in(0,a^*)\). 
Let \(N_\om\geq3\) be integer-valued, set
\(\delta_\om:=2\pi/N_\om\).
Then, for all sufficiently large $\om$ and every $0<\eta<1/4$, there exists \(\tau_{u_\om}\in[0,\delta_\om)\)  such that,  with
\[
\chi_{\om,j}(\theta):=\chi_j^{\tau_{u_\om}}(\theta),
\qquad
\mu_{\Omega,j}:= \lf\|\xi_{1,\Omega}(r)\chi_{\Omega,j}(\theta)\gs(r,\theta) \ri\|_2^2,
\qquad j=1,\ldots,N_\om,
\]
we have
\[
\sum_{j=1}^{N_\om}\mu_{\om,j}=1+o(1)
\]
and
\begin{equation}\label{eq:endpoint-fplb-uniform}
\frac{\energy(u_\om)-E_\om^{\rm int}}{\om}
\geq
\sum_{j=1}^{N_\om}
\mu_{\om,j}\mathfrak e_{a\mu_{\om,j}}
-C\eta-C\frac{N_\om^2}{\eta^2\om}-o(1).
\end{equation}
Here $C>0$ is independent of $\Omega$, $N_\om$ and $\eta$, and \(o(1)\to0\) as $\om\to\infty$ uniformly for $N_\om\geq 3$ and $0<\eta<1/4$.
\end{lem}

\begin{proof}[\normalfont\bfseries Proof]
Set \(v_\om:=\xi_{1,\om} u_\om\).  By  
\eqref{eq:endpoint-radial-mass-tail} and 
\eqref{eq:endpoint-radial-L4-tail}, we have

\[
\|v_\om\|_2^2=1+o_\om(1),\qquad
\frac1\om\int_{\mathbb R^2}\bigl(|u_\om|^4-|v_\om|^4\bigr)\,dx
=o_\om(1).
\]
Moreover, \eqref{eq:endpoint-coercivity} gives \( \lf\|\gsom \ri\|^4 \leq C\Omega\).

By Lemma~\ref{lem:fsa-partition},  applied to \(v_\om\) with $\delta_\om=2\pi/N_\om$, $0<\eta<1/4$, there exists
\(\tau_{u_\om}\in[0,\delta_\om)\) such that, with
\(\chi_{\om,j}=\chi_j^{\tau_{u_\om}}\),
\[
\int_{\mathbb R^2}|v_\om|^4\,dx
\le
\sum_{j=1}^{N_\om}\int_{\mathbb R^2}|\chi_{\om,j}v_\om|^4\,dx
+
C\eta\int_{\mathbb R^2}|v_\om|^4\,dx .
\]
Define
\(w_{\om,j}:=\chi_{\om,j}v_\om\).  Since
\(\sum_j|\chi_{\om,j}|^2=1\),
\[
\sum_{j=1}^{N_\om}\mu_{\om,j}=\|v_\om\|_2^2=1+o_\om(1).
\]

For the linear part, set \(\psi_0:=\xi_{2,\om}\) and
\(\psi_j:=\xi_{1,\om}\chi_{\om,j}\),
\(j=1,\cdots,N_\om\). Then
\(\psi_0^2+\sum_{j=1}^{N_\om}|\psi_j|^2=1\).  The magnetic IMS formula
\cite[Theorem~3.2]{Cycon1987} gives
\[
\mathcal Q_\om(u_\om)
=
\mathcal Q_\om(\xi_{2,\om} u_\om)
+
\sum_{j=1}^{N_\om}\mathcal Q_\om(w_{\om,j})
-
\int_{\mathbb R^2}
\Big(
|\nabla\xi_{2,\om}|^2+
\sum_{j=1}^{N_\om}|\nabla(\xi_{1,\om}\chi_{\om,j})|^2
\Big)|u_\om|^2\,dx.
\]
Since
$|\partial_r \xi_{1,\om}|+|\partial_r\xi_{2,\om}|
\le C\sqrt\om/\log\om$ and $|\partial_\theta\chi_{\om,j}|\leq CN_\om/\eta$, and  at most two terms in $\sum_{j=1}^{N_\om}|\nabla(\xi_{1,\om} \chi_{\om,j})|^2
$ are nonzero for each fixed \(\theta\), 
we conclude that
\[
\int_{\mathbb R^2}
\Big(
|\nabla\xi_{2,\om}|^2+
\sum_{j=1}^{N_\om}|\nabla(\xi_{1,\om} \chi_{\om,j})|^2
\Big)|u_\om|^2\,dx \leq C\frac{\om}{(\log\om)^2}+C\frac{N_\om^2}{\eta^2}.
\]
Since \(E_\om^{\rm int}\) is the bottom of the linear quadratic form,
\[
\mathcal Q_\om(\xi_{2,\om} u_\om)
-
E_\om^{\rm int}\|\xi_{2,\om} u_\om\|_2^2\ge0.
\]
Using \(\|\xi_{2,\om} u_\om\|_2^2+\sum_{j=1}^{N_\om}\mu_{\om,j}=1\), we obtain
\begin{equation}\label{eq:endpoint-linear-IMS-final}
	\mathcal Q_\om(u_\om)-E_\om^{\rm int}
	\ge
	\sum_{j=1}^{N_\om}
	\Big(
	\mathcal Q_\om(w_{\om,j})
	-
	E_\om^{\rm int}\mu_{\om,j}
	\Big)
	-C\frac{\om}{(\log\om)^2}
	-C\frac{N_\om^2}{\eta^2}.
\end{equation}

For the nonlinear part, the choice of \(\tau_{u_\om}\)  
and \(\int |u_\om|^4\le C\om\) imply
\begin{equation}\label{eq:endpoint-partition-ub}
\int_{\mathbb R^2}|u_\om|^4\,dx
\le\sum_{j=1}^{N_\om}\int_{\mathbb R^2}|w_{\om,j}|^4\,dx+C\eta\om+
o(\om).
\end{equation}
Combining \eqref{eq:endpoint-linear-IMS-final} and
\eqref{eq:endpoint-partition-ub}, we get
\bmln{
\enom(u_\om)-E_\om^{\rm int}
 \ge\sum_{j=1}^{N_\om}
\lf[\mathcal Q_\om(w_{\om,j})
-E_\om^{\rm int}\mu_{\om,j}
-\frac a2\int_{\mathbb R^2}|w_{\om,j}|^4\,dx
\ri]\\
 \qquad-C\eta\om-C\frac{\om}{(\log\om)^2}
-C\frac{N_\om^2}{\eta^2}-o(\om).
}

We next establish the local lower bound \eqref{eq:endpoint-piece-lb} for each piece \(w_{\Omega,j}\), with an error uniform in \(j\), and then sum these estimates to obtain \eqref{eq:endpoint-fplb-uniform}. The global energy bound \eqref{eq:endpoint-coercivity} does not ensure the uniform bound on the linear energy  for each \(w_{\Omega,j}\) required by \eqref{eq:oct22} in Lemma~\ref{lem:llbp}. We therefore distinguish two cases according to the size of \(\mathcal Q_\Omega(w_{\Omega,j})-E_\Omega^{\rm int}\mu_{\Omega,j}\) relative to \(\Omega\mu_{\Omega,j}\).

Set
\[
A_{\om,j}:=
\mathcal Q_\om(w_{\om,j})
-E_\om^{\rm int}\mu_{\om,j}\geq0.
\]
Fix \(K>C/\delta\), where \(\delta\) is given by
\eqref{eq:endpoint-stability-GN}.  We first consider the case where 
\(A_{\om,j}\leq K\om\mu_{\om,j}\) and
\(\mu_{\om,j}>0\). \eqref{eq:diamagnetic}, the
Gagliardo--Nirenberg inequality \eqref{eq:critical-mass}, and
\eqref{eq:eldef} give
\[
\int_{\mathbb R^2}|w_{\om,j}|^4\,dx
\leq C\mu_{\om,j}\mathcal Q_\om(w_{\om,j})
\leq C\om\mu_{\om,j}^2.
\]
Thus Lemma~\ref{lem:llbp} applies uniformly and yields
\[
\frac1\om
\left[
A_{\om,j}
-\frac a2\int_{\mathbb R^2}|w_{\om,j}|^4\,dx
\right]
\geq
\mu_{\om,j}
\left[
\mathfrak e_{a\mu_{\om,j}}-o(1)
\right].
\]
On the other hand, if \(A_{\om,j}>K\om\mu_{\om,j}\) and $\mu_{\om,j}>0$, combining the stability estimate
\eqref{eq:endpoint-stability-GN} and $\mathfrak e_{a\mu_{\om,j}}<0$ gives
\[
\frac1\om
\left[
A_{\om,j}
-\frac a2\int_{\mathbb R^2}|w_{\om,j}|^4\,dx
\right]
\geq\frac1\om\big[\delta A_{\om,j}-C\om\mu_{\om,j}\big]\geq0\geq \mu_{\om,j}
\left[
\mathfrak e_{a\mu_{\om,j}}-o(1)
\right].
\]
Finally, if \(\mu_{\Omega,j}=0\), then \(w_{\Omega,j}=0\) almost everywhere, and the local lower bound holds with both sides equal to zero.  Since $\sum_{j=1}^{N_\om}\mu_{\Omega,j}\leq1$, the total error is $o(1)$.
Consequently,
\[
\frac{\energy(u_\om)-E_\om^{\rm int}}{\om}
\geq
\sum_{j=1}^{N_\om}
\mu_{\om,j}\mathfrak e_{a\mu_{\om,j}}
-C\eta-C\frac{N_\om^2}{\eta^2\om}-o(1),
\]
which proves \eqref{eq:endpoint-fplb-uniform}.
\end{proof}

\subsection{Proof of Theorem~\ref{thm:ep-concentration}}

\begin{proof}[\normalfont\bfseries Proof of Theorem~\ref{thm:ep-concentration}]
The proof is divided into three steps.

\medskip
\noindent{\em Step 1: Energy asymptotics.}
In this step, we prove \eqref{eq:endpoint-main-energy}. Fix \(R>1\) and let \(\eta>0\) be sufficiently small.  Set
\(N_\om:=\lfloor \sqrt\om/R\rfloor\).  Then \(N_\om\ge3\) for
\(\om\) large, and the corresponding angular mesh satisfies
\[
\delta_\om:=\frac{2\pi}{N_\om}\le C\frac{R}{\sqrt\om}.
\]
Choose \(\tau_{\gs}\) and \(\mu_{\Omega,j}\) as in
Lemma~\ref{lem:endpoint-fplb}.  Since
\(N_\om^2/\om\leq C/R^2\),
\eqref{eq:endpoint-fplb-uniform} gives
\begin{equation}\label{eq:micro-binding-new}
\frac{\energy(u_\om)-E_\om^{\rm int}}{\om}
\geq
\sum_{j=1}^{N_\om}
\mu_{\om,j}\mathfrak e_{a\mu_{\om,j}}
-C\eta-\frac{C}{\eta^2R^2}-
o_\om(1).
\end{equation}
By the concavity of \(b\mapsto\mathfrak e_b\) and \(\mathfrak e_0=0\),
\[
\mathfrak e_{a\mu_{\Omega,j}}
\geq\mu_{\Omega,j}\mathfrak e_a,
\]
we obtain from \eqref{eq:micro-binding-new} that
\begin{equation}\label{eq:endpoint-fplb-coarse}
\frac{\energy(u_\om)-E_\om^{\rm int}}{\om}
\geq
\mathfrak e_a\sum_{j=1}^{N_\om}\mu_{\om,j}^2
-C\eta-\frac{C}{\eta^2R^2}-o_\om(1).
\end{equation}
Following Lemma~\ref{lem:endpoint-fplb}, the constant \(C>0\) in \eqref{eq:endpoint-fplb-coarse} is independent of \(N_\om=\lfloor \sqrt\om/R\rfloor\) and \(\eta\). We choose \(\eta=R^{-1/2}\) in \eqref{eq:endpoint-fplb-coarse}, so that
\[
C\eta+\frac{C}{\eta^2R^2}
=C R^{-1/2}+C R^{-1} \xrightarrow[R \to \infty]{} 0.
\]
Since \(\mathfrak e_a<0\) and \(\sum_j\mu_{\Omega,j}^2\leq1\), letting first \(\Omega\to\infty\) and then \(R\to\infty\) in \eqref{eq:endpoint-fplb-coarse} gives
\[
\liminf_{\Omega\to\infty}
\frac{\gse-E_\Omega^{\rm int}}{\Omega}
\geq\mathfrak e_a.
\]
Together with Lemma~\ref{lem:endpoint-upper-bound}, this proves \eqref{eq:endpoint-main-energy}.

\medskip
\noindent\emph{Step 2:  Construction of a minimizing sequence for \(\mathfrak e_a\).}
For \(\rho>0\) and \(\theta_0\in S^1\), define
\[
\mathcal W_{\om,\rho}(\theta_0)
:=
\Big\{(r,\theta):
\sqrt\om\,|r-1|\leq \rho,\quad
\sqrt\om\,d_{S^1}(\theta,\theta_0)\leq \rho
\Big\}.
\]
We next establish the following estimate
\begin{equation}\label{eq:centered-micro-tightness}
	\lim_{R\to\infty}\liminf_{\om\to\infty}
	\int_{\mathcal W_{\om,R}(\theta_\om)}|u_\om|^2\,dx=1.
\end{equation}

Recall from Lemma~\ref{lem:endpoint-fplb} that
\(\mu_{\Omega,j}=\|\xi_{1,\om} \chi_{\Omega,j}\gs\|_2^2\) for $j=1,\cdots,N_{\om}$ is the mass of the \(j\)-th localized piece. These masses satisfy \(\sum_{j=1}^{N_\om}\mu_{\om,j}=1+o(1)\). Since \(\mathfrak e_a<0\),
\[\sum_{j=1}^{N_\om}\mu_{\om,j}^2\leq
\max_{j=1,\ldots,N_\om}\mu_{\om,j}\sum_{j=1}^{N_\om}\mu_{\om,j},\] 
combining Lemma~\ref{lem:endpoint-upper-bound}
with \eqref{eq:endpoint-fplb-coarse} gives
\begin{equation}\label{eq:Aug-1}
\liminf_{\om\to\infty}
\max_{1\le j\le N_\om}\mu_{\om,j}
\ge
1-C\eta-\frac{C}{\eta^2R^2}.
\end{equation}
Let \(j_\om\) be an index attaining the maximum and let
\(\theta_{\om,R}\) be the center of the support interval of
\(\chi_{\om,j_\om}\).
Setting \(\eta=R^{-1/2}\) in \eqref{eq:Aug-1} gives
\begin{equation}\label{eq:oct23}
\lim_{R\to\infty}\liminf_{\Omega\to\infty}\mu_{\Omega,j_\Omega}=1.
\end{equation}
Furthermore, by \eqref{eq:shifted-angular-intervals} and \eqref{eq:shifted-angular-partition}, the support of \(\chi_{\Omega,j_\Omega}\) is contained in an interval centered at \(\theta_{\Omega,R}\) of length
\[
(1+2\eta)\delta_\Omega
=\frac{2\pi(1+2\eta)}{N_\Omega}
=O(R/\sqrt\Omega).
\]
By \eqref{eq:endpoint-radial-mass-tail},
\[
1\geq
\int_{\mathcal W_{\Omega,CR}(\theta_{\Omega,R})}|\gs|^2\,dx
\geq\mu_{\Omega,j_\Omega}-\frac{C}{R^2},
\]
which, together with \eqref{eq:oct23}, gives
\begin{equation}\label{eq:oct24}
\lim_{R\to\infty}\liminf_{\om\to\infty}
\int_{\mathcal W_{\om,CR}(\theta_{\om,R})}|u_\om|^2\,dx=1.
\end{equation}

We now choose centers \(\theta_\Omega\), independent of \(R\),
satisfying \eqref{eq:centered-micro-tightness}. Choose \(R_0\)  large such that the liminf above as \(\om\to\infty\)  exceeds \(3/4\), and set
\(\theta_\om:=\theta_{\om,R_0}\).  For every sufficiently
large fixed \(R\), both
\(\mathcal W_{\om,CR_0}(\theta_\om)\) and
\(\mathcal W_{\om,CR}(\theta_{\om,R})\) have mass greater than
\(1/2\) for all large \(\om\).  Their angular projections therefore
intersect, and hence
\[
d_{S^1}(\theta_\om,\theta_{\om,R})
\leq \frac{C(R+R_0)}{\sqrt\om}\quad\hbox{for all large}\,\ \om.
\]
Consequently,
\[
\mathcal W_{\om,CR}(\theta_{\om,R})\subset
\mathcal W_{\om,C'(R+R_0)}(\theta_\om).
\]
Together with the preceding concentration estimate \eqref{eq:oct24}, this proves
\eqref{eq:centered-micro-tightness}.  

By compactness of \(S^1\), after
passing to a subsequence we may further assume that
\(\theta_\om\to\theta_*\in S^1\) in \eqref{eq:centered-micro-tightness}.
Let \(W_\om\) be the function in
\eqref{eq:endpoint-rescaling} associated with the centers chosen
above, i.e., 
\begin{equation*}
		W_\om(t,s):= \tfrac{1}{\om^{1/2}}
		\lf(1+\tfrac{t}{\sqrt\om} \ri)^{1/2}
		e^{-i n_\om(\theta_\om+\tfrac{s}{\sqrt\om})} u_\om \lf(1+\tfrac{t}{\sqrt\om},
		\theta_\om+\tfrac{s}{\sqrt\om} \ri).
\end{equation*}
Since \(W_\Omega\) need not vanish on the boundary, its zero extension need not lie in \(\mathscr D(\mathfrak q)\).  We therefore introduce an interior cutoff
and show that the resulting normalized profiles form a minimizing
sequence for \(\mathfrak e_a\).

The change of variables gives
\( \lf\|W_\om \ri\|_2=1\), and
\eqref{eq:centered-micro-tightness} gives
\begin{equation}\label{eq:W-endpoint-tightness}
\lim_{R\to\infty}\liminf_{\om\to\infty}
\int_{[-R,R]^2}|W_\om(t,s)|^2\,dt\,ds=1.
\end{equation}
Set \(R_\Omega:=\sqrt{\log\Omega}\). By
\eqref{eq:W-endpoint-tightness},
\begin{equation}\label{eq:W-diagonal-tail}
	\int_{\mathbb R^2\setminus[-R_\Omega/2,R_\Omega/2]^2}
	|W_\Omega|^2\,dt\,ds\longrightarrow0.
\end{equation}
Let \(\xi_1,\xi_2 \in C^\infty(\R)\) be cutoff functions given by \eqref{eq:oct25}.  Define
\[
\begin{aligned}
\chi_\om^{\rm in}(r,\theta)
:= \xi_1 \lf(
\frac{\om[(r-1)^2+(\theta-\theta_\om)^2]}{4R_\om^2}
\ri),\qquad
\chi_\om^{\rm out}(r,\theta)
:=\xi_2 \lf(
\frac{\om[(r-1)^2+(\theta-\theta_\om)^2]}{4R_\om^2}
\ri),
\end{aligned}
\]
where \(\theta-\theta_\om\) is represented in \((-\pi,\pi]\).  Then
\((\chi_\om^{\rm in})^2+(\chi_\om^{\rm out})^2=1\), and
\eqref{eq:W-diagonal-tail} gives
\[
\lf\|\chi_\om^{\rm out}u_\om \ri\|_2=o(1),
\qquad
\lf\|\chi_\om^{\rm in}u_\om \ri\|_2\longrightarrow1.
\]
The magnetic IMS formula and
\(1-(\chi_\om^{\rm in})^4-(\chi_\om^{\rm out})^4
=2(\chi_\om^{\rm in})^2(\chi_\om^{\rm out})^2\) give
\bml{\label{eq:endpoint-cutoff-splitting}
 \enom(u_\om)-E_\om^{\rm int}\|u_\om\|_2^2
 = \enom(\chi_\om^{\rm in}u_\om)
-E_\om^{\rm int}\|\chi_\om^{\rm in}u_\om\|_2^2
+\energy(\chi_\om^{\rm out}u_\om)
-E_\om^{\rm int}\|\chi_\om^{\rm out}u_\om\|_2^2
\\
-\int_{\mathbb R^2}\left(
|\nabla\chi_\om^{\rm in}|^2
+|\nabla\chi_\om^{\rm out}|^2\right)|u_\om|^2\,dx
-a\int_{\mathbb R^2}
(\chi_\om^{\rm in})^2(\chi_\om^{\rm out})^2
|u_\om|^4\,dx .
}
By \eqref{eq:oct25}, the IMS error satisfies
\begin{equation}\label{eq:oct26}
\int_{\mathbb R^2}\left(
|\nabla\chi_\om^{\rm in}|^2
+|\nabla\chi_\om^{\rm out}|^2\right)|u_\om|^2\,dx=O(\om R_\om^{-2})=o(\om).
\end{equation}   
The product $(\chi_\om^{\rm in})^2(\chi_\om^{\rm out})^2$ is supported where
\begin{equation}\label{eq:oct27}
2R_\om^2\leq\om[(r-1)^2+
(\theta-\theta_\om)^2]\leq4R_\om^2.
\end{equation}
Set
\[
\widetilde \gsom(r,\theta):=\xi_2\!\left(
\frac{\Omega[(r-1)^2+(\theta-\theta_\Omega)^2]}
{2R_\Omega^2}
\right)\gsom(r,\theta).
\]
This function equals \(\gsom\) on the region \eqref{eq:oct27}.
\eqref{eq:W-diagonal-tail} gives
\(\|\widetilde \gsom\|_2^2=o(1)\).
Since the cutoff gradient is bounded by \(C\sqrt\Omega/R_\Omega\),
\eqref{eq:eldef} and \eqref{eq:endpoint-coercivity} imply
\[
\mathcal Q_\Omega(\widetilde \gsom)
\leq 2\mathcal Q_\Omega(\gs)+C\Omega R_\Omega^{-2}
= \OO(\Omega).
\]
Applying \eqref{eq:GN-diamagnetic-new}, we obtain
\begin{equation}\label{eq:oct28}
\begin{aligned}
\frac1\Omega
\int_{\{2R_\Omega^2\leq\Omega[(r-1)^2+
(\theta-\theta_\Omega)^2]\leq4R_\Omega^2\}}
|\gsom|^4\,dx
&\leq\frac1\Omega \lf \|\widetilde \gsom \ri\|_4^4\leq C \lf\|\widetilde \gsom \ri\|_2^2
\frac{\mathcal Q_\Omega(\widetilde \gsom)}{\Omega}
=o(1).
\end{aligned}
\end{equation}
Moreover, \eqref{eq:endpoint-stability-GN} gives
\begin{equation}\label{eq:oct29}
\frac{\energy(\chi_\om^{\rm out}u_\om)
-E_\om^{\rm int}\|\chi_\om^{\rm out}u_\om\|_2^2}{\om}
\geq-C\|\chi_\om^{\rm out}u_\om\|_2^2=o(1).
\end{equation}
Consequently, \eqref{eq:oct26}--\eqref{eq:oct29} and
\eqref{eq:endpoint-main-energy} imply
\begin{equation}\label{eq:Aug-3}
\frac{\energy(\chi_\om^{\rm in}u_\om)
-E_\om^{\rm int}\|\chi_\om^{\rm in}u_\om\|_2^2}{\om}
\leq\mathfrak e_a+o(1).
\end{equation}

Let \(Z_\om\) be the rescaled profile defined by
\eqref{eq:endpoint-rescaling}, with \(u_\om\) replaced by
\(\chi_\om^{\rm in}u_\om\) and with the same center
\(\theta_\om\).  Explicitly,
\begin{equation}\label{eq:W-form-cutoff}
Z_\om(t,s):=
\xi_1 \lf(
\frac{t^2+s^2}{4R_\om^2}
\ri) W_\om(t,s).
\end{equation}
Set \(\widehat Z_\om:=Z_\om/\|Z_\om\|_2\).
Its support is compactly contained in the domain of \(W_\om\), so its
zero extension belongs to \(\mathscr D(\mathfrak q)\).  Moreover,
\eqref{eq:W-diagonal-tail} gives
\begin{equation}\label{eq:W-cutoff-L2-error}
\lf\| Z_\om \ri\|_2\longrightarrow1,
\qquad
\lf\|\widehat Z_\om-W_\om \ri\|_2\longrightarrow0.
\end{equation}
Since \(\|\chi_\om^{\rm in}u_\om\|_2\to1\), the preceding
energy upper bound  \eqref{eq:Aug-3} and \eqref{eq:endpoint-stability-GN} show that
normalization changes   the scaled energy by \(o(1)\).  Since  \(\widehat Z_\om\) satisfies
\eqref{eq:endpoint-coercivity} and the compact support condition in Lemma~\ref{lem:5.1},  
we obtain from Lemma~\ref{lem:5.1} that
\[
\mathfrak q(\widehat Z_\om)
-\frac a2\int_{\mathbb R^2}|\widehat Z_\om|^4\,dt\,ds
\leq\mathfrak e_a+o(1).
\]
The reverse inequality follows from the definition of \(\mathfrak e_a\),
and therefore
\begin{equation}\label{eq:W-cutoff-minimizing}
\mathfrak q(\widehat Z_\om)
-\frac a2\int_{\mathbb R^2}|\widehat Z_\om|^4\,dt\,ds
=\mathfrak e_a+o(1).
\end{equation}

\medskip
\noindent\emph{Step 3: Convergence of the rescaled profiles.}
Following \eqref{eq:W-cutoff-minimizing}, \(\widehat Z_\om\) is a  minimizing sequence for
\(\mathfrak e_a\) in \(\mathscr D(\mathfrak q)\),  and hence \(\|\widehat Z_\Omega\|_{\mathscr D(\mathfrak q)}\leq C\)
uniformly for \(\Omega\).
Passing to a subsequence if necessary, there exists
\(V_a\in\mathscr D(\mathfrak q)\) such that
\[
\widehat Z_\om \xrightarrow[\om \to \infty]{\mathrm{w}-D(\mathfrak q)} V_a.
\]
On every fixed square,  the above minimizing sequence is bounded in \(H^1(\R^2)\)  uniformly for \(\om\).
Thus
\begin{equation}\label{eq:endpoint-cutoff-compactness}
\widehat Z_\om \xrightarrow[\om \to \infty]{L^2_{\rm loc}(\mathbb R^2)} V_a.
\end{equation}
Moreover, \eqref{eq:W-endpoint-tightness} and
\eqref{eq:W-cutoff-L2-error} imply
\[
\lim_{R\to\infty}\liminf_{\om\to\infty}
\int_{[-R,R]^2}|\widehat Z_\om|^2\,dt\,ds=1.
\]
Passing first to the limit in \(\om\) on each fixed square by
\eqref{eq:endpoint-cutoff-compactness}, and then letting
\(R\to\infty\), we obtain \(\|V_a\|_2=1\).  Since
\(\|\widehat Z_\om\|_2=1\), the weak convergence consequently improves to
\begin{equation}\label{eq:endpoint-cutoff-strong-L2}
\widehat Z_\om \xrightarrow[\om \to \infty]{L^2(\mathbb R^2)} V_a.
\end{equation}

The uniform form bound \(\|\widehat Z_\Omega\|_{\mathscr D(\mathfrak q)}\leq C\), \eqref{eq:diamagnetic}, and the
Gagliardo--Nirenberg inequality \eqref{eq:critical-mass} 
then give \(\widehat Z_\om\to V_a\) strongly in \(L^4(\mathbb R^2)\).
By lower semicontinuity of \(\mathfrak q\) and
\eqref{eq:W-cutoff-minimizing},
\[
\mathfrak q(V_a)-\frac a2\int_{\mathbb R^2}|V_a|^4\,dt\,ds
\leq\mathfrak e_a.
\]
The reverse inequality follows from \(\|V_a\|_2=1\) and the definition of
\(\mathfrak e_a\). Hence \(V_a\) is a minimizer.  Finally,
\eqref{eq:W-cutoff-L2-error} and
\eqref{eq:endpoint-cutoff-strong-L2} give
\(W_\om\to V_a\) strongly in \(L^2(\mathbb R^2)\), which proves
\eqref{eq:endpoint-rescaled-profile}.
\end{proof}

\appendix

\section{Uniform Radial Control}
\label{app:bounded-q-radial}

\begin{proof}[\normalfont\bfseries Proof of Lemma~\ref{lem:bounded-q-radial-control}]
We first prove \eqref{eq:bounded-q-low-energy}.
Let \(h_{\om,\nu}\) denote the quadratic form associated with
\(H_{\om,\nu}\) on the Hilbert space \(\domain_\om\) given by  
\eqref{eq:rescaled-radial-operator}.  We use the following form of the min--max principle:
\begin{equation}\label{eq:bounded-q-minmax}
e_{\om,\nu}^{(\ell)}
=
\inf_{X\subset\mathscr D(h_{\om,\nu}), \dim X=\ell}
\ \sup_{0\neq v\in X}
\frac{h_{\om,\nu}[v]}{\|v\|_{\domain_\om}^2}.
\end{equation}
Since \(t>-\sqrt\om\),
\(2+t/\sqrt\om\geq1\), and therefore
\begin{equation}\label{eq:proof-basic-confinement}
h_{\om,\nu}[v]
\geq
\int_{I_\om}
\lf(|v'|^2+\tfrac{1}{8}t^2|v|^2 \ri)\rho_\om\,dt.
\end{equation}
To obtain an upper bound for the \(\ell\)-th eigenvalue, fix \(R>0\)
and an \(\ell\)-dimensional test space
\(X_\ell\subset C_c^\infty((-R,R))\).  Since \(X_\ell\) is
finite-dimensional, equivalence of norms on \(X_\ell\) gives
\begin{equation*}
\|v'\|_{L^2(-R,R)}^2
\leq C_\ell\|v\|_{L^2(-R,R)}^2
\qquad (v\in X_\ell).
\end{equation*}
For all sufficiently large \(\om\), the weight \(\rho_\om\) is
uniformly bounded on \((-R,R)\), while the potential in
\eqref{eq:rescaled-potential-q} is uniformly bounded whenever
\(|q_{\om,\nu}|\leq A\).  Consequently,
\begin{equation*}
\sup_{0\neq v\in X_\ell}
\frac{h_{\om,\nu}[v]}{\|v\|_{\domain_\om}^2}
\leq C_{A,\ell}.
\end{equation*}
Using \(X_\ell\) as a test space in \eqref{eq:bounded-q-minmax} therefore
gives
\(e_{\om,\nu}^{(\ell)}\leq C_{A,\ell}\).
Applying \eqref{eq:proof-basic-confinement} to the normalized eigenfunction
\(\psi_{\om,\nu}^{(\ell)}\) proves \eqref{eq:bounded-q-low-energy}.
Moreover, for every fixed \(m\geq1\), we have
\begin{equation}\label{eq:bounded-q-mass-tightness}
\sup_{|q_{\om,\nu}|\leq A}
\int_{I_\om\cap\{|t|>R\}}
|\psi_{\om,\nu}^{(m)}|^2\rho_\om\,dt
\leq\frac{C_{A,m}}{R^2}
\qquad(R>0).
\end{equation}

We next prove \eqref{eq:bounded-q-eigenlimit} and \eqref{eq:q-zero-low-mode-conv}.  For \(q\in[-A,A]\), define
\[
H_q:=-\partial_t^2+(q-2t)^2+\tfrac12t^2
=-\partial_t^2+\tfrac92 \lf(t-\tfrac{4q}{9} \ri)^2+\tfrac{q^2}{9} 
\]
on its form
domain \(H^1(\mathbb R)\cap L^2(\mathbb R,t^2\,dt)\), and let \(h_q\) be its quadratic form.    By the
harmonic oscillator spectral formula
\cite[Theorem~7.4]{Borthwick2020}, its \(m\)-th eigenvalue is
\begin{equation}\label{eq:pros}
\mu_m(q)=(2m-1)\beta+\tfrac{q^2}{9}\quad\hbox{where}\,\ \beta=\tfrac{3}{\sqrt2}.
\end{equation}

We first prove that, along every sequence such that
\(q_{\om,\nu}\to q\), the eigenvalues of \(H_{\om,\nu}\) converge to those of \(H_q\).  Let
\(\Omega_k\to\infty\) and choose \(\nu_k\) such that
\[
q_k:=q_{\Omega_k,\nu_k}\longrightarrow q\in[-A,A].
\]
Write \(I_k:=I_{\Omega_k}\), \(\rho_k:=\rho_{\Omega_k}\),
\(h_k:=h_{\Omega_k,\nu_k}\), and
\(e_k^{(m)}:=e_{\Omega_k,\nu_k}^{(m)}\).  On every compact interval,
\(\rho_k\to1\) and the potential in \(h_k\) converges uniformly to the
potential in \(h_q\). 
We claim that
\begin{equation}\label{eq:ek-convergence}
e_k^{(\ell)}\longrightarrow\mu_\ell(q).	
\end{equation}

To prove \eqref{eq:ek-convergence}, we first establish the following upper bound
\begin{equation}\label{eq:bounded-q-eigenvalue-limsup}
	\limsup_{k\to\infty}e_k^{(\ell)}
	\leq
	\lim_{R\to\infty}
	\sup_{0\neq v\in Y_R}\frac{h_q[v]}{\|v\|_{L^2(\R)}^2}
	=\mu_\ell(q).
\end{equation}
Let \(\zeta_1,\cdots,\zeta_\ell\) be the first \(\ell\) orthonormal
eigenfunctions of \(H_q\).  Choose \(\eta_R\in C_c^\infty(\mathbb R)\)
such that
\[
\eta_R=1 \quad\hbox{on }[-R,R],\qquad
\operatorname{supp}\eta_R\subset[-2R,2R],\qquad
|\eta_R'|\leq \frac{C}{R},
\]
and define
\[
Y_R:=\operatorname{span}\{\eta_R\zeta_1,\cdots,\eta_R\zeta_\ell\}.
\]
Since the functions \(\zeta_1,\cdots,\zeta_\ell\) decay exponentially,
\(\dim Y_R=\ell\) for all sufficiently large \(R\).
Fix such an \(R\), then  for all sufficiently large \(k\), every function
in \(Y_R\) is supported in \(I_k\).  The uniform convergence of the
coefficients on \([-2R,2R]\) gives
\[
\lim_{k\to\infty}
\sup_{0\neq v\in Y_R}
\frac{h_k[v]}{\|v\|_{\domain_{\Omega_k}}^2}
=
\sup_{0\neq v\in Y_R}\frac{h_q[v]}{\|v\|_{L^2(\R)}^2}.
\]
Using \(Y_R\) as a test space in the min--max formula therefore yields
\[
\limsup_{k\to\infty}e_k^{(\ell)}
\leq
\sup_{0\neq v\in Y_R}\frac{h_q[v]}{\|v\|_{L^2(\R)}^2}.
\]
Finally, exponential decay implies
\(\eta_R\zeta_m\to\zeta_m\) as $R\to\infty$ in the form norm of \(h_q\) for
\(1\leq m\leq\ell\).  Hence \eqref{eq:bounded-q-eigenvalue-limsup} holds true.

We next prove the following matching lower bound
\begin{equation}\label{eq:bounded-q-eigenvalue-liminf}
\mu_\ell(q)
\leq\sup_{0\neq u\in\operatorname{span}\{u_1,\cdots,u_\ell\}}
\frac{h_q[u]}{\|u\|_{L^2(\R)}^2}
\leq\liminf_{k\to\infty}e_k^{(\ell)}.
\end{equation}
First pass to a subsequence,
not relabeled, such that
\(
e_k^{(\ell)}\rightarrow
\liminf_{j\to\infty}e_j^{(\ell)}.
\)
For \(1\leq m\leq\ell\), let
\(\psi_k^{(m)}:=\psi_{\Omega_k,\nu_k}^{(m)}\), with the eigenfunctions
chosen orthonormally in \(\domain_{\Omega_k}\).  Applying \eqref{eq:bounded-q-low-energy} for
\(m=1,\cdots,\ell\) yields uniform bounds in \(H^1(-R,R)\) for every
fixed \(R>0\).
After passing to a further subsequence and using a diagonal argument,
there are
\(u_1,\cdots,u_\ell\in H^1_{\mathrm{loc}}(\mathbb R)\) such that as $k\to\infty$
\[
\psi_k^{(m)} \xrightarrow[k \to \infty]{\mathrm{w}-H^1_{\mathrm{loc}}(\mathbb R)} u_m,
\qquad
\psi_k^{(m)} \xrightarrow[k \to \infty]{L^2_{\mathrm{loc}}(\mathbb R)} u_m.
\]
Since \(\rho_k\to1\) uniformly on compact intervals,
the local convergence and
\eqref{eq:bounded-q-mass-tightness}, by passing first
to the limit \(k\to\infty\) on \([-R,R]\) and then
letting \(R\to\infty\), give
\[
\braketr{u_m}{u_n}_{L^2(\mathbb R)}=\delta_{m,n}.
\]
For \(c=(c_1,\cdots,c_\ell)\in\mathbb C^\ell\), set
\(u_k^c:=\sum_{m=1}^\ell c_m\psi_k^{(m)}\) and
\(u^c:=\sum_{m=1}^\ell c_mu_m\).  
For every fixed \(R>0\), \([-R,R]\subset I_k\) for all
sufficiently large \(k\). Weak lower semicontinuity on
\((-R,R)\), together with the local uniform convergence
of the coefficients and the nonnegativity of the potentials,
gives
\[
\int_{-R}^{R}
\left(
|(u^c)'|^2+
\left[(q-2t)^2+\frac12t^2\right]|u^c|^2
\right)\,dt
\leq\liminf_{k\to\infty}h_k[u_k^c].
\]
Letting \(R\to\infty\) and using monotone convergence, we obtain
\[
h_q[u^c]\leq\liminf_{k\to\infty}h_k[u_k^c].
\]
Since \(u_k^c\) belongs to the span of the first \(\ell\) eigenfunctions,
\[
\|u_k^c\|_{\domain_{\Omega_k}}^2
=\|u^c\|_{L^2(\R)}^2=\sum_{m=1}^\ell|c_m|^2,
\qquad
h_k[u_k^c]
=\sum_{m=1}^\ell e_k^{(m)}|c_m|^2
\leq e_k^{(\ell)}\|u_k^c\|_{\domain_{\Omega_k}}^2,
\]
where the first identity uses the orthonormality of both families.
Consequently, \(u^c\in\mathscr D(h_q)\) and, for every \(c\neq0\),
\[
\frac{h_q[u^c]}{\|u^c\|_{L^2(\R)}^2}
\leq
\liminf_{k\to\infty}
\frac{h_k[u_k^c]}{\|u_k^c\|_{\domain_{\Omega_k}}^2}
\leq
\liminf_{k\to\infty}e_k^{(\ell)}.
\]
The min--max principle for \(H_q\), applied to the \(\ell\)-dimensional
space \(\operatorname{span}\{u_1,\cdots,u_\ell\}\),   gives \eqref{eq:bounded-q-eigenvalue-liminf}.
Combining \eqref{eq:bounded-q-eigenvalue-limsup} and
\eqref{eq:bounded-q-eigenvalue-liminf} proves the claim \eqref{eq:ek-convergence}.

We now upgrade the sequential convergence
\eqref{eq:ek-convergence} to the uniform convergence
\eqref{eq:bounded-q-eigenlimit}.  Suppose that
\eqref{eq:bounded-q-eigenlimit} fails.  Then there are
\(\varepsilon>0\), \(\Omega_k\to\infty\), and \(\nu_k\) with
\(|q_{\Omega_k,\nu_k}|\leq A\) such that
\[
\Big|e_{\Omega_k,\nu_k}^{(\ell)}
-\mu_\ell(q_{\Omega_k,\nu_k})\Big|\geq\varepsilon.
\]
After passing to a subsequence if necessary, we assume that
\(q_{\Omega_k,\nu_k}\to q\in[-A,A]\).  The sequential convergence just
proved gives
\[
e_{\Omega_k,\nu_k}^{(\ell)}\longrightarrow\mu_\ell(q),
\]
whereas the continuity of \(\mu_\ell\) gives
\(\mu_\ell(q_{\Omega_k,\nu_k})\to\mu_\ell(q)\).  Hence
\[
\begin{aligned}
\Big|e_{\Omega_k,\nu_k}^{(\ell)}
-\mu_\ell(q_{\Omega_k,\nu_k})\Big|
&\leq
\Big|e_{\Omega_k,\nu_k}^{(\ell)}-\mu_\ell(q)\Big|
+\Big|\mu_\ell(q)-\mu_\ell(q_{\Omega_k,\nu_k})\Big|
\longrightarrow0,
\end{aligned}
\]
contradicting the lower bound by \(\varepsilon\).
This proves \eqref{eq:bounded-q-eigenlimit}.
Suppose that \(q_{\om,\nu}\to0\).
The eigenvalue convergence in \eqref{eq:bounded-q-eigenlimit} shows that \eqref{eq:q-zero-low-mode-conv} holds true. 

Finally, we prove the eigenfunction convergences \eqref{eq:q-zero-low1} and \eqref{eq:oct30}.
For any sequence \(\Omega_k\to\infty\) with
\(q_{\Omega_k,\nu_k}\to0\), set
\(\psi_k^{(m)}:=\psi_{\Omega_k,\nu_k}^{(m)}\).
Applying the same compactness argument used to prove
\eqref{eq:bounded-q-eigenvalue-liminf} with \(q=0\) and \(\ell=2\),  we may pass to a subsequence such that as $k\to\infty$,
\[
\psi_k^{(m)} \xrightarrow[k \to \infty]{\mathrm{w}-H^1_{\mathrm{loc}}(\mathbb R)} u_m,
\qquad
\psi_k^{(m)} \xrightarrow[k \to \infty]{L^2_{\mathrm{loc}}(\mathbb R)} u_m,
\]
where
\[
 \braketr{u_m}{u_n}_{L^2(\mathbb R)}=\delta_{m,n},
\qquad
h_0[u_m]\leq e_{m-1},
\qquad m,n=1,2.
\]
The first variational principle and the simplicity of \(e_0\) give
\(u_1=\pm\chi_0\).  Since \(u_2\perp u_1\), the variational principle
on \(\{\chi_0\}^\perp\) and the simplicity of \(e_1\) give
\(u_2=\pm\chi_1\).  The sign convention in the lemma selects
\(u_1=\chi_0\) and \(u_2=\chi_1\).  The local convergence,
\eqref{eq:bounded-q-mass-tightness}, and the Gaussian decay of
\(\chi_0,\chi_1\) give
\[
\lf\|\psi_k^{(1)}-\chi_0 \ri\|_{\domain_{\Omega_k}}
+\lf\|\psi_k^{(2)}-\chi_1 \ri\|_{\domain_{\Omega_k}}
\xrightarrow[R \to \infty]{} 0.
\]
Since every subsequence has a further subsequence with the same limit,
the subsequence principle proves \eqref{eq:q-zero-low1} for the full
sequence.

It remains to prove  \eqref{eq:oct30}.  Fix a cutoff
\(\eta\in C_c^\infty(\mathbb R)\) to localize the eigenvalue equation
to a fixed compact interval.  Thus, as $\om\to\infty$,  
\begin{equation}\label{eq:oct31}
\rho_\om\to1 \quad\hbox{and}\quad
\Big(
\frac{q_{\om,\nu}-2t-t^2/\sqrt\om}{\rho_\om}
\Big)^2
+\frac18\Big(2t+\frac{t^2}{\sqrt\om}\Big)^2
\to\frac92t^2\quad\hbox{uniformly in}\,\  \operatorname{supp}\eta.
\end{equation}
Testing the eigenvalue equation with
\(\eta^2\psi_{\om,\nu}^{(m)}\) gives
\[
h_{\om,\nu}[\eta\psi_{\om,\nu}^{(m)}]
=e_{\om,\nu}^{(m)}
\|\eta\psi_{\om,\nu}^{(m)}\|_{\domain_\om}^2
+\int_{I_\om}|\eta'|^2
|\psi_{\om,\nu}^{(m)}|^2\rho_\om\,dt.
\]
The \(L^2\) convergence proved above, together with
\(e_{\om,\nu}^{(m)}\to e_{m-1}\) and
\(H_0\chi_{m-1}=e_{m-1}\chi_{m-1}\), shows that the right-hand side
converges to \(h_0[\eta\chi_{m-1}]\).  Combining the energy convergence, \eqref{eq:oct31} and the local weak
\(H^1\) convergence, 
we obtain $(\eta\psi_{\om,\nu}^{(m)})'
\longrightarrow(\eta\chi_{m-1})'$ strongly in $L^2(\R)$. Thus we obtain from \eqref{eq:q-zero-low1} that
\[
\eta\psi_{\om,\nu}^{(m)}
\xrightarrow[\om \to \infty]{H^1(\mathbb R)} \eta\chi_{m-1}.
\]
Choosing \(\eta=1\) on \([-R,R]\) proves \eqref{eq:oct30}
and completes the proof.
\end{proof}

\section{The Sharp Magnetic Interpolation Inequality on the Circle}
\label{app:sharp-circle-interpolation}

\begin{proof}[\normalfont\bfseries Proof of Lemma~\ref{lem:sharp-circle-interpolation}]
By replacing \(h\) with \(\overline h\) if necessary, it suffices to consider
\(\alpha\in[0,1/2)\).  Put
\[
\rho:=|h|,\qquad
A:=\int_{S^1}|\rho'|^2\,d\theta,\qquad
D:=\int_{S^1}\rho^4\,d\theta-\frac1{2\pi}.
\]
We first claim that
\begin{equation}\label{eq:appendix-nonmagnetic-sharp}
	D\leq\frac2\pi A.
\end{equation}

Indeed, set
\(q:=\rho^2-(2\pi)^{-1}\).  Since \(\|\rho\|_{L^2(S^1)}=1\), the function \(q\)
has zero average, and
\begin{equation}\label{eq:density-square-identity}
	\int_{S^1}\rho^4\,d\theta-\frac1{2\pi}
	=
	\int_{S^1}q^2\,d\theta .
\end{equation}
Let \(\phi\) be the zero mean periodic solution of
\(-\phi''=q\) on \(S^1\).  By \eqref{eq:density-square-identity},
\(D=\int_{S^1}q^2\,d\theta\).
Using the zero mean condition on \(q\), integration by parts, and
Cauchy--Schwarz, we obtain
\[
D
=
\int_{S^1}q\rho^2\,d\theta
=
2\int_{S^1}\phi'\rho\rho'\,d\theta,
\qquad
D^2
\le
4A\int_{S^1}\rho^2|\phi'|^2\,d\theta .
\]
Since \(\rho^2=(2\pi)^{-1}-\phi''\), periodicity yields
\[
	\int_{S^1}\rho^2|\phi'|^2\,d\theta
	=
	\frac1{2\pi}\int_{S^1}|\phi'|^2\,d\theta,
	\qquad
	\int_{S^1}\phi''|\phi'|^2\,d\theta
	=
	\frac13\int_{S^1}\bigl((\phi')^3\bigr)'\,d\theta
	=0.
\]
The Poincar\'e inequality for the zero mean function \(\phi'\) now gives
\[
	\int_{S^1}|\phi'|^2\,d\theta
	\le
	\int_{S^1}|\phi''|^2\,d\theta
	=D.
\]
This proves \eqref{eq:appendix-nonmagnetic-sharp}.

Next, with the convention \(S=+\infty\) if \(\rho^{-1}\notin L^2(S^1)\),
set
$$S:=\int_{S^1}\rho^{-2}\,d\theta.$$
We claim that
\begin{equation}\label{eq:appendix-half-flux-bound}
	A\geq1/4-\pi^2/S.
\end{equation}
Recall that $A=\int_{S^1}|\rho'|^2\,d\theta$. If \(S<\infty\), define a periodic phase \(\psi\) by
\[
\psi'=\frac12-\frac{\pi}{S\rho^2}.
\]
Indeed, \(\int_{S^1}\psi'=0\).  The spectral lower bound for
\((-i\partial_\theta-\frac12)^2\) gives
\[
\frac14
\leq
\int_{S^1}
\Big|(-i\partial_\theta-\tfrac12)(\rho e^{i\psi})\Big|^2\,d\theta
=A+\frac{\pi^2}{S},
\]
which proves \eqref{eq:appendix-half-flux-bound}.  If \(S=+\infty\), then
\(\rho\) has a zero.  Cutting the circle at such a zero and applying the
Dirichlet Poincar\'e inequality on an interval of length \(2\pi\) gives
\(A\geq1/4=1/4-4\pi^2/S\). This proves the claim \eqref{eq:appendix-half-flux-bound}.

We next prove 
\begin{equation}\label{eq:appendix-phase-lower-bound}
\int_{S^1}|(-i\partial_\theta-\alpha)h|^2\,d\theta-\alpha^2
\geq A-\alpha^2[1-4\pi^2/S].
\end{equation}
If \(\rho>0\) on $S^1$, write
\(h=\rho e^{i\varphi}\), where
\(\varphi(2\pi)-\varphi(0)=2\pi k\) for some \(k\in\mathbb Z\). Then
\[
\int_{S^1}|(-i\partial_\theta-\alpha)h|^2\,d\theta
=\int_{S^1}\big[|\rho'|^2+\rho^2(\varphi'-\alpha)^2\big]\,d\theta=A+\int_{S^1}\big[\rho^2(\varphi'-\alpha)^2\big]\,d\theta.
\]
Weighted Cauchy--Schwarz and \(0\leq\alpha<1/2\) give
\[
\int_{S^1}\rho^2(\varphi'-\alpha)^2\,d\theta
\geq
\frac{4\pi^2(k-\alpha)^2}{S}
\geq\frac{4\pi^2\alpha^2}{S}.
\]
This proves \eqref{eq:appendix-phase-lower-bound} in the case where $\rho>0$ on $S^1$.
If \(\rho(\theta_0)=0\) for some \(\theta_0\in S^1\), then,
in a neighborhood of $\theta_0$, 
\[
\rho(\theta)^2
=\left|\int_{\theta_0}^{\theta}\rho'(s)\,ds\right|^2
\leq A|\theta-\theta_0|.
\]
Hence \(\rho^{-2}\) is not integrable near \(\theta_0\),
so \(S=+\infty\). Moreover, the diamagnetic inequality \eqref{eq:diamagnetic} gives
\[
\int_{S^1}|(-i\partial_\theta-\alpha)h|^2\,d\theta
\geq \int_{S^1}|\rho'|^2\,d\theta
=A.
\]
Thus \eqref{eq:appendix-phase-lower-bound} also holds when
\(\rho\) has a zero on \(S^1\).
This proves the claim \eqref{eq:appendix-phase-lower-bound}.

Based on \eqref{eq:appendix-nonmagnetic-sharp}, 
\eqref{eq:appendix-half-flux-bound} and \eqref{eq:appendix-phase-lower-bound}, we now prove \eqref{eq:scis}. By the definition of \(D\) and
\eqref{eq:appendix-phase-lower-bound}, to prove
\eqref{eq:scis} it suffices to show that
\[
A-\alpha^2\big(1-4\pi^2/S\big)
\geq \frac{\pi}{2}(1-4\alpha^2)D.
\]
Indeed,
\[
\begin{aligned}
A-\alpha^2[1-4\pi^2/S]-\frac{\pi}{2}(1-4\alpha^2)D
={}&
(1-4\alpha^2)\Big(A-\frac{\pi}{2}D\Big)\\
&+4\alpha^2\big(A-[1/4-\pi^2/S]\big)
\end{aligned}
\]
is nonnegative by \eqref{eq:appendix-nonmagnetic-sharp} and
\eqref{eq:appendix-half-flux-bound}.  This proves \eqref{eq:scis}.

To see that the constant in \eqref{eq:scis} is sharp, take
\[
\eta_\alpha(\theta):=
\frac{(1+2\alpha)e^{i\theta}+(1-2\alpha)e^{-i\theta}}
{2\sqrt{2\pi}},
\qquad
h_\varepsilon:=
\frac{h_0+\varepsilon\eta_\alpha}
{\sqrt{1+\varepsilon^2\|\eta_\alpha\|_{L^2(S^1)}^2}} .
\]
Then
\[
\int_{S^1}|(-i\partial_\theta-\alpha)h_\varepsilon|^2\,d\theta-\alpha^2
=\frac{1-4\alpha^2}{2}\varepsilon^2+O(\varepsilon^4),
\]
whereas
\[
\int_{S^1}|h_\varepsilon|^4\,d\theta-\frac1{2\pi}
=\frac1\pi\varepsilon^2+O(\varepsilon^4).
\]
Thus the ratio of the left-hand side of \eqref{eq:scis} to the magnetic
energy in parentheses converges to \(2/[\pi(1-4\alpha^2)]\) as $\varepsilon\to 0$.
\end{proof}

\section{An \(L^4\) Estimate in the Lowest Eigenspace}
\label{app:first-band-L4}

\begin{lemma}\label{lem:first-band-L4}
For all sufficiently large \(\Omega\), let \(A\subset\mathcal J_\Omega=\{j\in\mathbb Z:|j-\alpha_\om|\leq \sqrt{\om}\}\),
and \(d_j\in\mathbb C\) for \(j\in A\).  Set
\[
M:=\sum_{j\in A}|d_j|^2,
\qquad
K:=\sum_{j\in A}(1+|j-\alpha_\om|^2)|d_j|^2,
\]
and define
\[
v_A(r,\theta)
:=
\frac1{\sqrt{2\pi}}
\sum_{j\in A}d_j\phi_{\om,n_\om+j}^{(1)}(r)
e^{i(n_\om+j)\theta},
\]
where \(\phi_{\Omega,n_\Omega+j}^{(1)}\) is the normalized
first eigenfunction of \(\mathcal L_{\Omega,n_\Omega+j}\),
defined in \eqref{eq:continuous-radial-operator}.
Then
\begin{equation}\label{eq:P-in-precoeff-new}
\|v_A\|_4^4
\leq C\sqrt\om\,M^{3/2}K^{1/2},
\end{equation}
where \(C>0\) is independent of \(\Omega\), \(A\), and \((d_j)_{j\in A}\).
\end{lemma}

\begin{proof}[\normalfont\bfseries Proof]
For every finitely supported sequence \((b_j)_{j\in\mathbb Z}\),
the Hausdorff--Young inequality on \(S^1\), with conjugate exponents \(4/3\) and \(4\), gives
\begin{equation}\label{eq:circle-Hausdorff-Young-new}
\int_0^{2\pi}
\Big|
\frac1{\sqrt{2\pi}}\sum_{j\in\mathbb Z}b_je^{ij\theta}
\Big|^4\,d\theta
\leq
C\Big(\sum_{j\in\mathbb Z}|b_j|^{4/3}\Big)^3.
\end{equation}
After factoring out \(e^{in_\Omega\theta}\), we apply
\eqref{eq:circle-Hausdorff-Young-new} for each fixed \(r\), with
\(b_j=d_j\phi_{\Omega,n_\Omega+j}^{(1)}(r)\) for \(j\in A\)
and \(b_j=0\) otherwise.  Minkowski's inequality in \(r\) and
\eqref{eq:radial-L4-bound-new} then give
\[
\begin{aligned}
\|v_A\|_4^4
&\leq
C\Big(
\sum_{j\in A}|d_j|^{4/3}
\Big(
\int_0^\infty
|\phi_{\om,n_\om+j}^{(1)}(r)|^4r\,dr
\Big)^{1/3}
\Big)^3\leq
C\sqrt\om
\Big(\sum_{j\in A}|d_j|^{4/3}\Big)^3.
\end{aligned}
\]
It remains to prove
\begin{equation}\label{eq:discrete-Sobolev-new}
\Big(\sum_{j\in A}|d_j|^{4/3}\Big)^3
\leq CM^{3/2}K^{1/2}.
\end{equation}
If \(M=0\), then every \(d_j=0\), so \eqref{eq:discrete-Sobolev-new} holds.
For \(M>0\), put \(R=(K/M)^{1/2}\geq1\).
Since \(\{|j-\alpha_\Omega|\leq R\}\) contains at most \(CR\) integers,
H\"older's inequality gives
\begin{equation}\label{eq:oct32}
\begin{split}
\sum_{\substack{j\in A\\|j-\alpha_\Omega|\leq R}}
|d_j|^{4/3}
&\leq
\Big(\sum_{j\in A}|d_j|^2\Big)^{2/3}
\Big(\#\{j\in\mathbb Z:|j-\alpha_\Omega|\leq R\}\Big)^{1/3}\\
&\leq CM^{2/3}R^{1/3}=CM^{1/2}K^{1/6}.
\end{split}
\end{equation}
For the complementary sum, 
H\"older's inequality with exponents \(3/2\) and \(3\) yields
\begin{equation}\label{eq:oct33}
\begin{split}
\sum_{\substack{j\in A\\|j-\alpha_\Omega|>R}}
|d_j|^{4/3}
&\leq
\Big(
\sum_{j\in A}|j-\alpha_\Omega|^2|d_j|^2
\Big)^{2/3}
\Big(
\sum_{|j-\alpha_\Omega|>R}|j-\alpha_\Omega|^{-4}
\Big)^{1/3}\\
&\leq CK^{2/3}R^{-1}=CM^{1/2}K^{1/6},
\end{split}
\end{equation}
where the last inequality follows from the fact that
\(
\sum_{|j-\alpha_\Omega|>R}|j-\alpha_\Omega|^{-4}
\leq CR^{-3}.
\)
Adding the two bounds in \eqref{eq:oct32}, \eqref{eq:oct33}, and then cubing the resulting inequality, gives \eqref{eq:discrete-Sobolev-new},
and hence \eqref{eq:P-in-precoeff-new}.
\end{proof}

\end{document}